\documentclass[acmlarge]{acmart}
\AtBeginDocument{%
  }

\setcopyright{acmlicensed}
\copyrightyear{2018}
\acmYear{2018}
\acmDOI{XXXXXXX.XXXXXXX}
\usepackage{algorithm}     
\usepackage{algorithmicx}  
\usepackage{algpseudocode} 
\usepackage{makecell} 
\usepackage{xcolor} %
\usepackage{subcaption}
\usepackage{array}
\acmJournal{POMACS}
\acmVolume{37}
\acmNumber{4}
\acmArticle{111}
\acmMonth{8}

\begin{document}

        \title{Collaborative Scheduling of Time-dependent UAVs,Vehicles and Workers for Crowdsensing in Disaster Response}

\author{Lei Han}
\email{hanlei@xidian.edu.cn}
\affiliation{%
	\institution{Xidian University}
	\city{Xi'an}
	\country{China}
}

\author{Jinhao Zhang}
\email{z1028406171@163.com}
\affiliation{%
	\institution{Xidian University}
	\city{Xi'an}
	\country{China}
}

\author{Jinhui Liu}
\authornote{Corresponding author.}
\email{jhliu@mail.xidian.edu.cn}
\affiliation{%
	\institution{Xidian University}
	\city{Xi'an}
	\country{China}
}

\author{Zhiyong Yu}
\email{yuzhiyong@fzu.edu.cn}
\affiliation{%
	\institution{Fuzhou University}
	\city{Fuzhou}
	\country{China}
}

\author{Liang Wang}
\email{liangwang@nwpu.edu.cn}
\affiliation{%
	\institution{Northwestern Polytechnical University}
	\city{Xi'an}
	\country{China}
}

\author{Quan Wang}
\email{qwang@xidian.edu.cn}
\affiliation{%
	\institution{Xidian University}
	\city{Xi'an}
	\country{China}
}

\author{Zhiwen Yu}
\email{zhiwenyu@nwpu.edu.cn}
\affiliation{%
	\institution{Harbin Engineering University}
	\city{Harbin}
	\country{China}
}


\begin{abstract}
Frequent natural disasters cause significant losses to human society, and timely, efficient collection of post-disaster environmental information is the foundation for effective rescue operations. Due to the extreme complexity of post-disaster environments, existing sensing technologies such as mobile crowdsensing suffer from weak environmental adaptability, insufficient professional sensing capabilities, and poor practicality of sensing solutions. Therefore, this paper explores a heterogeneous multi-agent online collaborative scheduling algorithm, HoCs-MPQ, to achieve efficient collection of post-disaster environmental information. HoCs-MPQ models collaboration and conflict relationships among multiple elements through weighted undirected graph construction, and iteratively solves the maximum weight  independent set based on multi-priority queues, ultimately achieving collaborative sensing scheduling of time-dependent UAVs, vehicles, and workers. Specifically, (1) HoCs-MPQ constructs weighted undirected graph nodes based on collaborative relationships among multiple elements and quantifies their weights, then models the weighted undirected graph based on conflict relationships between nodes; (2) HoCs-MPQ solves the maximum weight independent set based on iterated local search, and accelerates the solution process using multi-priority queues. Finally, we conducted detailed experiments based on extensive real-world and simulated data. The experiments show that, compared to baseline methods (e.g., HoCs-GREEDY, HoCs-K-WTA, HoCs-MADL, and HoCs-MARL), HoCs-MPQ improves task completion rates by an average of 54.13\%, 23.82\%, 14.12\%, and 12.89\% respectively, with computation time for single online autonomous scheduling decisions not exceeding 3 seconds.

\end{abstract}

\keywords{Mobile crowdsensing, Multi-agent collaborative sensing, Collaborative sensing scheduling, Weighted undirected graph, Maximum weight independent set.}

\maketitle

\section{Introduction}

\hspace{1em}According to the 2023 Global Natural Disaster Assessment Report~\cite{ADREM2023}, there were 326 large-scale natural disasters globally in 2023, causing 86,473 deaths and direct economic losses of up to \$202.652 billion. To reduce the severe impact of natural disasters, it is crucial to organize post-disaster rescue operations efficiently and promptly. Accurate and real-time situational awareness is the foundation for effective rescue actions~\cite{McEntire2021}. 

Currently, mainstream situational awareness approaches include wireless sensor networks (WSNs)~\cite{GULATI2022161, SUN2025111081} and mobile crowdsensing (MCS)~\cite{10529209,KANG2025111189}. However, these technologies have limitations in post-disaster scenarios. WSNs rely on pre-deployed static sensors and lack adaptability in complex and dynamic environments. MCS uses people with portable smart devices as sensing units. Although it is more adaptable and requires less infrastructure, human mobility is limited in disaster settings, and portable devices lack sensing professionalism.

In recent years, UAVs have played an increasingly important role in disaster response. UAVs are highly mobile, easy to deploy, and capable of carrying professional sensors~\cite{zhou2018mobile,liu2020curiosity,wang2022h,Wang_2024}. However, current studies on UAV-based data collection often rely on two overly optimistic assumptions: (1) UAVs can autonomously collect data; however, the complex post-disaster environment severely constrains the UAVs' mobility, which  drastically impairs their sensing capabilities.(2) UAVs can automatically recharge at charging stations. In reality, such stations are rare in cities and may be damaged after disasters. 
It is also difficult for UAVs to connect to power without assistance in outdoor environments.

\begin{figure}[htbp]
    \centering
    \setlength{\abovecaptionskip}{0.2cm}
    \setlength{\belowcaptionskip}{-0.25cm}
    \begin{subfigure}[b]{0.48\linewidth} 
        \centering
    \includegraphics[height=5.5cm,keepaspectratio]{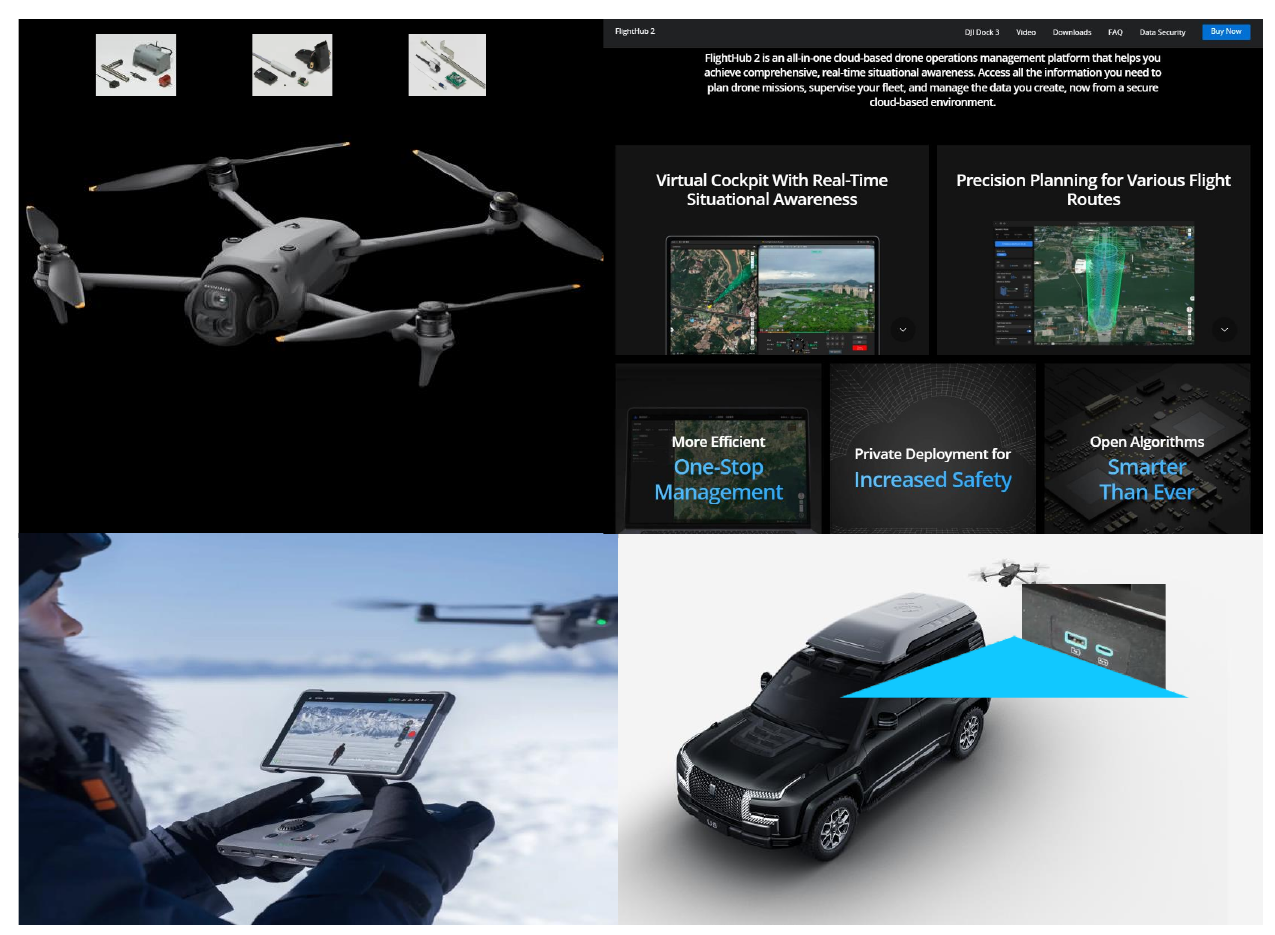} 
        \caption{Illustration of UAVs, vehicles, and workers with a takeover platform}
        \label{realUse}
    \end{subfigure}
    \hfill 
    \begin{subfigure}[b]{0.48\linewidth} 
        \centering
        \includegraphics[height=5.5cm,keepaspectratio]{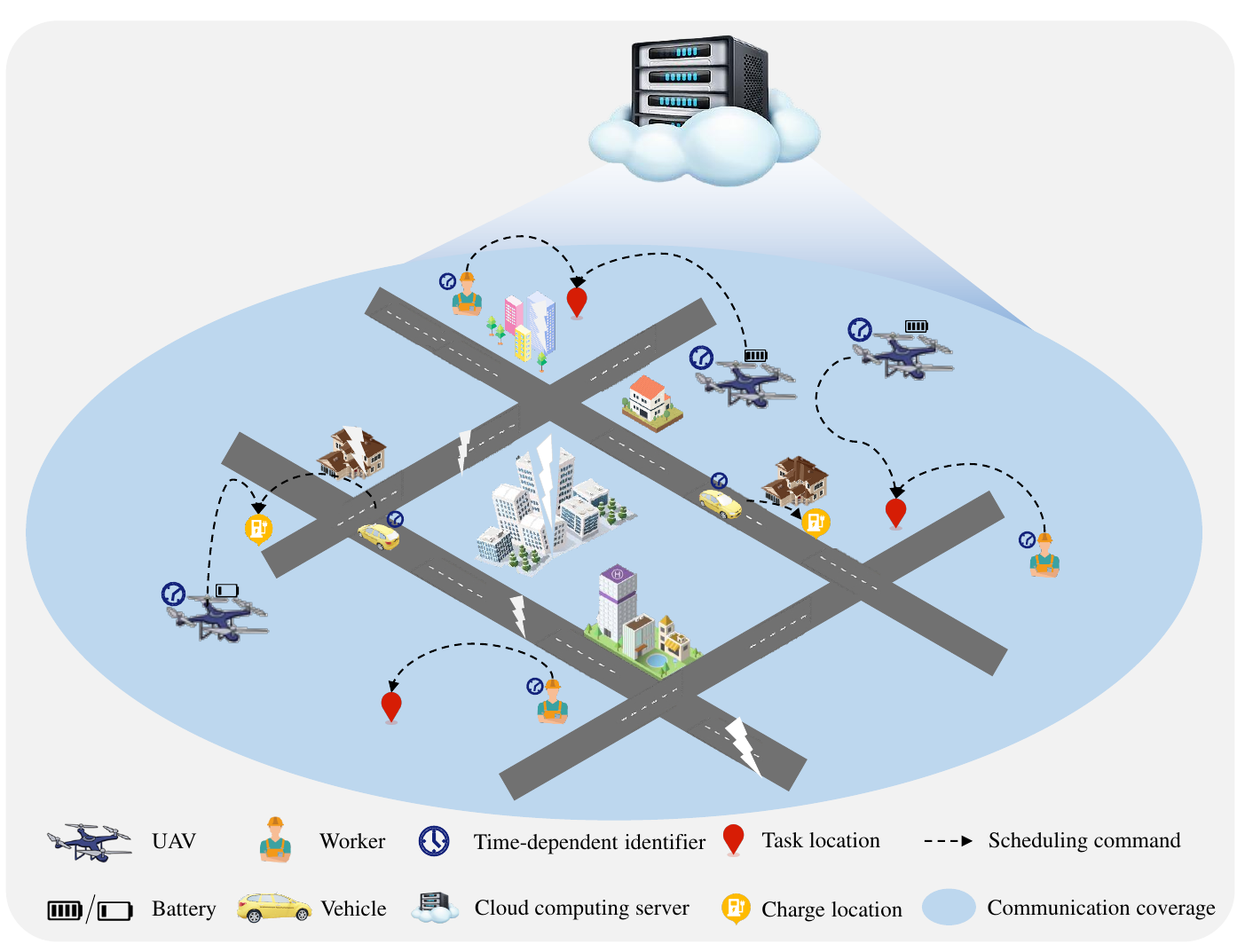} 
        \caption{Example of collaboration among time-dependent UAVs, vehicles, and workers.}
        \label{figure1}
    \end{subfigure}
    \caption{Collaboration Among Time-Dependent UAVs, Vehicles, and Workers.}
\end{figure}

In recent years, with the widespread adoption and application of UAVs, new energy vehicles, and portable smart terminals, their complementary advantages have offered potential solutions to the aforementioned problems. Specifically, as shown in Fig. ~\ref{realUse}, UAVs possess controlled autonomy, superior mobility, and the flexibility to carry professional sensing equipment\cite{mohsan2022towards}. Combined with the charging and endurance capabilities of new energy vehicles \cite{wang2023air} and the agile control capabilities of portable smart terminals \cite{herdel2022above}, and further aided by UAV flight control takeover platforms \cite{dji_flighthub2}, we can effectively achieve efficient post-disaster environmental data collection\cite{Guo2022HCPS}.Building upon this collaborative approach, a recent study~\cite{10.1109/TNET.2024.3395493} leverages the heterogeneity and collaboration of UAVs, vehicles, and workers to collect environmental data efficiently. As shown in Fig. ~\ref{figure1}, workers can use portable devices to control UAVs for agile data collection. Vehicles can act as mobile charging stations for UAVs with low power. However, this study still has two limitations: (1) Since natural disasters are highly unpredictable, it is hard to model training environments. Learning-based solutions are difficult to apply in real post-disaster scenarios. (2) The proposed solution assumes fixed coupling among UAVs, vehicles, and workers, while in reality, their relationships are time-dependent.

This paper studies the collaborative scheduling of time-dependent UAVs, vehicles, and workers for crowdsensing in disaster response. We have two goals: (1) Propose an online heterogeneous multi-agent collaborative scheduling algorithm that does not require modeling training environments. (2) Effectively schedule time-dependent UAVs, vehicles, and workers to complete post-disaster data collection tasks efficiently.
To realize the aforementioned goals, we face two main challenges: (1) The five-dimensional matching problem involving UAVs, vehicles, workers, task points, and charging points is very complex. It is a challenge to model the problem accurately to reflect the complex logical relationships. (2) Solving the five-dimensional matching problem requires high computation and must consider resource and time constraints to ensure timely decisions in dynamic environments.

To address these challenges, we propose a heterogeneous multi-agent online collaborative scheduling algorithm based on weighted undirected graph modeling and maximum weight independent set solution, named HoCs-MPQ. (1) We model the five-dimensional matching problem as a maximum weight independent set (MWIS) problem on a weighted undirected graph. The nodes and edges represent collaborative and conflictual relationships among agents (e.g., UAVs, vehicles, and workers). By solving the MWIS, we maximize the collaboration among UAVs, vehicles, and workers. (2) As the graph is dense, we use multiple queues to store the core nodes, reducing the size of the graph while retaining key information. We then iteratively read nodes from the queues to optimize the MWIS solution and shorten computation time. 

Our main contributions are as follows:

\begin{itemize}
  \item To the best of our knowledge, this work is the first research addressing the collaborative scheduling of time-dependent UAVs, vehicles, and workers for crowdsensing in disaster response. We also prove that the problem is NP-Hard.
  \item To tackle the aforementioned issues, we propose an online heterogeneous multi-agent collaborative scheduling algorithm, named HoCs-MPQ. (1) HoCs-MPQ models collaborative and conflictual relationships among agents through a weighted undirected graph. (2) HoCs-MPQ uses multi-priority queues to speed up the solution of the maximum weight independent set as the result of heterogeneous multi-agent collaborative scheduling.
  \item We conducted detailed experiments using real-world and simulated data. Compared with the baseline algorithms HoCs-GREEDY, HoCs-K-WTA, HoCs-MADL, and HoCs-MARL, HoCs-MPQ improves the task completion rate by 54.13\%, 23.82\%, 14.12\%, and 12.89\%, respectively. The single matching decision time does not exceed \texttt{3} seconds on any dataset. The experimental code and data examples for this study can be referenced in \cite{CocaColaZero}.
\end{itemize}

The contents of this paper are arranged as follows: Section~\ref{section2} discusses some related works; Section~\ref{section3} formulates the problem of collaborative scheduling of time-dependent UAVs, vehicles and workers for crowdsensing in disaster response; Section~\ref{section4} provides a comprehensive and detailed description of the proposed HoCs-MPQ method; then the experiments are conducted in Section~\ref{section5}; finally, conclusions and future work are summarized in Section~\ref{section6}. 
\section{Related Work} 
\label{section2}

\hspace{1em}The properties and functions of heterogeneous agents such as UAVs, vehicles, and workers vary significantly. Their collaboration modes are also diverse. Existing studies on heterogeneous multi-agent collaboration can be classified into three types: (1) Heterogeneous multi-agent collaborative sensing based on spatial division; (2) Heterogeneous multi-agent collaborative sensing based on task-phase division; (3) Heterogeneous multi-agent collaborative sensing based on fine-grained coordination.

\subsection{Heterogeneous Multi-Agent Collaborative Sensing Based on Spatial Division} 
\hspace{1em}In this collaborative sensing mode, heterogeneous agents collect data at different spatial locations according to their mobility differences. This collaboration can be divided into three types.First, the most common type is the collaboration between UAVs and vehicles based on spatial separation. For example, Wu et al.~\cite{wu2020cooperative} planned joint paths for UAVs and vehicles, aiming to minimize global sensing time through high-altitude and ground-level deployment.In the field of navigation, Zuo et al.~\cite{zuo2023real} used UAVs to collect ground information from the air to assist vehicle navigation in complex environments.Zang et al.~\cite{zang2024coordinated} aimed to reduce collision risks among large-scale vehicle fleets by using UAVs to collect ground-level information from high altitudes.In terms of collaborative mapping between UAVs and vehicles, some works used UAVs to provide large-scale aerial data while vehicles filled in ground-level details~\cite{christie2016radiationsearchoperationsusing, michael2014collaborative, kruijff2014designing, kalaitzakis2021marsupial}.Second, some studies focus on UAV/vehicle and human collaboration based on spatial separation. For example, Ding et al.~\cite{ding2021crowdsourcing} constructed a low-cost urban sensing system by leveraging the spatial distribution differences between human crowds and autonomous vehicles.Zheng et al.~\cite{zheng2019evolutionary} studied how UAVs and humans could search different areas simultaneously to efficiently locate fugitives.In emergency response scenarios, Minaeian et al.~\cite{minaeian2015vision} explored how UAVs at high altitudes could assist ground patrols by providing real-time feedback on suspicious targets.Finally, a few works investigate the spatial collaboration among UAVs, vehicles, and humans. For example, Wu et al.~\cite{wu2022task} dispatched small UAVs, intelligent connected vehicles, and human groups to different areas to search for multiple ground targets.

\subsection{Heterogeneous Multi-Agent Collaborative Sensing Based on Task-phase Division} 
\hspace{1em}In this collaborative sensing mode, the task is divided into multiple sub-tasks at different stages, and heterogeneous agents complete each sub-task in sequence according to their functional characteristics.First, the most common mode is the collaboration between UAVs and vehicles in task-phase division. For example, studies~\cite{russell2016artificial, plocher2017german} used UAVs to perform fast area exploration and mark points of interest, and then used unmanned vehicles to conduct precise target detection.Wang et al.~\cite{wang2023air} focused on how to use vehicles as UAV carriers to extend their range, in order to perform long-duration monitoring tasks.Based on the strong transport capacity of trucks and the precise delivery ability of UAVs, Luo et al.~\cite{luo2024collaborative} realized a two-stage delivery system using trucks and UAVs.Similarly, Fu et al.~\cite{fu2022energy} used UAVs to collect data and ground vehicles to transmit the collected data, aiming to balance high-efficiency data collection and processing.Second, some studies focus on UAV/vehicle and worker collaboration in task-phase division. For example, Niroui et al.~\cite{niroui2019deep} used UAVs to capture aerial images of disaster areas and transmit them in real time to human-operated devices. Then, humans marked key targets using augmented reality (AR).In the field of construction engineering, Krizmancic et al.~\cite{krizmancic2020cooperative} used UAVs to perform aerial scanning tasks, while humans conducted semantic analysis and material transport.In power inspection scenarios, Zheng et al.~\cite{zheng2020evolutionary} used UAVs to detect transmission faults and inform operators for repair of power facilities.To enable rapid search for missing tourists, Xu et al.~\cite{xu2024iterated} proposed a two-stage model where UAVs conduct detection and rescue workers carry out search and rescue, greatly improving the efficiency of the mission.Finally, some studies involve three-agent collaboration across phases. Ghassemi et al.~\cite{ghassemi2019decentralized} proposed a three-stage model: UAVs scan the area, vehicles deliver supplies, and workers coordinate operations.

\subsection{Heterogeneous Multi-Agent Collaborative Sensing Based on Fine-grained Coordination}
\hspace{1em}In this collaborative sensing mode, heterogeneous agents achieve fine-grained cooperation on specific tasks by leveraging their diverse capabilities and synergies.First, the most common fine-grained collaboration mode is between UAVs and vehicles. Hu et al.\cite{hu2023collaboration} proposed a 3D cooperative detection framework, CoCa3D, which enables real-time sharing of critical data between UAVs and vehicles through deep uncertainty estimation. Wang et al.\cite{wang2022multi} jointly optimized the trajectories of UAVs and vehicles to maintain line-of-sight tracking, addressing the problem of UAV-based monitoring being blocked by obstacles in urban environments. Zhao et al.~\cite{zhao2024energy} proposed a novel goal-oriented MADRL framework named gMADRL-VCS, where UGVs act as UAV carriers for road transport and charging, while UAVs navigate to collect post-disaster data. The system simultaneously learns the routing policy of UAVs and the navigation strategy of UGVs.Second, some studies explore fine-grained collaboration between UAVs/vehicles and humans. To address the challenge of heterogeneous multi-agent sensing caused by communication redundancy and transmission delay, one study~\cite{yang2023how2comm} used UAV-based semantic communication to transmit high-quality feature information to humans, supporting real-time collaborative decision-making. In addition,  one study~\cite{manjunatha2020using} innovatively introduced brain-computer interfaces into air-ground swarm control, enabling dynamic mapping between human intentions and UAV behaviors.Finally, a few works investigate fine-grained collaboration among UAVs, vehicles, and humans. Miller et al.\cite{miller2022stronger} designed an LLM-driven air-ground collaborative system based on the efficient reasoning ability of LLMs. It dynamically adjusts task priorities through natural language interaction between UAVs, UGVs, and humans. In terms of data collection, Han et al.\cite{10.1109/TNET.2024.3395493} explored a collaborative path planning method involving UAVs, vehicles, and human groups to achieve fine-grained coordination in post-disaster response scenarios.

\subsection{Issues of Existing Research} 
\hspace{1em}The heterogeneous multi-agent collaboration model that rely on spatial division or task-phase division generally exhibit limited coordination. The former typically assigns agents to different spatial regions based on their mobility capabilities, while the latter often decomposes a large task into sequential subtasks executed by different agents. Fine-grained collaboration better utilizes the strengths of each agent type and enables complementary cooperation. However, reasoning capabilities of LLMs are limited for highly complex tasks. Learning-based methods need large training datasets, which are hard to collect in disaster scenarios. Moreover, learning models are sensitive to agent coupling. Changes in agent numbers or status can significantly affect performance~\cite{li2024reinforcement}. Therefore, further research is needed on collaborative scheduling of time-dependent UAVs, vehicles, and workers in disaster sensing. 
\section{Problem Formulation}
\label{section3}
\hspace{1em}In this section, we define the basic concepts, analyze the constraints of the studied problem, and present the optimization objective.

\textbf{{\itshape Definition 1.}} UAV set $UAV = \{u_0, \ldots, u_i, \ldots\}$, where $u_i = \langle {uLoc}_i, {uRge}_i, {Fullpower}_i, {uPow}_i, {U\_uptime}_i, \\{U\_downtime}_i \rangle$ denotes the $i$-th UAV in $UAV$. ${uLoc}_i$ represents the current location of $u_i$; ${uRge}_i$ denotes the moving speed of $u_i$; ${Fullpower}_i$ is the maximum distance $u_i$ can move with full power; ${uPow}_i \in [0, {Fullpower}_i]$ indicates the current movable distance of $u_i$; ${U\_uptime}_i$ and ${U\_downtime}_i$ represent the online and offline times of $u_i$, modeling its time-dependent behavior. UAVs refer to unmanned aerial vehicles collected from the disaster scenario that are capable of performing sensing tasks.

\textbf{{\itshape Definition 2.}} Worker set $Worker = \{w_0, \ldots, w_j, \ldots\}$, where $w_j = \langle {wLoc}_j, {wRge}_j, {W\_uptime}_j, {W\_downtime}_j \rangle$ denotes the $j$-th worker in $Worker$. ${wLoc}_j$ is the current location of $w_j$; ${wRge}_j$ denotes the moving speed of $w_j$; ${W\_uptime}_j$ and ${W\_downtime}_j$ represent the online and offline times of $w_j$, modeling its time-dependent behavior. Workers are personnel in the disaster scenario who are skilled in UAV operation and collaboratively control UAVs to complete sensing tasks.

\textbf{{\itshape Definition 3.}} Vehicle set $Vehicle = \{v_0, \ldots, v_k, \ldots\}$, where $v_k = \langle {vLoc}_k, {vRge}_k, {V\_chargePow}_k, {V\_uptime}_k,  \\{V\_downtime}_k \rangle$ denotes the $k$-th vehicle in $Vehicle$. ${vLoc}_k$ is the current location of $v_k$; ${vRge}_k$ denotes the moving speed of $v_k$; ${V\_chargePow}_k$ represents the charging power provided to UAVs per unit time; ${V\_uptime}_k$ and ${V\_downtime}_k$ indicate the online and offline times of $v_k$, modeling its time-dependent behavior. Vehicles refer to electric vehicles collected from the disaster scenario that can provide charging for UAVs. The drivers can manually connect UAVs to the power supply based on the electric vehicles’ energy resources.

\textbf{{\itshape Definition 4.}} Sensing task set $Task = \{{\rm task}_0, \ldots, {\rm task}_x, \ldots\}$, where ${\rm task}_x = \langle {tLoc}_x, {T\_costPow}_x \rangle$ denotes the $x$-th task in $Task$. ${tLoc}_x$ represents the location of ${\rm task}_x$; ${T\_costPow}_x$ is the distance that a UAV needs to move in order to complete ${\rm task}_x$. Tasks are locations in the disaster scenario where workers, leveraging their proficiency in precisely maneuvering UAVs, operate them to complete sensing tasks.

\textbf{{\itshape Definition 5.}} Charging point set $Charge = \{{\rm charge}_0, \ldots, {\rm charge}_y, \ldots\}$, where ${\rm charge}_y = \langle {chLoc}_y \rangle$ denotes the $y$-th charging point in $Charge$. ${chLoc}_y$ represents the location of ${\rm charge}_y$. Charging points are open areas in the disaster scenario where UAVs can be recharged by vehicles under stationary and safe operation conditions. When UAVs and vehicles meet at a charging point, the driver can assist in connecting the UAVs to recharge.

Before defining the problem of this paper, we need to introduce the following constraints.
\begin{itemize}
\item[(1)] When UAV $u_i$ and worker $w_j$ meet at location ${tLoc}_x$, the task ${\rm task}_x$ can be completed, and ${uPow}_i$ decreases by ${T\_costPow}_x$:
\begin{equation}
{\rm task}_x = \emptyset \, \& \, {uPow}_i = {uPow}_i - {T\_costPow}_x,  \text{if } {uLoc}_i = {wLoc}_j = {tLoc}_x
\end{equation}
\item[(2)] When $u_i$ and $w_j$ perform task ${\rm task}_x$, the time consumed by both of them is ${T\_costPow}_x / {uRge}_i$.
\item[(3)] When $u_i$ and $v_k$ meet at charging point ${\rm charge}_y$, $v_k$ can charge $u_i$ until it is fully charged:
\begin{equation}
{uPow}_i = {Fullpower}_i,  \text{if } {uLoc}_i = {vLoc}_k = {chLoc}_y
\end{equation}
\item[(4)] When $v_k$ charges $u_i$ at ${\rm charge}_y$, the time cost for both is $({Fullpower}_i - {uPow}_i)/{V\_chargePow}_k$. Note: If multiple UAVs need to be charged, $v_k$ follows a first-come-first-served principle.
\item[(5)] When $u_i$ chooses charging point ${\rm charge}_y$, it must ensure:
\begin{equation}
Dis({chLoc}_y, {uLoc}_i) \leq {uPow}_i
\end{equation}
\item[(6)] When $u_i$ selects to execute ${\rm task}_x$, it must ensure:
\begin{equation}
Dis({uLoc}_i, {tLoc}_x) + {T\_costPow}_x + Dis({tLoc}_x, {Opt\_charge}_y) \leq {uPow}_i,  {Opt\_charge}_y \in Charge
\end{equation}
\end{itemize}

\textbf{Problem 1 }(Collaborative Scheduling of Time-dependent UAVs, Vehicles and Workers for Crowdsensing in Disaster Response): Given UAV set $UAV$, Worker set $Worker$, Vehicle set $Vehicle$, Task set $Task$, and Charging point set $Charge$, determine the scheduling actions $\{{{A\_u}_i^{t \times Interval}}\}, \{{{A\_w}_j^{t \times Interval}}\}, \{{{A\_v}_k^{t \times Interval}}\}$ for $t=0,1,2,\ldots$ and $t \times Interval \in [0, LimitTime]$, so as to maximize the number of completed sensing tasks $Cplt\_Tasks$:
\begin{small}
\begin{equation}
\begin{array}{l}
{\rm{\textbf{confirm}}}
\
\{ {A\_u}_i^{t \times Interval} \},
\{ {A\_w}_j^{t \times Interval} \},
\{ {A\_v}_k^{t \times Interval} \}, 
(t = 0, 1, 2, \ldots,\ t \times Interval \in [0, LimitTime]) \\
\qquad \qquad\qquad\qquad\qquad \qquad  \qquad \qquad  
{\rm{\textbf{maximize}}}
\
Cplt\_Tasks \\
\qquad \qquad\qquad\qquad \qquad \qquad \qquad 
s.t.\
\text{constraints (1)(2)(3)(4)(5)(6)}
\end{array}
\end{equation}
\end{small}

We prove in Appendix A that the \textbf{Problem 1} is NP-hard.
\section{Methodology}
\label{section4}

\begin{figure}[htbp]
  \centering
  \setlength{\abovecaptionskip}{0.2cm}
  \setlength{\belowcaptionskip}{-0.25cm}
  \includegraphics[width=\linewidth]{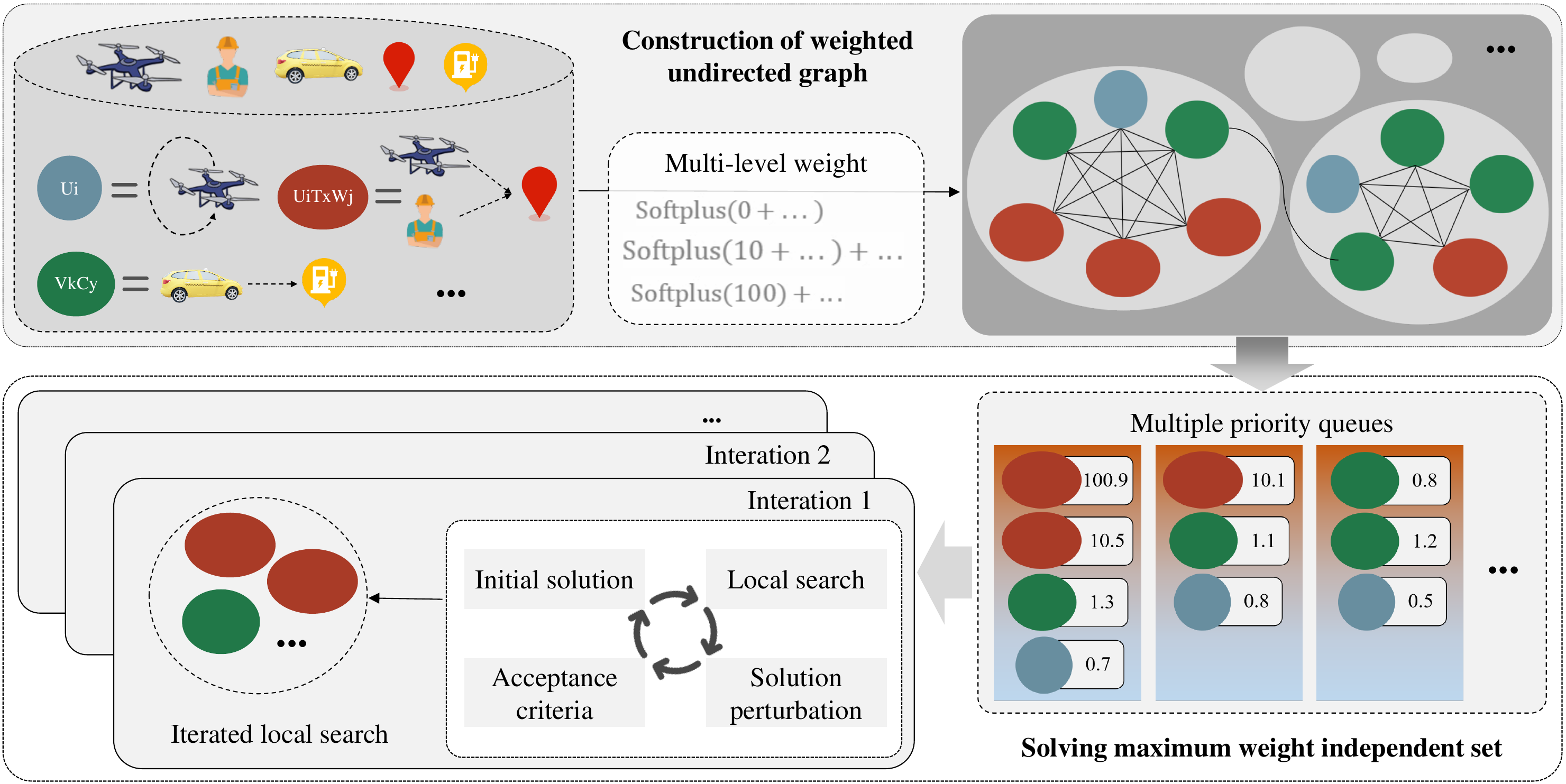}
  \caption{Framework of HoCs-MPQ.}
  \Description{Framework diagram of HoCs-MPQ.}
  \label{figure2}
\end{figure}

\hspace{1em}To solve \textbf{Problem 1},we transform \textbf{Problem 1} into the MWIS problem on an undirected graph $G=(\text{Nd}, \text{Eg})$. Here, each node $\text{nd} \in \text{Nd}$ represents a weighted, feasible multi-agent collaborative action. For any two nodes ${\text{nd}}_m, {\text{nd}}_n \in \text{Nd}$, an edge $({\text{nd}}_m, {\text{nd}}_n) \in \text{Eg}$ is established if their corresponding scheduling behaviors conflict. An independent set $S \subseteq Nd$ in graph $G$ implies that for any ${nd}_m', {nd}_n' \in S$, $({nd}_m', {nd}_n') \notin Eg$, which, by our edge definition, means their corresponding scheduling behaviors do not conflict. Therefore, an independent set directly represents a set of feasible and non-conflicting actions for \textbf{Problem 1}. This paper sets node weights as the expected benefit per time interval of scheduling behaviors, which is positively correlated with the optimization objective of \textbf{Problem 1}. Identifying the maximum weighted independent set $S^*$ (the independent set with the largest sum of node weights) then directly ensures the selection of the most beneficial set of compatible actions, thereby maximizing the overall mission objective. Based on this weighted undirected graph G, we propose a heterogeneous multi-agent online collaborative scheduling algorithm based on weighted undirected graph modeling and maximum weight independent set solution, named HoCs-MPQ. Its framework, as shown in Fig. ~\ref{figure2}, consists of two main modules: weighted undirected graph construction and maximum weight independent set solution. (1) Weighted undirected graph construction involves three parts: weighted undirected graph node construction, graph node weight evaluation and quantification, and weighted undirected graph generation. (2) Maximum weight independent set solution includes two parts: Heterogeneous multi-agent online collaborative scheduling algorithm based on iterated local search for maximum weight independent set (HoCs-ILS) and solution process acceleration based on multi-priority queues.

\subsection{Weighted Undirected Graph Construction}

\subsubsection{Weighted Undirected Graph Node Construction}

\begin{figure}[htbp]
  \centering
  \setlength{\abovecaptionskip}{0.2cm}
  \setlength{\belowcaptionskip}{-0.25cm}
  \includegraphics[width=12cm]{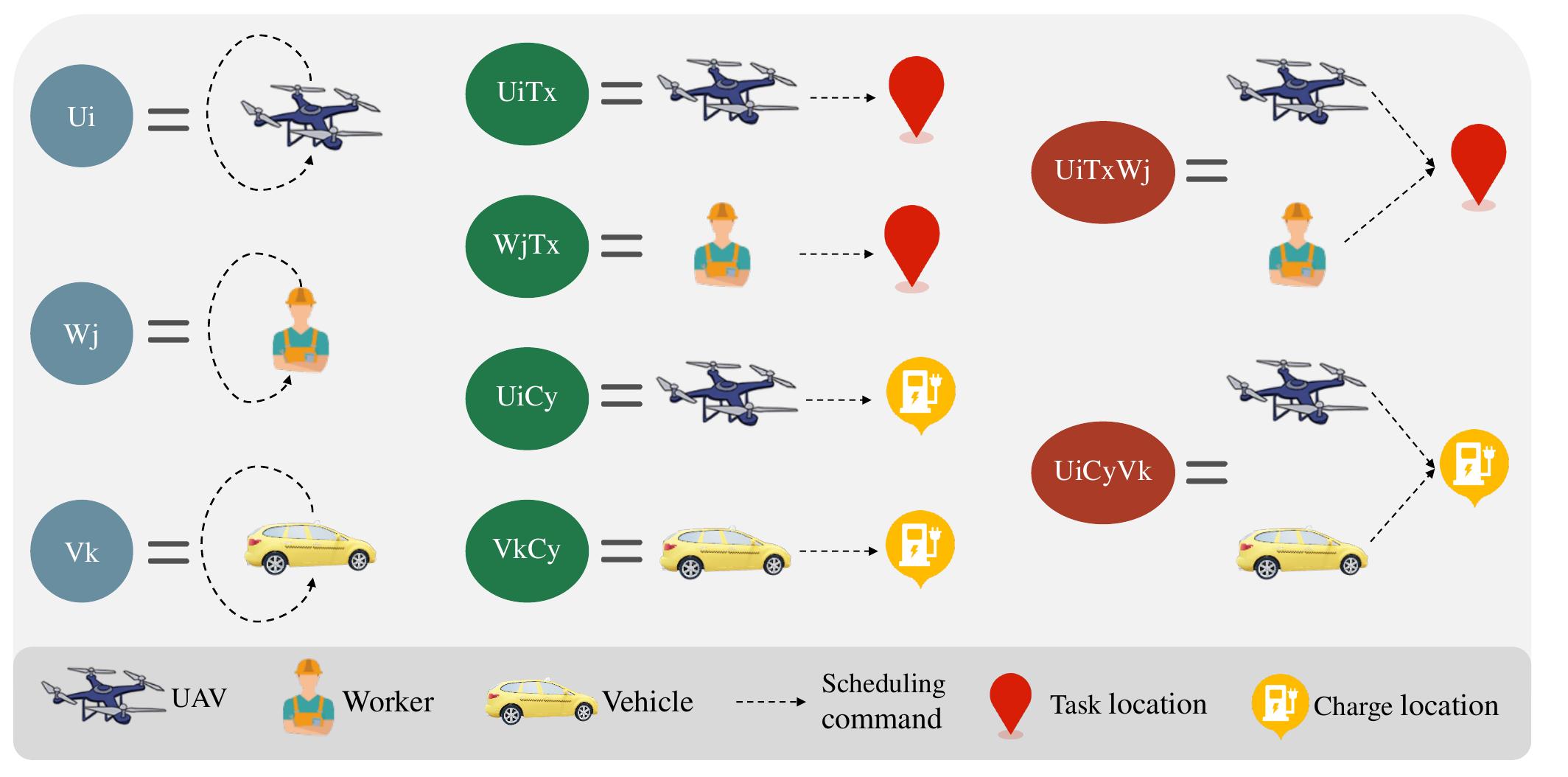}
  \caption{Types of weighted undirected graph nodes.}
  \Description{A diagram showing nine types of weighted undirected graph nodes used in the algorithm.}
  \label{figure3}
\end{figure}

To construct a weighted undirected graph, we need to first clarify its node set $Nd$. Collaborative scheduling of time-dependent UAVs, vehicles, and workers can result in 9 possible scenarios, meaning the set $Nd$ should include 9 types of graph nodes, as shown in Fig. ~\ref{figure3}.
\begin{itemize}
\item$Ui=(u_i,{uLoc}_i)$ represents UAV $u_i$ staying at its current location without movement.

\item$Wj=(w_j,{wLoc}_j)$ represents worker $w_j$ staying at its current location without movement.

\item${Vk=(v}_k,{vLoc}_k)$ represents vehicle $v_k$ staying at its current location without movement.

\item$UiTx=(u_i,{tLoc}_x)$ represents UAV $u_i$ moving to the location ${tLoc}_x$ of task ${\rm task}_x$.

\item$WjTx=(w_j,{tLoc}_x)$ represents worker $w_j$ moving to the location ${tLoc}_x$ of task ${\rm task}_x$.

\item$UiCy=(u_i,{chLoc}_y)$ represents UAV $u_i$ moving to the location ${chLoc}_y$ of charging point ${\rm charge}_y$.

\item$VkCy=(v_k,{chLoc}_y)$ represents vehicle $v_k$ moving to the location ${chLoc}_y$ of charging point ${\rm charge}_y$.

\item$UiTxWj=(u_i,{tLoc}_x,w_j)$ represents both UAV $u_i$ and worker $w_j$ moving to the location ${tLoc}_x$ of task ${\rm task}_x$.

\item$UiCyVk=(u_i,{chLoc}_y,v_k)$ represents both UAV $u_i$ and vehicle $v_k$ moving to the location ${chLoc}_y$ of charging point ${\rm charge}_y$.
\end{itemize}
\subsubsection{Graph Node Weight Evaluation and Quantification}

To construct a weighted undirected graph, we need to assign appropriate weights to the above 9 types of graph nodes. The results of these nodes can be classified into three levels: (1) $UiTxWj=(u_i,{tLoc}_x,w_j)$ leads to the completion of task ${\rm task}_x$ and has potential impact on the subsequent scheduling of $u_i$ and $w_j$; (2) $UiCyVk=(u_i,{chLoc}_y,v_k)$ leads to UAV $u_i$ being fully charged by vehicle $v_k$ and has potential impact on the subsequent scheduling of $u_i$ and $v_k$; (3) The other 7 types of nodes besides $UiTxWj$ and $UiCyVk$ only have potential impact on the subsequent scheduling of agents such as UAV $u_i$, worker $w_j$, and vehicle $v_k$. Therefore, we implement a hierarchical weight evaluation and quantification for these nodes. It is crucial to ensure that the quantified weights at different levels exhibit distinct orders of magnitude, thereby effectively differentiating between nodes of varying importance. Given that the optimization objective of \textbf{Problem 1} is to maximize the number of completed tasks, we assign a foundational weight of 100 to $UiTxWj$ nodes. Furthermore, recognizing the positive impact of enhanced UAV endurance on subsequent task performance, we set the foundational weight for $UiCyVk$ nodes to 10. Consequently, the remaining seven node types, excluding $UiTxWj$ and $UiCyVk$, are assigned a foundational weight of 0. Building upon these foundational weight assignments, our next step involves evaluating the potential impact of individual nodes to precisely quantify their overall weights within the graph. We define the benefit derived from this potential impact as the reduction rate in the expected matching time for the agents involved.
First, we need to evaluate the correlation ${\mathrm{E\_tL}}_i$ between the power of UAV $u_i$ and its task execution.
\begin{equation}
{\mathrm{E\_tL}}_i=\frac{\left(1-e^{-({uPow}_i/{\mathrm{FullPower}}_i)}\right)}{(1-e^{-1})}, \, u_i\ \text{s.t.}\ \text{constraints}\ (6)
\label{eq:4-1}
\end{equation}

${\mathrm{E\_tL}}_i$ should satisfy the following properties:
\begin{itemize}
\item[(1)]The larger ${uPow}_i/{\mathrm{FullPower}}_i$ is, the larger ${\mathrm{E\_tL}}_i$ is, indicating that the more power UAV $u_i$ has, the higher the possibility of executing tasks.
\item[(2)] As ${uPow}_i/{\mathrm{FullPower}}_i$ decreases, the rate of decrease in ${\mathrm{E\_tL}}_i$ increases, indicating that the lower the power of UAV $u_i$, the greater the urgency for charging, and the possibility of executing tasks should decrease at an accelerated rate.
\end{itemize}

Let $f\left({uPow}_i/{\mathrm{FullPower}}_i\right)={\mathrm{E\_tL}}_i$, ${uPow}_i/{\mathrm{FullPower}}_i\in[0,1]$, then
\begin{equation}
f^\prime\left({uPow}_i/{\mathrm{FullPower}}_i\right)=\frac{e^{-\left({uPow}_i/{\mathrm{FullPower}}_i\right)}}{1-e^{-1}}
\label{eq:4-2}
\end{equation}

For any ${uPow}_i/{\mathrm{FullPower}}_i\in[0,1]$, $e^{-\left({uPow}_i/{\mathrm{FullPower}}_i\right)}>0$, i.e., $f^\prime\left({uPow}_i/{\mathrm{FullPower}}_i\right)>0$, so \\ $f\left({uPow}_i/{\mathrm{FullPower}}_i\right)$ is strictly monotonically increasing on $[0,1]$, satisfying property (1).

\begin{equation}
{f}^{\prime\prime}\left({uPow}_i/{\mathrm{FullPower}}_i\right)=-\frac{e^{-\left({uPow}_i/{\mathrm{FullPower}}_i\right)}}{1-e^{-1}}
\label{eq:4-3}
\end{equation}

For any ${uPow}_i/{\mathrm{FullPower}}_i\in[0,1]$, ${-e}^{-\left({uPow}_i/{\mathrm{FullPower}}_i\right)}<0$, i.e., $f^{\prime\prime}\left({uPow}_i/{\mathrm{FullPower}}_i\right)<0$. This means that as ${uPow}_i/{\mathrm{FullPower}}_i$ approaches 0, $f^\prime\left({uPow}_i/{\mathrm{FullPower}}_i\right)$ becomes larger, and $f\left({uPow}_i/{\mathrm{FullPower}}_i\right)$ decreases faster, satisfying property (2).

Next, we need to evaluate the correlation $E\_{chL}_i$ between the power of UAV $u_i$ and its charging needs. $E\_{chL}_i$ should be negatively correlated with ${\mathrm{E\_tL}}_i$: (1) The higher the charging urgency of UAV $u_i$, the lower the possibility of executing tasks; (2) The faster the charging urgency of UAV $u_i$ increases, the faster the possibility of executing tasks decreases.

\begin{equation}
E\_{chL}_i=\frac{\left(1-e^{-({1-{uPow}_i}/{\mathrm{FullPower}}_i)}\right)}{(1-e^{-1})}, \, u_i\ \text{s.t.}\ \text{constraints}\ (5)
\label{eq:4-4}
\end{equation}

The expected cost for a task ${\rm task}_x$ to be matched by a UAV is as follows:
\begin{equation}
{\mathrm{Du\_tL}}_x=\frac{1}{|UAV|}\sum_{u_i\in UAV}\frac{ceil(Dis({uLoc}_i,{tLoc}_x)/{uRge}_i, Interval)}{E\_{tL}_i}, \, u_i\ \text{s.t.}\ \text{constraints}\ (6)
\label{eq:4-5}
\end{equation}
where $Dis()$ represents the distance calculation function; $ceil()$ represents the ceiling function.

Since workers do not need to be charged, the expected cost for a task ${\rm task}_x$ to be matched by a worker is as follows:
\begin{equation}
{\mathrm{Dw\_tL}}_x=\frac{1}{|Worker|}\sum_{w_j\in Worker}ceil(\frac{Dis({wLoc}_j,{tLoc}_x)}{{wRge}_j}, Interval)
\label{eq:4-6}
\end{equation}

The expected cost for a charging point ${chLoc}_y$ to be matched by a UAV is as follows:
\begin{equation}
\mathrm{Du\_chL}_y=\frac{1}{|UAV|}\sum_{u_i\in UAV}\frac{ceil(Dis({uLoc}_i,{chLoc}_y)/{uRge}_i, Interval)}{E\_{chL}_i}, \, u_i\ \text{s.t.}\ \text{constraints}\ (5)
\label{eq:4-7}
\end{equation}

Since vehicles do not need to be charged, the expected cost for a charging point ${chLoc}_y$ to be matched by a vehicle is as follows:
\begin{equation}
\mathrm{Dv\_chL}_y=\frac{1}{|Vehicle|}\sum_{v_k\in Vehicle}ceil(\frac{Dis({vLoc}_k,{chLoc}_y)}{{vRge}_k}, Interval)
\label{eq:4-8}
\end{equation}

The greater the expected time for worker $w_j$ to move to task location ${tLoc}_x$, the lower should be the possibility of choosing task ${\rm task}_x$. Therefore, the probabilities for worker $w_j$ at location ${wLoc}_j$ to choose different tasks are as follows:
\begin{equation}
\{...,pw_x,...\}=\text{softmax}(...,-ceil(\frac{Dis({wLoc}_j,{tLoc}_x)}{{wRge}_j}, Interval),...)
\label{eq:4-9}
\end{equation}
where $\text{softmax}()$ is the normalized exponential function \cite{boltzmann1868studien, bridle1990probabilistic}.

The expected benefit per time interval for worker $w_j$ moving from location  ${wLoc}_j$ to a new location  ${wLoc}_j'$ is as follows:
\begin{equation}
icmWj=\frac{\text{Softplus}\left((\sum_{x}({\rm pw}_x\cdot{\mathrm{Du\_tL}}_x)-\sum_{x}({\rm pw}_x'\cdot{\mathrm{Du\_tL}}_x))/{\sum_{x}({\rm pw}_x\cdot{\mathrm{Du\_tL}}_x)}\right)}{ceil(Dis({wLoc}_j,{wLoc}_j')/{wRge}_j, Interval)}
\label{eq:4-10}
\end{equation}
where $\text{Softplus}()$ is a smoothed version of the rectified linear unit function \cite{dugas2000incorporating}, and its purpose is to avoid negative expected benefits per time interval.

The greater the expected time for UAV $u_i$ to move to task location ${tLoc}_x$, the lower should be the possibility of choosing task ${\rm task}_x$. Therefore, the probabilities for UAV $u_i$ at location ${uLoc}_i$ to choose different tasks are as follows:
\begin{equation}
\{...,pu_x,...\}=\text{softmax}(...,-ceil(\frac{Dis({uLoc}_i,{tLoc}_x)}{{uRge}_i}, Interval),...)
\label{eq:4-11}
\end{equation}

The expected benefit per time interval for UAV $u_i$ moving from location  ${uLoc}_i$ to a new location  ${uLoc}_i'$ is as follows:
\begin{equation}
icmUiT=\frac{\text{Softplus}\left((\sum_{x}({\rm pu}_x \cdot {\mathrm{Dw\_tL}}_x) - \sum_{x}({\rm pu}_x' \cdot {\mathrm{Dw\_tL}}_x))/{\sum_{x}({\rm pu}_x \cdot {\mathrm{Dw\_tL}}_x})\right)}{ceil(Dis({uLoc}_i,{uLoc}_i')/{uRge}_i, Interval)}
\label{eq:4-12}
\end{equation}

The greater the expected time for UAV $u_i$ to move to charging point ${chLoc}_y$, the lower should be the possibility of choosing charging point ${\rm charge}_y$. Therefore, the probabilities for UAV $u_i$ at location ${uLoc}_i$ to choose different charging points are as follows:
\begin{equation}
\{...,{\rm pu}_y,...\}=\text{softmax}(...,-ceil(\frac{Dis({uLoc}_i,{chLoc}_y)}{{uRge}_i}, Interval),...)
\label{eq:4-13}
\end{equation}

The expected benefit per time interval for UAV $u_i$ moving from ${uLoc}_i$ to a new location ${uLoc}_i'$ is as follows:
\begin{equation}
icmUiC=\frac{\text{Softplus}\left((\sum_{y}({\rm pu}_y\cdot \mathrm{Dv\_chL}_y)-\sum_{y}({\rm pu}_y'\cdot \mathrm{Dv\_chL}_y))/{\sum_{y}({\rm pu}_y\cdot \mathrm{Dv\_chL}_y)}\right)}{ceil(Dis({uLoc}_i,{uLoc}_i')/{uRge}_i, Interval)}
\label{eq:4-14}
\end{equation}

Therefore, the comprehensive expected benefit per time interval for UAV is:
\begin{equation}
icmUi=\frac{{uPow}_i'}{{\mathrm{FullPower}}_i}\cdot icmUiT+\frac{({{\mathrm{FullPower}}_i-{uPow}_i'})}{{\mathrm{FullPower}}_i}\cdot icmUiC
\label{eq:4-15}
\end{equation}

The greater the expected time for vehicle $v_k$ to move to charging point ${chLoc}_y$, the lower should be the possibility of choosing charging point ${\rm charge}_y$. Therefore, the probabilities for vehicle $v_k$ at location ${chLoc}_y$ to choose different charging points are as follows:
\begin{equation}
\{...,{\rm pv}_y,...\}=\text{softmax}(...,-ceil(\frac{Dis({vLoc}_k,{chLoc}_y)}{{vRge}_k}, Interval),...)
\label{eq:4-16}
\end{equation}

The expected benefit per time interval for vehicle $v_k$ moving from location  ${chLoc}_y$ to a new location  ${chLoc}_y'$ is as follows:
\begin{equation}
icmVk=\frac{\text{Softplus}\left((\sum_{y}({\rm pv}_y\cdot \mathrm{Du\_chL}_y)-\sum_{y}({\rm pv}_y'\cdot \mathrm{Du\_chL}_y))/{\sum_{y}({\rm pv}_y\cdot \mathrm{Du\_chL}_y)}\right)}{ceil(Dis({vLoc}_k,{vLoc}_k')/{vRge}_k, Interval)}
\label{eq:4-17}
\end{equation}

The weight calculation process for $UiTxWj$ is as follows:
\begin{equation}
icmUiWj=\frac{\text{Softplus}(100)}{\text{max}(ceilU,ceilW)}+icmWj+icmUi
\label{eq:4-18}
\end{equation}
where $ceilU=ceil(Dis({uLoc}_i,{uLoc}_i')/{uRge}_i, Interval)$; $ceilW=ceil(Dis({wLoc}_j,{wLoc}_j')/{wRge}_j, Interval)$;\text{max()} represents taking the larger of two numbers.

The weight calculation process for $UiCyVk$ is as follows:
\begin{equation}
icmUiVk=\frac{\text{Softplus}\left(10+e^{(1-{uPow}_i/{\mathrm{FullPower}}_i)}\right)}{\text{max}(ceilU,ceilV)}+icmVk+icmUi
\label{eq:4-19}
\end{equation}
where $e^{(1-{uPow}_i/{\mathrm{FullPower}}_i)}$ is used to quantify the additional benefit of increasing UAV power; $ceilV=ceil(\\ Dis({vLoc}_k,{vLoc}_k')/{vRge}_k, Interval)$.

The weight calculation process for nodes $Ui/UiTx/UiCy$ refers to Equation (\ref{eq:4-15}); the weight calculation process for $Wj, WjTx$ refers to Equation (\ref{eq:4-10}); the weight calculation process for $Vk, VkCy$ refers to Equation (\ref{eq:4-17}).

\subsubsection{Weighted Undirected Graph Generation}

To construct a weighted undirected graph $G=(Nd,Eg)$, we define its nodes and edges based on the problem's requirements. The generation process of this weighted undirected graph $G$ is shown in \textbf{Algorithm 1.}
\begin{algorithm}[htbp]
	\caption{: Weighted Undirected Graph Construction Algorithm (WUGCA)}
	\label{algorithm1}
        \begin{flushleft}
		\hspace*{0in} \textbf{Input:} 
        	$UAV$,$Worker$, $Vehicle$, $Task$, $Charge$\\
		\hspace*{0in} \textbf{Output:}
        	Graph $G$
	\end{flushleft}
	\begin{algorithmic}[1]
		\State $G \leftarrow$ empty graph
		\State Compute ${\mathrm{Du\_tL}}_x$, $\mathrm{Du\_chL}_y$, $\mathrm{Du\_chL}_y$, $\mathrm{Dv\_chL}_y$
		\For{each $u_i\in UAV$}
			\State $G.Nd \leftarrow G.Nd \cup \{Ui\}$ with $w(Ui) = \text{softPlus}(0)$
			\For{each ${\rm task}_x \in Task$ \textbf{such that} $u_i\ \text{s.t.}\ \text{constraints}\ (6)$}
				\State $G.Nd \leftarrow G.Nd\cup\{UiTx\}$, $w(UiTx) = icmUiTx$
				\For{each $w_j\in Worker$
                \textbf{such that} $u_i\ \text{s.t.}\ \text{constraints}\ (6)$}
					\State $G.Nd \leftarrow G.Nd \cup \{UiTxWj\}$, $w(UiTxWj) = icmUiWj$
				\EndFor
			\EndFor
			\For{each ${\rm charge}_y \in Charge$ \textbf{such that} ${u}_i\ \text{s.t.}\ \text{constraints}\ (5)$}
				\State $G.Nd \leftarrow G.Nd \cup \{UiCy\}$, $w(UiCy) = icmUiCy$
				\For{each $v_k \in Vehicle$}
					\State $G.Nd \leftarrow G.Nd\cup\{UiCyVk\}$, $w(UiCyVk)=icmUiVk$
				\EndFor
			\EndFor
		\EndFor
		\For{each $w_j\in Worker$}
			\State $G.Nd \leftarrow G.Nd \cup \{Wj\}$ with $w(Wj) = \text{softPlus}(0)$
			\For{each ${\rm task}_x \in Task$}
				\State $G.Nd \leftarrow G.Nd \cup \{WjTx\}$ with $w(WjTx) = icmWj$
			\EndFor
		\EndFor
		\For{each $v_k \in Vehicle$}
			\State $G.Nd \leftarrow G.Nd \cup \{V_k\}$ with $w(V_k) = \text{softPlus}(0)$
			\For{each ${\rm charge}_y\in Charge$}
				\State $G.Nd \leftarrow G.Nd \cup \{VkCy\}$ with $w(VkCy) = icmVk$
			\EndFor
		\EndFor
		\If{HasConflict(${nd}_m, {nd}_n$)}  \Comment{Connect nodes when a resource conflict exists}
            \State Add edge between ${nd}_m$ and ${nd}_n$
            \EndIf
		\State \Return $G$
	\end{algorithmic}
\end{algorithm}

Based on the above weighted undirected graph generation process, we analyze the graph structure. Assume that the current number of online UAVs, vehicles, and workers are $|u|$, $|v|$, and $|w|$ respectively, the number of tasks is $|task|$, and the number of charging points is $|charge|$. The number of $Ui=(u_i,{uLoc}_i)$ is $|u|$; the number of $Wj=(w_j,{wLoc}_j)$ is $|w|$; the number of ${Vk=(v}_k,{vLoc}_k)$ is $|v|$; in the worst case, the number of $UiTx=(u_i,{tLoc}_x)$ is $|u|*|task|$; the number of $WjTx=(w_j,{tLoc}_x)$ is $|w|*|task|$; in the worst case, the number of $UiCy=(u_i,{chLoc}_y)$ is $|u|*|charge|$; in the worst case, the number of $VkCy=(v_k,{chLoc}_y)$ is $|v|*|charge|$; the number of $UiTxWj=(u_i,{tLoc}_x,w_j)$ is $|u|*|task|*|w|$; in the worst case, the number of $UiCyVk=(u_i,{chLoc}_y,v_k)$ is $|u|*|charge|*|v|$. If the cumulative sum of the above nodes is denoted as $N$, then the order of magnitude of the number of nodes in graph $G$ is $O(N)$, and the order of magnitude of the number of edges in graph $G$ is $O(N^2)$. Therefore, with the linear growth of $|u|$, $|w|$, $|v|$, $|task|$, and $|charge|$, the node scale of graph $G$ shows cubic polynomial growth, and the edge scale of graph $G$ shows sixth-order polynomial growth, both of which are high-order polynomial growth.

\begin{figure}[htbp]
  \centering
  \setlength{\abovecaptionskip}{0.2cm}
  \setlength{\belowcaptionskip}{-0.25cm}
  \includegraphics[width=12cm]{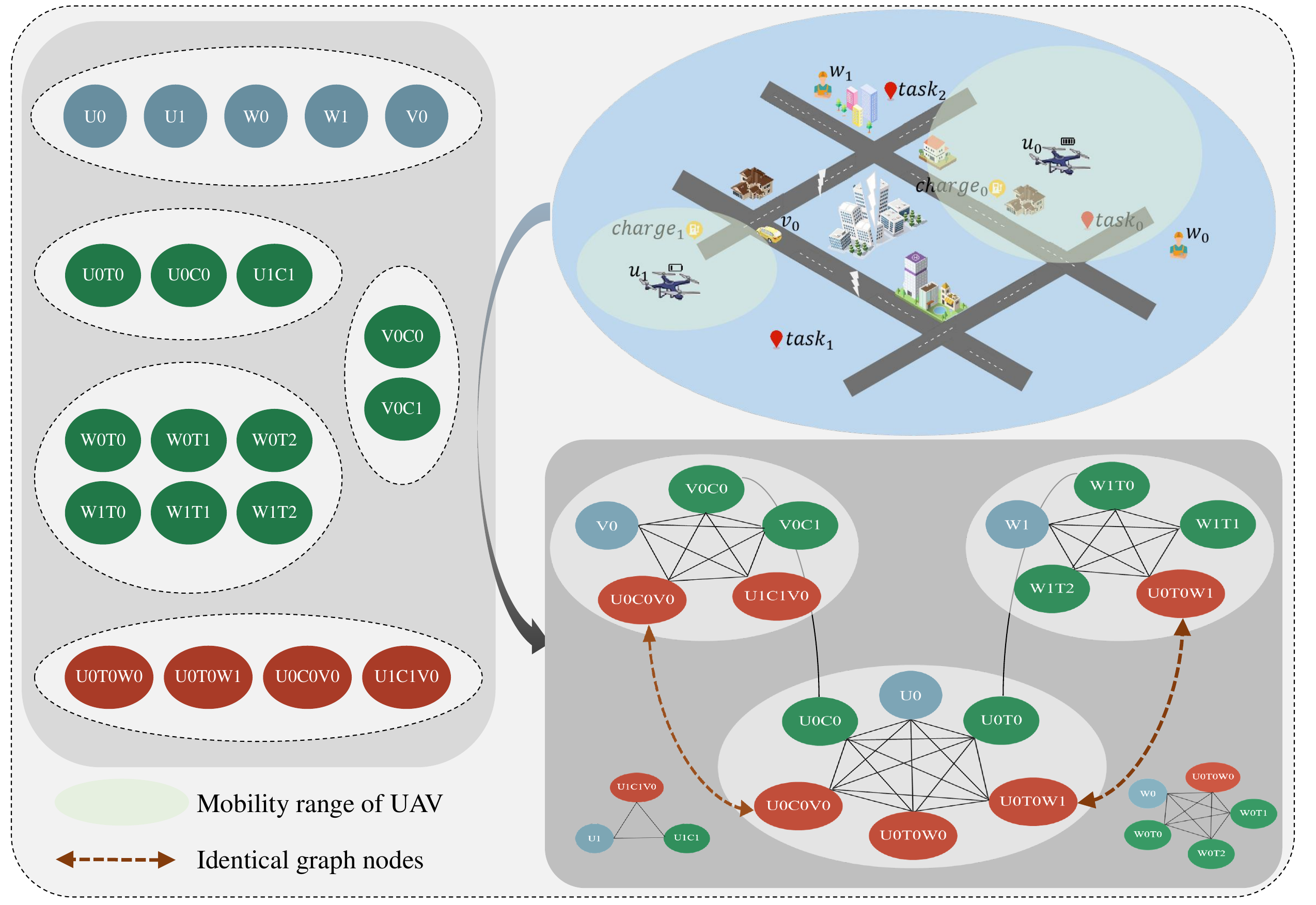}
  \caption{Example of weighted undirected graph generation.}
  \Description{An example illustrating how weighted undirected graph is generated with UAVs, workers, vehicles, tasks and charging points.}
  \label{figure4}
\end{figure}

Fig. ~\ref{figure4} shows an example of weighted undirected graph generation, which includes two UAVs $\{u_0,u_1\}$, two workers $\{w_0,w_1\}$, one vehicle $v_0$, four  tasks $\{task_0,...,task_2\}$, and two charging points $\{charge_0,charge_1\}$. Based on constraint (6), UAVs, workers, and tasks are combined to form three-dimensional matching nodes $UiTxWj$; based on constraint (6), UAVs and tasks are combined to form two-dimensional matching nodes $UiTx$; workers and tasks are freely combined to form two-dimensional matching nodes $WjTx$; based on constraint (5), UAVs, vehicles, and charging points are combined to form three-dimensional matching nodes $UiCyVk$; based on constraint (5), UAVs and charging points are combined to form two-dimensional matching nodes $UiCy$; vehicles and charging points are freely combined to form two-dimensional matching nodes $VkCy$. UAVs, vehicles, and workers can remain at their original positions unchanged, forming nodes $Ui$, $Wj$, and $Vk$ respectively. Next, weights are assigned to all the above nodes based on the weighted graph node weight evaluation and quantification process (see 4.1.2). Finally, any two nodes are selected from all the above nodes, and if the scheduling behaviors corresponding to the selected nodes have conflicts, an edge connection is set between the two nodes.

\subsection{Maximum Weight Independent Set Solution}

\hspace{1em}The solution process of the maximum weight independent set $S^\ast$ of the weighted undirected graph $G$ mainly includes two parts: (1) Due to the large scale of graph $G$, exact algorithms are difficult to find its maximum weight independent set $S^\ast$ in a short time. In order to find high-quality approximate solutions in a limited time, we design a maximum weight independent set solution algorithm based on iterative search. (2) The edges of graph $G$ are too dense, which leads to excessive computational time overhead of the solution algorithm, making it difficult to apply to online scenarios. To reduce the computational time overhead of the algorithm, we implemented solution process acceleration based on multi-priority queues.

\subsubsection{HoCs-ILS}

For the weighted undirected graph constructed above, we have developed a MWIS solution algorithm based on the principles of Iterated Local Search (ILS) \cite{lourencco2018iterated}, as detailed in \textbf{Algorithm 2}. This algorithm is structured around four core modules: initial solution generation, local search, perturbation, and acceptance criterion. The initial solution is constructed using a greedy approach, leveraging a hierarchical weight-based strategy. Within the local search module, we quantify the potential value of nodes and maintain an evaluation queue to expedite the discovery of local optimal solutions. Pertaining to the perturbation strategy, our algorithm adaptively adjusts its intensity based on the quality of the current local optimal solution, thereby expanding the solution search space. Finally, the acceptance criterion module facilitates the acceptance of suboptimal solutions, enhancing the diversity of explored local solutions. Guided by this acceptance criterion, the algorithm progressively refines the final solution through a continuous cycle of local search and adaptive perturbation, until either the maximum number of iterations is reached or a predefined termination condition is satisfied.

\begin{algorithm}[htb]
	\caption{: HoCs-ILS}
	\label{algorithm3}
        \begin{flushleft}
		\hspace*{0in} \textbf{Input:} 
        	$G$, $maxIter$\\
		\hspace*{0in} \textbf{Output:}
        	$S^*$
	\end{flushleft}
	\begin{algorithmic}[1]
		\State $S \leftarrow \text{InitialSolution}(G)$
		\State $S \leftarrow \text{LocalSearch}(G, S)$
		\State $S^* \leftarrow S$
		\State $iter,noImprovCount \leftarrow 0$
		\State $acceptThreshold \leftarrow 0.95$
		\While{$iter < maxIter$}
			\State $iter \leftarrow iter + 1$
			\State $S' \leftarrow \text{Copy}(S)$
			\State $pStrength \leftarrow \text{AdaptPerturbStrength}(S, noImprovCount)$
			\State $S' \leftarrow \text{Perturbation}(G, S', pStrength)$
			\State $S' \leftarrow \text{LocalSearch}(G, S')$
			\If{$\text{AcceptSolution}(S, S', H, acceptThreshold)$}
				\State $S \leftarrow S'$
				\If{$w(S) > w(S^*)$}
					\State $S^* \leftarrow S$
				\EndIf
			\Else
				\State $noImprovCount \leftarrow noImprovCount + 1$
			\EndIf
			\If{$noImprovCount > maxIter / 4$}
				\State $acceptThreshold \leftarrow acceptThreshold \times 0.99$
			\EndIf
		\EndWhile
		\State \Return $S^*$
	\end{algorithmic}
\end{algorithm}

The local search process is shown in \textbf{Algorithm 3}. We introduce a node evaluation function $\sigma({nd}_m)$ to quantify the improvement value of vertex ${nd}_m$ to the current solution:
\begin{equation}
\sigma\left({nd}_m\right)=w\left({nd}_m\right)-\sum_{{nd}_n\in N\left({nd}_m\right)\cap S} w\left({nd}_n\right)
\end{equation}
where $w({nd}_m)$ is the weight of vertex ${nd}_m$, $N({nd}_m)$ is the set of neighbors of ${nd}_m$, and $S$ is the current solution, i.e., the independent set. When $\sigma({nd}_m) > 0$, including ${nd}_m$ into the current solution $S$ can bring positive benefits. The algorithm maintains a candidate queue $Q$ and continuously outputs vertices that satisfy $\sigma({nd}_m) > 0$ from $Q$.

\begin{algorithm}[htb]
	\caption{: Local Search Optimization Algorithm (LSOA)}
	\label{algorithm4}
        \begin{flushleft}
		\hspace*{0in} \textbf{Input:} 
        	$G$, $S'$\\
		\hspace*{0in} \textbf{Output:}
        	$S''$
	\end{flushleft}
	\begin{algorithmic}[1]
		\State $improved \leftarrow true$
		\State $S'' \leftarrow S'$
		\While{$improved$}
			\State $improved \leftarrow false$
			\If{$(nd_m \leftarrow Q.\mathrm{pop}()) \neq nil$}
				\State RemoveNeighbors($nd_m, S'', Eg$)  \Comment{Remove from $S''$ all nodes adjacent to $nd_m$ in $Eg$}
				\State $S''.Add({nd}_m)$
				\State $improved \leftarrow true$
			\EndIf
		\EndWhile
		\State \Return $S''$
	\end{algorithmic}
\end{algorithm}

Generally, local search should not undo perturbation, otherwise it would affect the expansion of the solution space. Based on this, we propose a weight-aware structural perturbation strategy, as shown in \textbf{Algorithm 4}.
\begin{algorithm}[htb]
	\caption{: Stratified Cyclic Perturbation Algorithm (SCPA)}
	\label{algorithm5}
        \begin{flushleft}
		\hspace*{0in} \textbf{Input:} 
        	$G$, $S'$, $pStrength$\\
		\hspace*{0in} \textbf{Output:}
        	$S$
	\end{flushleft}
	\begin{algorithmic}[1]
		\State $S_0 \leftarrow \{nd_m \in S' | nd_m \text{ is of type Other}\}$
		\State $S_1 \leftarrow \{nd_m \in S' | nd_m \text{ is of type } UiCyVk\}$
		\State $S_2 \leftarrow \{nd_m \in S' | nd_m \text{ is of type } UiTxWj\}$
		\State $l \leftarrow iter \bmod 3$
		\State $R \leftarrow \emptyset,C \leftarrow \emptyset$
		\If{$l = 0$ and $|S_0| > 0$}
			\State $R \leftarrow$ randomly select $min(k, |S_0|)$ nodes from $S_0$
		\ElsIf{$l = 1$ and $|S_1| > 0$}
			\State $R \leftarrow$ randomly select $min(k, |S_1|)$ nodes from $S_1$
		\ElsIf{$l = 2$ and $|S_2| > 0$}
			\State $R \leftarrow$ randomly select $min(max(1, \lfloor k / 2 \rfloor), |S_2|)$ nodes from $S_2$
		\EndIf
		\State FindDifferentLabelNeighbors(R, S', C) \Comment{Add neighbors       of R with different labels to set C}
            \State S ← RemoveNodesFromSolution(S', R) \Comment{Remove nodes in set R from solution S'}
		\State Shuffle$(C)$
		\State $S \leftarrow \text{AddIndependentNodes}(S, C, Eg)$ \Comment{Add nodes with no edge connection}
		\State \Return $S$
	\end{algorithmic}
\end{algorithm}
First, since the graph node weights are quantified at different levels (see 4.1.2), the weight hierarchical distribution characteristics are significant, which provides the implementation basis for structural perturbation. We define the function $L({nd}_m)$ to assign weight levels to nodes in the graph according to their types:
\begin{equation}
L({nd}_m) = 
\begin{cases}
2, & {nd}_m = UiTxWj \\
1, & {nd}_m = UiCyVk \\
0, & {nd}_m = \text{Other}
\end{cases}
\end{equation}

Secondly, the core of this perturbation strategy is to create structural barriers to prevent local search from easily undoing the perturbation effect. Specifically, construct a node set $R$ to be removed and select node ${nd}_m$ from it, then construct a node set $C$ to be added and select node ${nd}_n$ from it to be added to set $S$, to construct structural barrier $A(R,C)$, which specifically includes the following constraints: ${nd}_m \in Nd \setminus S$; $\exists {nd}_m \in R, ({nd}_n,{nd}_m) \in Eg$; $L({nd}_m) \neq L({nd}_n)$; $\forall s \in S, ({nd}_n,s) \notin Eg$. The node ${nd}_n$ selected based on the above constraints only conflicts with node ${nd}_m$, and does not conflict with other nodes in the current solution $S$, thus constructing a structural barrier.

\subsubsection{Solution Process Acceleration Based on Multi-Priority Queues}
The inherent density of the weighted undirected graph $G$ poses a significant computational challenge for directly solving the MWIS problem using traditional approaches like HoCs-ILS, leading to excessive time consumption. To address this, and by specifically leveraging the structural characteristics of graph $G$, we designed the HoCs-MPQ algorithm. Its primary objective is to substantially reduce the time required to find the MWIS of the weighted undirected graph $G$. Specifically, HoCs-MPQ builds upon the foundation of the HoCs-ILS algorithm by employing multi-priority queues to classify and store nodes associated with different agents, thereby effectively reducing the operative graph scale during the solution process.
As depicted in Fig. ~\ref{figure4}, the weighted undirected graph $G$ exhibits distinct structural characteristics:(1) Clique Structures are Prevalent: The generation process of the weighted undirected graph reveals that edges (representing conflict relationships between nodes) are triggered by shared resources among the combined elements (nodes). Consequently, all nodes that share the same element (e.g., a specific agent like worker $w_0$) form a clique (i.e., a complete subgraph). For instance, any task ${\rm task}x \in \text{Task}$ can form a node $W0Tx=(w_0,{tLoc}x)$ with worker $w_0$. Therefore, all nodes in the set $\{W0T0,...,W0Tx,...\}$ conflict with each other, forming a prominent clique structure.
(2) Dense within Cliques and Sparse between Cliques: The graph demonstrates a pattern of dense connections within cliques and sparse connections between cliques. Within each clique, nodes generally exhibit pervasive conflict relationships, leading to exceptionally dense edge connections. Conversely, nodes belonging to different clique structures typically share no conflict relationships, resulting in very sparse edge connections between them. Therefore, despite the overall density of the weighted undirected graph $G$, it possesses clearly discernible sparse regions that HoCs-MPQ exploits.

Leveraging these identified structural characteristics of graph $G$, we propose an incremental Maximum Weighted Independent Set solution algorithm based on multi-priority queues (HoCs-MPQ), as detailed in \textbf{Algorithm 5}. Initially, nodes are logically grouped based on the specific agent types they involve. Within each group, nodes are then sorted by output priority in descending order of their associated weights. A noteworthy aspect is that if a node incorporates both a UAV and other agent types, its primary grouping reference is solely the UAV category. Subsequently, nodes are dequeued from each multi-priority queue following an elastic expansion rule. As nodes are dequeued, the weighted undirected graph $G'$ (the current working graph for the MWIS problem) is dynamically updated to reflect the newly considered nodes and their relationships. The refined subgraph $G'$ is then loaded into the HoCs-ILS algorithm, which computes the Maximum Weighted Independent Set for this updated subgraph. Finally, the algorithm assesses whether the Maximum Weighted Independent Set obtained in the current iteration round is superior to that from the preceding round. If no improvement is observed, the node dequeuing speed is adaptively accelerated to explore other parts of the solution space more quickly; otherwise, the dequeuing speed remains at its standard rate.There are three noteworthy points about the HoCs-MPQ algorithm:

\begin{enumerate}
\item Incremental Warm Start of the Solution (refer to line 6 of \textbf{Algorithm 5}):
By gradually expanding the subgraph $G'$, the optimal solution $S^{sub}$ obtained from each stage is consistently utilized as the initial solution for the subsequent iteration of the HoCs-ILS algorithm. This operation serves two critical purposes: Firstly, it ensures the monotonic non-decreasing property of the solution's total weight sum, meaning that as the subgraph $G'$ expands, the accumulated weight of the solution $S^{sub}$ is guaranteed to gradually increase. Secondly, by initiating the HoCs-ILS algorithm from an already high-quality solution, this strategy effectively reduces ineffective exploration of the search space, thereby accelerating convergence.
\item High-quality Subgraph Construction:
High-quality subgraphs are efficiently constructed by strategically prioritizing the output of high-weight nodes through the multi-priority queues. This operational principle primarily allows the subgraph $G'$ to preferentially cover areas that are dense with high-weight nodes, ensuring that the most impactful collaborative opportunities are considered early. Furthermore, this method explicitly models the resource competition relationship between agents by prioritizing combinations that yield higher benefits, thereby guiding the solution toward more efficient resource allocation.
\item Elastic Expansion Rule (refer to lines 13 of \textbf{Algorithm 5}):
Through the implementation of an exponential expansion strategy, this rule ensures that the algorithm effectively covers the dynamically constructed subgraph $G'$ within a finite number of steps. This particular operation enables the HoCs-ILS algorithm to quickly obtain reliable approximate solutions at a low computational cost during its early stages. Subsequently, it facilitates a controlled expansion to encompass the global solution space as dictated by the problem's requirements in the later stages, striking an optimal balance between initial efficiency and comprehensive solution discovery.
\end{enumerate}

\begin{figure}[hbtp]
  \centering
  \setlength{\abovecaptionskip}{0.2cm}
  \setlength{\belowcaptionskip}{-0.25cm}
  \includegraphics[width=12cm]{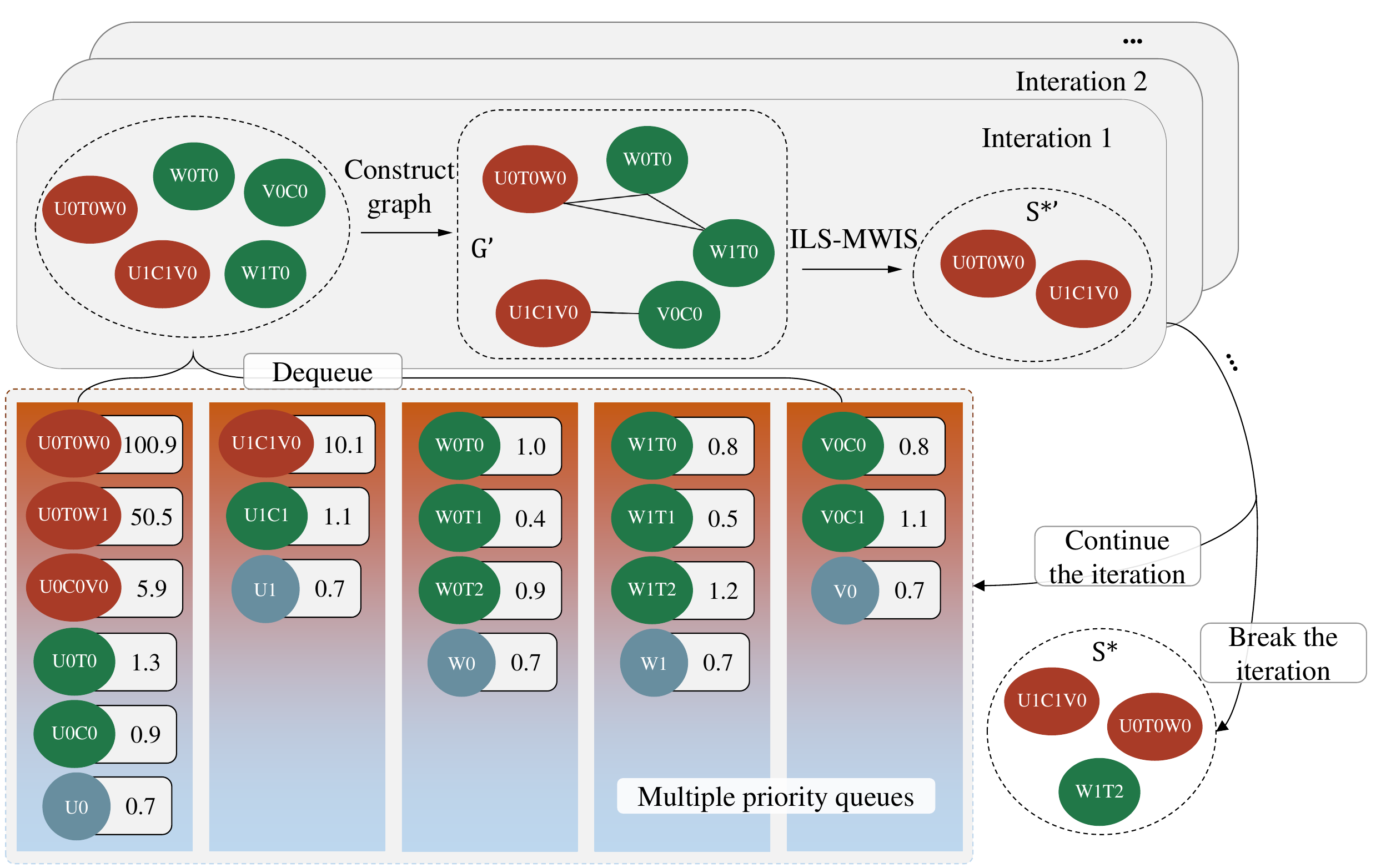}
  \caption{Example framework diagram of the HoCs-MPQ algorithm flow.}
  \Description{A diagram showing the example framework of the HoCs-MPQ algorithm flow.}
  \label{figure5}
\end{figure}

As shown in Fig. ~\ref{figure5}, we have provided an intuitive description of the algorithm process of HoCs-MPQ based on the weighted undirected graph example shown in Fig. ~\ref{figure4}. This weighted undirected graph example contains two UAVs $\{u_0,u_1\}$, two workers $\{w_0,w_1\}$, one vehicle $v_0$, four  tasks $\{{\rm task}_0,...,{\rm task}_3\}$, and two charging points $\{{\rm charge}_0,{\rm charge}_1\}$. First, all nodes in the example are classified according to the different agents they contain, and inserted into the corresponding queues in order of weight. Then, according to the elastic rule, nodes are output from all priority queues in sequence, and the weighted undirected graph $G'$ to be solved is updated. Next, graph $G'$ is loaded into the HoCs-ILS algorithm to solve the maximum weight independent set MWIS' of graph $G'$. Finally, it is determined whether the maximum weight independent set obtained in the current iteration round is better than that in the previous round. If not, the node dequeuing speed is accelerated, otherwise the dequeuing speed remains normal.

\begin{algorithm}[hbtp]
	\caption{: HoCs-MPQ}
	\label{algorithm6}
        \begin{flushleft}
		\hspace*{0in} \textbf{Input:} $G$ \\
            \hspace*{0in} \textbf{Output:} $S_{best}$ 
	\end{flushleft}
	\begin{algorithmic}[1]
		\State Initialize priority queues $Q_k$ for each agent category $k$
		\State $S_{best}, S_{current} \leftarrow \emptyset$
            \State$K \leftarrow 1$
		\While{$\exists Q_k \neq \emptyset$ \textbf{and} not all categories represented in solution}
			\State ExtractTopKNodes($K$, $\{Q_k\}$, $G_{sub}$) \Comment{Extract top-$K$ nodes from each queue}
			\State AddConflictEdges($G_{sub}$) \Comment{Add edges between conflicting nodes}
			\State $S_{candidate}, \Delta Z \leftarrow \text{ILS}(G_{sub}, S_{current})$ \Comment{Refer to Algorithm 2}
			\If{$\Delta Z > 0$}
				\State $S_{current} \leftarrow S_{candidate}$
				\If{$w(S_{candidate}) > w(S_{best})$}
					\State $S_{best} \leftarrow S_{candidate}$
				\EndIf
			\Else
				\State $K \leftarrow 2K$ \Comment{Double $K$ if no improvement}
			\EndIf
			\State RemoveNodesFromQueues($S_{current}$, $\{Q_k\}$) \Comment{Remove solution nodes from queues}
			\State CheckCategoryRepresentation($S_{current}$) \Comment{Check if every category has a node}
		\EndWhile
		\State \Return $S_{best}$
	\end{algorithmic}
\end{algorithm}

\section{EVALUATION}

\label{section5}

\subsection{Experimental Setup}

\subsubsection{Dataset Setup}
Our proposed algorithm's performance is comprehensively evaluated using both real-world datasets and randomly simulated datasets. A summary of these datasets is presented in Table \ref{table1} in APPENDIX B. The real-world datasets include BicycleOrder data \citep{li2015traffic, zheng2014urban}, TaxiTrajectories data \citep{yu2021object}, and DidiOrders data \citep{gaia_didi}. The randomly simulated datasets are generated through a parameterized process, with the detailed data generation procedures available in \citep{CocaColaZero}.

\subsubsection{Experiment Setup}
Based on these datasets, we established five groups of experiments to comprehensively evaluate the HoCs-MPQ algorithm. These experiments investigate its performance under varying "Interval", "Limit time", "area scale, tasks number, charging point number, and agents online time", "workers number, UAVs number, and vehicles number" (representing agents number), and "task cost power and charging power". Further details regarding these experimental groups and their controlled variables can be found in Table \ref{table2} in APPENDIX B.

\subsubsection{Comparison Algorithms}
In combinatorial optimization research, solution methods are typically categorized into heuristic \citep{DBLP:conf/kdd/LiC0W019, DBLP:journals/tmc/WangYYXHPG23,DBLP:journals/tmc/LiuLCQJ24, DBLP:journals/tnn/LiuZ25, DBLP:journals/tcyb/QiJLSL22}, meta-heuristic\cite{DBLP:journals/cn/SalimiAD25,DBLP:journals/nca/TanhaSR21}, and training-based approaches \citep{liu2019energy, liu2020energy, zhao2022cadre, wang2022human, ye2023exploring}. For comparative analysis, we selected HoCs-GREEDY, HoCs-KWTA, HoCs-MADL, and HoCs-MARL as comparison algorithms. Their characteristics and implementation details are summarized in Table \ref{table_comparison_algs_new} in Appendix B;notably, HoCs-MADL and HoCs-MARL heavily reference \citep{10.1109/TNET.2024.3395493}. Additionally, a comparative analysis of our proposed algorithm against several recent multi-agent deep learning algorithms, including QMIX, CW-QMIX\cite{NEURIPS2020_73a427ba}, OW-QMIX, QTRAN\cite{son2019qtran}, QPLEX\cite{wang2020qplex}, MF-QMIX\cite{10887881}, and SQIX \cite{zhang2024sqix}, is presented with detailed experimental results in Appendix C. While recent efforts have explored leveraging large language models (LLMs) \citep{li2024urbangpt, yuan2024unist} for such problems, including our own attempts\cite{CocaColaZero} (detailed prompt in Appendix D), current large language models exhibit significant limitations in handling multi-constraint scheduling tasks in complex post-disaster environments, often failing to generate reasonable scheduling plans due to a lack of logical consistency and reliable reasoning.

\subsection{Experimental Results}

\hspace{1em}To control variables and accurately assess the impact of different factors on algorithm performance, all parameters are set to default values, except for those explicitly varied in each experimental group. The experiment is conducted in a 30×30 area scale, with 120 tasks, 20 charging points, and a 60-minute agent online time. The agents consist of 50 workers, 30 UAVs, and 20 vehicles. Each task incurs a task power cost of 3, and each vehicle has a charging power of 10. Each experiment group only changes the target variable while keeping other parameters at their default values. The following experimental results (task completion rate) are averaged over multiple experiments.

\subsubsection{Experimental Results Under Different  Intervals}

\begin{figure}[htbp]
	\centering
	\setlength{\abovecaptionskip}{0.2cm}
	\setlength{\belowcaptionskip}{-0.25cm}
	\includegraphics[width=\linewidth]{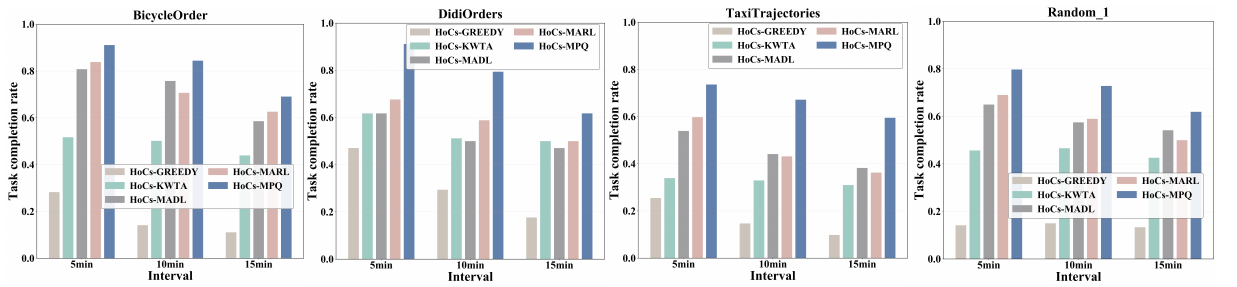}
	\caption{Experimental results under different intervals.}
	\Description{A graph showing the task completion rate of different algorithms under different intervals (5 min, 10 min, 15 min).}
	\label{figure7}
\end{figure}

As shown in Fig. \ref{figure7}, when the decision time interval increases from 5 minutes to 15 minutes, the task completion rate of the HoCs-MPQ algorithm consistently outperforms other baseline algorithms. Specifically, with a decision time interval of 5 minutes, HoCs-MPQ's average task completion rate is 83.92\%, which is 55.17\%, 35.65\%, 18.42\%, and 13.67\% higher than the HoCs-GREEDY, HoCs-KWTA, HoCs-MADL, and HoCs-MARL methods, respectively. With a decision time interval of 10 minutes, HoCs-MPQ maintains a task completion rate of 75.96\%, with improvements of 57.64\%, 30.73\%, 18.96\%, and 17.96\% over the baseline methods. With a decision time interval of 15 minutes, although HoCs-MPQ's average task completion rate decreases to 63.07\%, it still shows improvements of 50.10\%, 21.19\%, 13.57\%, and 13.32\% compared to the baseline methods. 

These results  indicate that the decision time interval has a significant impact on the  task completion rate, while HoCs-MPQ demonstrates significant advantages under different time constraints.

\subsubsection{Experimental Results Under Different Limit Times}

\begin{figure}[h]
	\centering
	\setlength{\abovecaptionskip}{0.2cm}
	\setlength{\belowcaptionskip}{-0.25cm}
	\includegraphics[width=\linewidth]{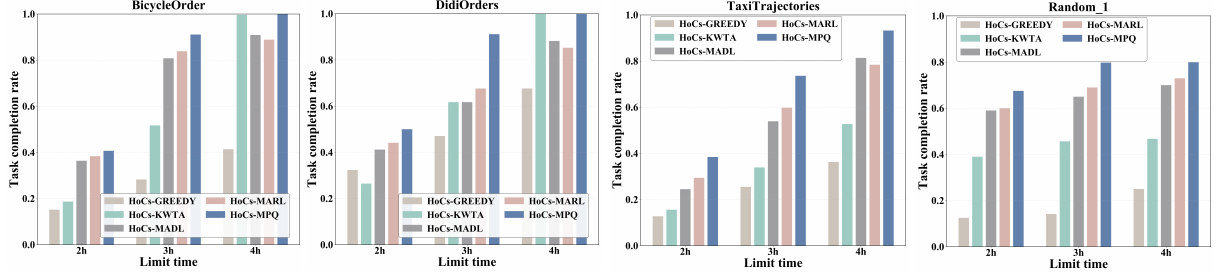}
	\caption{Experimental results under different limit times.}
	\Description{A graph showing the task completion rate of different algorithms under three duration times (2h, 3h, 4h).}
	\label{figure8}
\end{figure}

As shown in Fig. \ref{figure8}, as the duration time increases from 2 hours to 4 hours, the task completion rate of the HoCs-MPQ algorithm consistently outperforms other baseline algorithms. Specifically, with a duration time of 2 hours, HoCs-MPQ's average task completion rate is 49.16\%, which is 30.97\%, 24.22\%, 8.91\%, and 6.41\% higher than the HoCs-GREEDY, HoCs-KWTA, HoCs-MADL, and HoCs-MARL methods, respectively. With a duration time of 3 hours, HoCs-MPQ's average task completion rate increases to 83.92\%, with improvements of 55.17\%, 35.65\%, 14.17\%, and 12.92\% over the baseline methods. As the duration time increases to 4 hours, HoCs-MPQ's average task completion rate further increases to 93.29\%, still showing improvements of 50.71\%, 18.51\%, 10.79\%, and 12.04\% over the baseline methods. 

These results  indicates that as the duration time increases, the task completion rate of all algorithms improves, but the HoCs-MPQ algorithm maintains significant advantages under various duration time conditions, especially in performance within limited time.

\subsubsection{Experimental Results Under Different Background Factors}

\begin{figure}[htpb]
	\centering
	\setlength{\abovecaptionskip}{0.2cm}
	\setlength{\belowcaptionskip}{-0.25cm}
	\includegraphics[width=\linewidth]{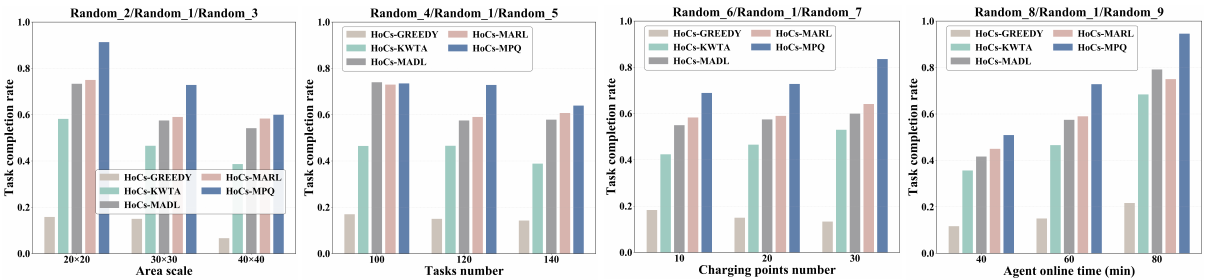}
	\caption{Experimental results under different background factors.}
	\Description{A set of four graphs showing the task completion rate of different algorithms under varying area scales, numbers of  tasks, numbers of charging points, and agent online durations.}
	\label{figure9}
\end{figure}

As shown in Fig. \ref{figure9}, under different conditions of area scale, number of  tasks, number of charging points, and agent online duration, the HoCs-MPQ algorithm demonstrates significant advantages. In the area scale experiment, HoCs-MPQ's average task completion rate is 74.72\%, which is 62.22\%, 26.91\%, 13.05\%, and 10.72\% higher than the HoCs-GREEDY, HoCs-KWTA, HoCs-MADL, and HoCs-MARL methods, respectively. In the  task quantity experiment, HoCs-MPQ's average task completion rate is 70.09\%, with improvements of 54.66\%, 26.11\%, 6.76\%, and 5.76\% over the baseline methods. In the charging point quantity experiment, HoCs-MPQ's average task completion rate reaches 75.11\%, with improvements of 59.56\%, 27.78\%, 17.44\%, and 14.78\% over the baseline methods. In the agent online duration experiment, HoCs-MPQ's average task completion rate is 72.78\%, with improvements of 56.67\%, 22.56\%, 13.11\%, and 13.11\% over the baseline methods. 

These results indicate that the HoCs-MPQ algorithm can maintain stable high performance under various environmental factor changes, with the most outstanding performance in the area scale and charging point quantity experiments.
\subsubsection{Experimental Results Under Different Agent Configurations}

\begin{figure}[htbp]
	\centering
	\setlength{\abovecaptionskip}{0.2cm}
	\setlength{\belowcaptionskip}{-0.25cm}
	\includegraphics[width=\linewidth]{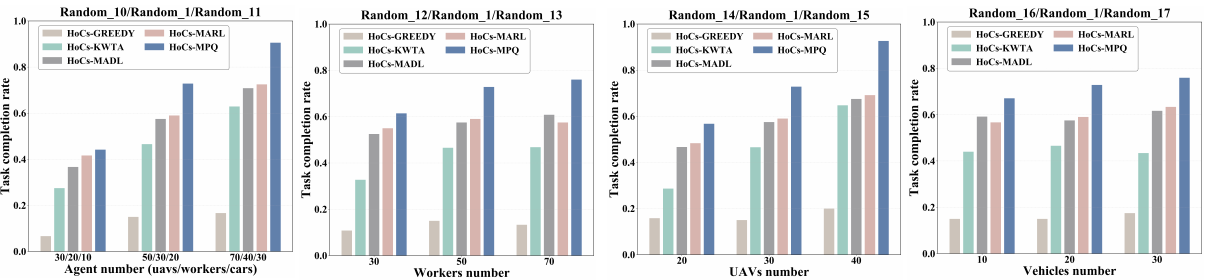}
	\caption{Experimental results under different agent configurations.}
	\Description{A set of four graphs showing the task completion rate of different algorithms under varying total numbers of agents, numbers of workers, numbers of UAVs, and numbers of vehicles.}
	\label{figure10}
\end{figure}

As shown in Fig. \ref{figure10}, the HoCs-MPQ algorithm demonstrates significant advantages under different agent quantity configurations. In the total agent quantity experiment, HoCs-MPQ's average task completion rate is 69.19\%, which is 56.41\%, 23.52\%, 13.86\%, and 11.19\% higher than the HoCs-GREEDY, HoCs-KWTA, HoCs-MADL, and HoCs-MARL methods, respectively. In the worker quantity experiment, HoCs-MPQ's average task completion rate is 70.08\%, with improvements of 57.03\%, 28.03\%, 12.75\%, and 12.75\% over the baseline methods. In the UAV quantity experiment, HoCs-MPQ's average task completion rate is 74.08\%, with improvements of 57.14\%, 27.41\%, 16.41\%, and 15.41\% over the baseline methods. In the vehicle quantity experiment, HoCs-MPQ's average task completion rate is 71.94\%, with improvements of 56.11\%, 27.27\%, 12.27\%, and 12.27\% over the baseline methods. 

These results indicate that the HoCs-MPQ algorithm can maintain stable high performance under various agent configurations and has strong adaptability to changes in the number of different types of agents.

\subsubsection{Experimental Results Under Different Power Factors}

\begin{figure}[htbp]
  \centering
  \setlength{\abovecaptionskip}{0.2cm}
  \setlength{\belowcaptionskip}{-0.25cm}
  \includegraphics[width=\linewidth]{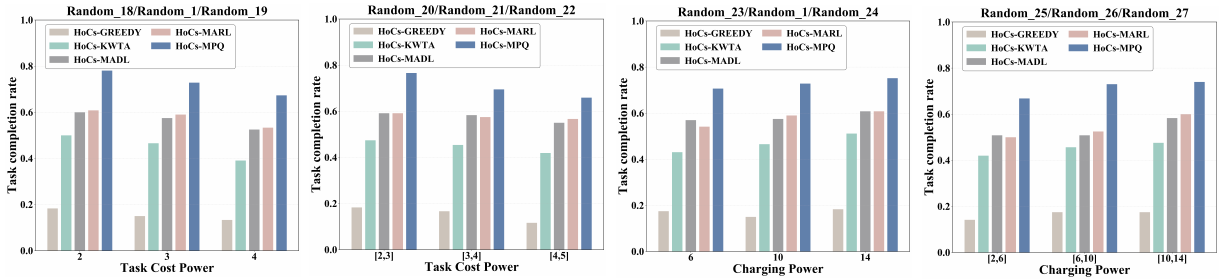}
  \caption{Experimental results under different power factors.}
  \Description{A set of two graphs showing the task completion rate of different algorithms under varying  task power consumption and charging power.}
  \label{figure11}
\end{figure}

As shown in Fig. \ref{figure11}, the HoCs-MPQ algorithm demonstrates significant advantages under different  task power consumption and charging power conditions. In the  task power consumption experiment, HoCs-MPQ's average task completion rate is 71.71\%, which is 56.15\%, 26.64\%, 14.54\%, and 13.88\% higher than the HoCs-GREEDY, HoCs-KWTA, HoCs-MADL, and HoCs-MARL methods, respectively. In the charging power experiment, HoCs-MPQ's average task completion rate is 71.75\%, with improvements of 55.08\%, 25.74\%, 15.75\%, and 15.58\% over the baseline methods. 

These results indicate that the HoCs-MPQ algorithm has strong adaptability to changes in energy-related parameters, especially showing extremely significant advantages compared to the HoCs-GREEDY algorithm, while maintaining stable performance when facing high power consumption tasks.

\subsection{Analysis and Discussion}
\subsubsection{Simulation Trajectory Analysis of Multiple Agents}
\begin{figure}[htbp]
  \centering
    \setlength{\abovecaptionskip}{0.2cm}
    \setlength{\belowcaptionskip}{-0.25cm}
    \begin{subfigure}[b]{0.48\linewidth} 
        \centering
       \includegraphics[height=4cm]{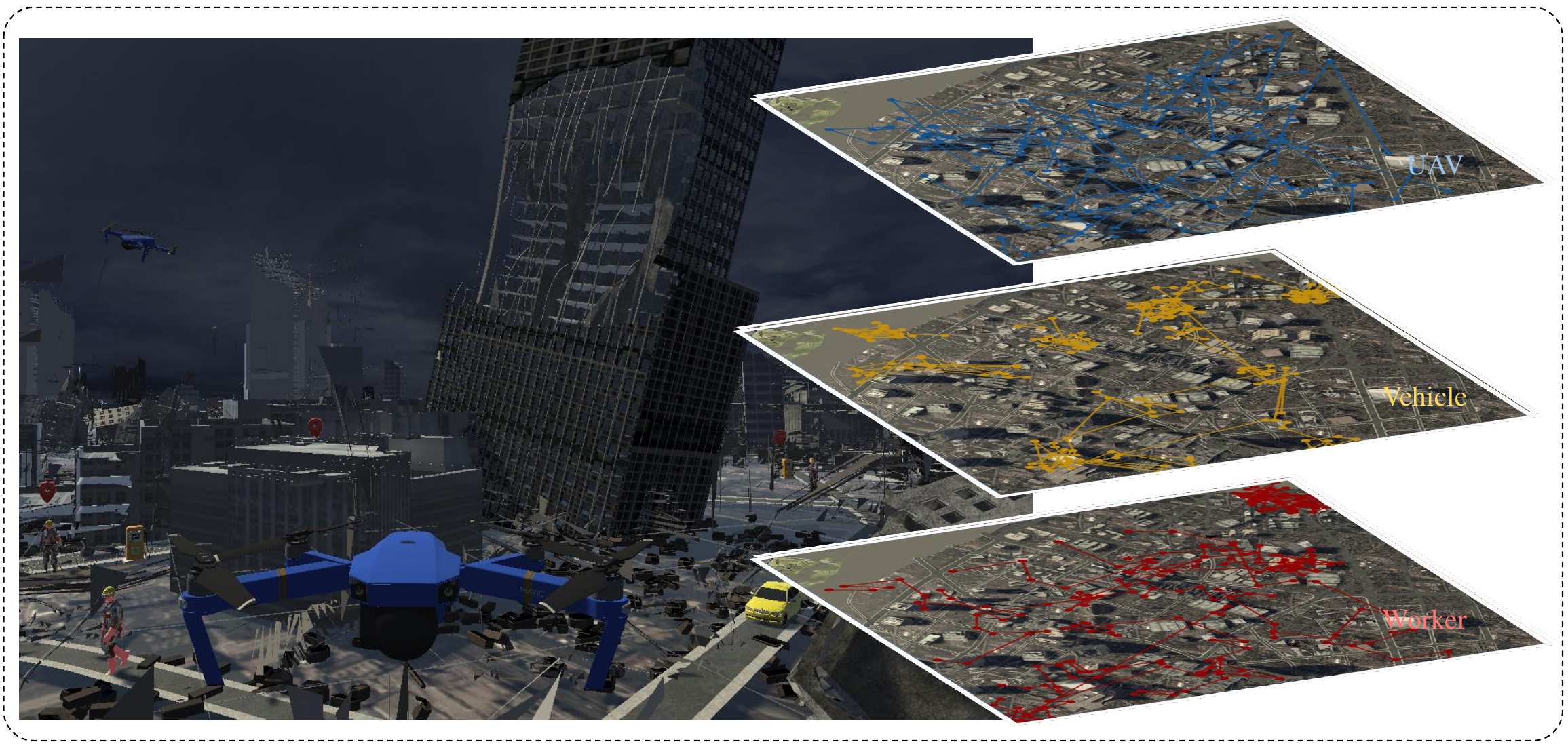} 
	\caption{Multi-agent coordination in 3D disaster scenarios.}
	\Description{Multi-agent coordination in 3D disaster scenarios.}
	\label{simulation}
    \end{subfigure}
    \hfill 
    \begin{subfigure}[b]{0.48\linewidth} 
        \centering
        \includegraphics[height=4cm]{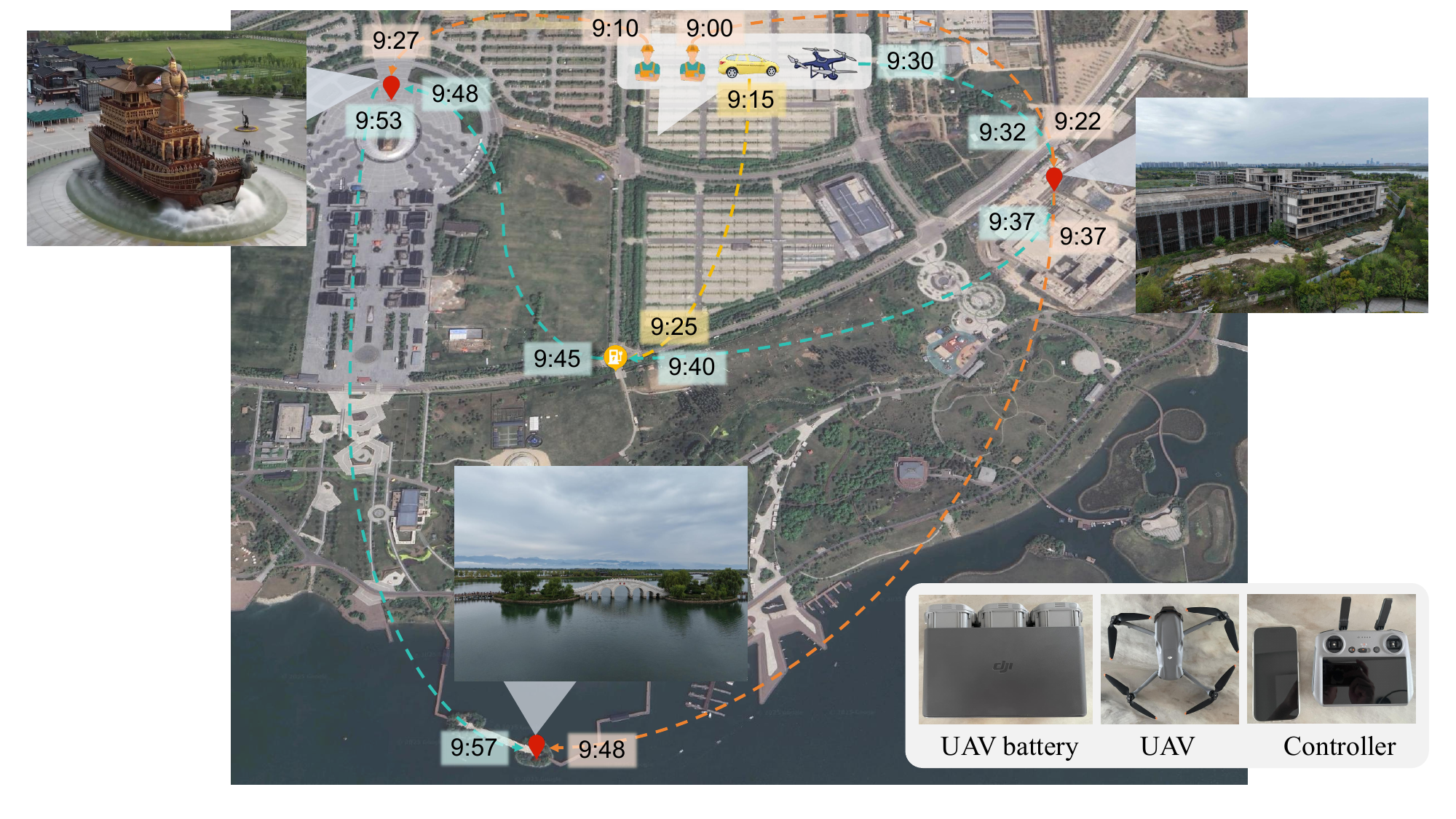} 
        \caption{Real-world experiment.} 
        \label{real-world experiment} 
    \end{subfigure}
    \caption{Overview of simulation and real-world experiment scenarios.} 
    \label{fig:combined_scenarios} 
\end{figure}

We constructed a 3D urban ruins scenario using the Unity engine\cite{unity}, in which the initial positions of 10 unmanned aerial vehicles (UAVs), 10 vehicles, and 15 workers were configured. The collaborative scheduling of these heterogeneous agents was carried out based on the proposed algorithm HoAs-MPQ. As shown in Fig.  \ref{simulation}, simulation results show that the trajectories of the UAVs, vehicles, and workers exhibit significant spatial and temporal overlap, indicating a high level of interaction and coordination. These findings suggest that HoAs-MPQ effectively facilitates collaborative behavior among the agents, thereby improving task efficiency and overall system performance.Furthermore, constrained by venue availability, policy regulations, and equipment limitations, we conducted a small-scale, restricted real-world scenario validation. As illustrated in Fig.~\ref{real-world experiment}, our experimental setup comprised a single UAV, a three-cell UAV battery pack, a mobile phone, and a UAV control handle. During the field test conducted at Xi'an Kunming Pool Qixi Park, the UAV successfully captured photographs of three different buildings, namely a large commemorative statue, an unfinished building, and an aquatic landscape, at 9:22, 9:48, and 9:57, with a battery replacement performed at 9:40.These real-world experiments demonstrate the practical applicability and operational viability of our proposed algorithm in authentic scenarios.

\subsubsection{Comparison Between HoCs-ILS and HoCs-MPQ Methods}

As shown in Table \ref{table3}, The experimental results show that HoCs-MPQ significantly improves computational efficiency on large-scale graph structures, with the advantage becoming more apparent as the number of nodes increases. On the TaxiTrajectories dataset, the decision time is reduced by 98.7\%, and on the Random\_1 dataset, it is reduced by 99.7\%. On the small-scale DidiOrders dataset, the two algorithms perform similarly, confirming that our algorithm's performance advantage positively correlates with the graph scale.The above experimental data fully demonstrates that the HoCs-MPQ algorithm exhibits excellent performance and time efficiency on various real-world scenarios and randomly generated datasets. As the scenario scale and problem complexity increase, the advantages of HoCs-MPQ become more apparent. While maintaining similar solution quality, HoCs-MPQ reduces computation time from tens of seconds to less than 1 second, which has important practical value for real-time scheduling systems.

\begin{table}[htbp]
	\centering
	\setlength{\abovecaptionskip}{0.2cm}
	\setlength{\belowcaptionskip}{-0.25cm}
	\caption{Comparison between HoCs-ILS and HoCs-MPQ methods.} 
	\label{table3}
	\begin{tabular}{c|cc|cc}
		\hline
		Dataset& \multicolumn{2}{c|}{Task completion rate} & \multicolumn{2}{c}{Decision time (s)} \\
		\cline{2-5}
		& HoCs-ILS & HoCs-MPQ & HoCs-ILS & HoCs-MPQ \\
		\hline
		TaxiTrajectories & 0.6569 & 0.6716 & 38.48 & 0.50 \\
		BicycleOrder & 0.8586 & 0.8444 & 32.23 & 0.58 \\
		DidiOrders & 0.8529 & 0.7941 & 0.10 & 0.10 \\
		Random\_1 & 0.7250 & 0.7283 & 56.38 & 0.18 \\
		\hline
	\end{tabular}
\end{table}

\subsubsection{Comparison Between Hierarchical Weights and Uniform Weights}

For the uniform weight experiment, Equation  (\ref{eq:4-18}) was mainly modified to:
\begin{equation}
icmUiWj=\frac{\text{Softplus}(1)}{max(ceilU,ceilW)}+icmWj+icmUi
\end{equation}

And Equation  (\ref{eq:4-19}) was modified to:
\begin{equation}
icmUiVk=\frac{\text{Softplus}\left((e^{(1-{uPow}_i/{\mathrm{FullPower}}_i)})/e\right)}{max(ceilU,ceilV)}+icmVk+icmUi
\end{equation}

As shown in Table \ref{table4}, the hierarchical weight scheme outperforms the uniform weight scheme mainly for the following reasons:
(1) Clear distinction of decision priorities: Hierarchical weights assign clear priorities to different types of nodes through order-of-magnitude differences, allowing the algorithm to first focus on task completion, then consider energy replenishment, and finally consider potential benefits from operations. This hierarchical design directly corresponds to the optimization objective (maximizing task completion), while uniform weights cannot effectively distinguish node importance.
(2) Optimization of resource allocation strategies: The hierarchical weight mechanism enables the algorithm to make more reasonable decisions when resources are limited. The experimental results show that on the BicycleOrder dataset, the task completion rate of the hierarchical weight scheme is 39.15\% higher than that of the uniform weight scheme, indicating that clear priority division can significantly improve system performance in complex scenarios.
(3) Consideration of long-term decision benefits: Hierarchical weights not only focus on immediate returns but also guide the algorithm to make decisions beneficial for long-term optimization objectives through weight design. For example, setting moderate weights for UAV charging nodes ensures the necessity of energy replenishment while avoiding the loss of task execution opportunities due to excessive charging.
(4) Adaptability to problem characteristics: The hierarchical weight scheme is designed according to the optimization objective of \textbf{Problem 1 }, assigning the highest weight to task completion nodes, fully reflecting the core focus of the problem, while the uniform weight scheme lacks this problem-specific adaptation.
\begin{table}[htbp]
	\centering
	\setlength{\abovecaptionskip}{0.2cm}
	\setlength{\belowcaptionskip}{-0.25cm}
	\caption{Comparison of task completion rate between HoCs-MPQ with uniform weights and hierarchical weights.}
	\label{table4}
	\begin{tabular}{c|cccc}
		\hline
		Dataset & BicycleOrder & DidiOrders & TaxiTrajectories & Random\_1 \\
		\hline
		HoCs-MPQ (Uniform) & 0.4529 & 0.7353 & 0.5707 & 0.5042 \\
    		HoCs-MPQ (Hierarchical) & 0.8444 & 0.7941 & 0.6716 & 0.7283 \\
		\hline
	\end{tabular}
\end{table}

\begin{figure}[htbp]
  \centering
  \setlength{\abovecaptionskip}{0.2cm}
  \includegraphics[width=1\linewidth,keepaspectratio]{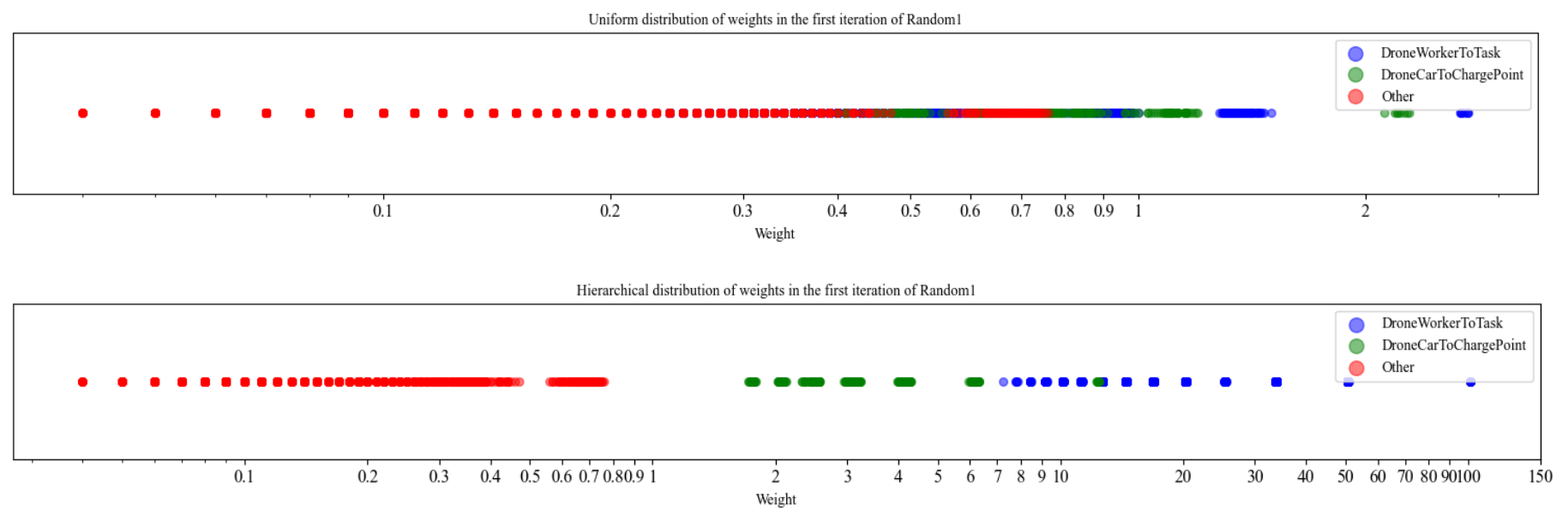}
  \caption{Distribution of weights in the first iteration of Random1.}
  \Description{Distribution of weights in the first iteration of Random1}
  \label{weight}
\end{figure}

Hierarchical weight design enables the scheduling algorithm to clearly distinguish the contributions of different node types to the overall scheduling objective, allowing the system to prioritize nodes that directly promote task completion, thus ensuring efficiency and rationality. As shown in Fig .\ref{weight}, the hierarchical weight setting results in a clear and non-overlapping distribution of the three types of node weights, further validating the critical role of hierarchical weights in improving system performance. Future work will further investigate multi-level adaptive weighting mechanisms to enhance the generalization and robustness of the algorithm in various application scenarios.

\subsubsection{Analysis of settings close to actual scenarios}

To thoroughly assess the practical applicability and resilience of our algorithm, we conducted a series of experiments incorporating various stochastic elements designed to mimic real-world operational challenges. Firstly, to account for the unpredictable effects of wind resistance on UAV operations, the energy consumed by UAVs per unit distance was randomized, drawing values from a uniform distribution within the range of [0.1, 0.4]. Secondly, considering the inevitable communication overhead in a distributed system, each scheduling decision incurred an additional random energy cost, sampled from [0.1, 0.5]. Thirdly, to evaluate the algorithm's robustness against system failures and network unreliability, agents were subjected to sudden failures, with each agent having a probability $p$ (between 0.01 and 0.05) of immediately going offline. Furthermore, acknowledging the potential for communication delays, data packet loss, or anomalous network fluctuations in practical scenarios, we simulated instances where successful matches could fail, with a probability ranging from 5\% to 10\%. The comparative outcomes under these challenging conditions are presented in Table \ref{tab:real_scenario_results}. These results unequivocally demonstrate that our algorithm consistently maintains high performance even under adverse and unpredictable circumstances, thereby validating its robust design and suitability for real-world deployments.

\begin{table}[htbp]
\centering
\setlength{\abovecaptionskip}{0.2cm}
\setlength{\belowcaptionskip}{-0.25cm}
\caption{Task completion rate under various realistic scenario settings.}
\label{tab:real_scenario_results}
\begin{tabular}{
    >{\raggedright\arraybackslash}p{2.5cm}|
    >{\centering\arraybackslash}p{1.2cm}
    >{\centering\arraybackslash}p{3.2cm}
    >{\centering\arraybackslash}p{2.2cm}
    >{\centering\arraybackslash}p{2.2cm}
    >{\centering\arraybackslash}p{1.5cm}
}
\hline
Datasets & base & wind resistance power cost & communication power cost & sudden agent failues & packet loss \\
\hline
BicycleOrder & 0.8444 & 0.8181 & 0.7878 & 0.7566 & 0.8101 \\
TaxiTrajectories & 0.6716 & 0.6578 & 0.6559 & 0.5882 & 0.6412 \\
DidiOrders & 0.7941 & 0.7618 & 0.7706 & 0.7324 & 0.7412 \\
Random\_1 & 0.7283 & 0.6833 & 0.6750 & 0.7008 & 0.6825 \\
\hline
\end{tabular}
\end{table}

\subsubsection{Discussion and Analysis of Scalability in Large-Scale Graphs}
To evaluate the scalability of our proposed HoCs-MPQ algorithm, we conducted experiments on distinct large-scale graph scenarios.Across the comparative experiments reported in Table~\ref{tab:scalability_results}, we isolate the distinct influences of agent count and spatial density (agent number/area scale) on HoCs-MPQ’s runtime. When agent, task, and charging point counts increase simultaneously (Row 1 vs. Row 2), decision time rises sharply (0.18 s → 673.2 s), indicating that sheer scale can render computation the principal bottleneck. With other parameters held constant (Row 2 vs. Row 3),  a decrease in spatial density reduces runtime, because denser deployments generate more coverage relations and produce a denser graph. 
In the large-scale probe (=1000 agents with proportionally scaled tasks, area, and charging points), holding spatial density approximately constant, a decrease in agent count reduces runtime (Row 4 vs. Row 5). Conversely, when both agent count and spatial density decrease (Row 4 vs. Row 6), the runtime also decreases, suggesting that sparser deployments with fewer agents lead to reduced graph complexity and lower computational cost.
Collectively, these results demonstrate that runtime is jointly determined by agent count and spatial density via their effect on graph sparsity and the number of coverage-dependent node combinations.

\begin{table*}[htbp] 
    \centering
    \setlength{\abovecaptionskip}{0.2cm}
    \setlength{\belowcaptionskip}{-0.25cm}
    \caption{HoCs-MPQ performance on large-scale graph scenarios.}
    \label{tab:scalability_results}
    \begin{tabular}{>{\raggedright\arraybackslash}p{1.4cm}|
    >{\centering\arraybackslash}p{1.2cm}
    >{\centering\arraybackslash}p{2.2cm}
    >{\centering\arraybackslash}p{2.2cm}
    >{\centering\arraybackslash}p{2.2cm}
    >{\centering\arraybackslash}p{2.8cm}
    >{\centering\arraybackslash}p{1.2cm}} 
        \hline
        Area scale
        & Tasks number 
        & Charging point number 
        & Agent number
        & Spatial density
        & Task completion rate 
        & Decision time (s) \\
        \hline
        30 $\times$ 30 & 120 & 20  & (50,30,20) & 1:9 & 0.9111 & 0.18 \\
        60 $\times$ 60 & 480 & 80  & (200,120,80) & 1:9 & 0.8313 & 673.2 \\
        60 $\times$ 60 & 480 & 80 & (80,48,32) & 1:22.5 & 0.3063 & 223.0 \\
        150 $\times$ 150 & 600 & 200 & (500,300,200) & 1:22.5 & 0.9317 &  433.6\\
        90 $\times$ 90 & 600 & 200 & (180,108,72) & 1:22.5 & 0.4967 & 291.3 \\
        150 $\times$ 150 & 600 & 200 & (300,180,120) & 1:37.5 & 0.4700 &  289.1\\
        \hline
    \end{tabular}
\end{table*}

Collectively, the comparative experiments reported in Table~\ref{tab:scalability_results} indicate that HoCs-MPQ’s computation time is governed not merely by the total number of agents but more critically by their spatial density. The root cause is the graph-construction stage: nodes and edges are instantiated only when a UAV's remaining energy suffices to cover a task or a charging point, so the effective graph size grows with coverage-dependent combinatorial terms. In particular, the number of graph elements scales with factors such as
(UAV count) $\times$ (worker count) $\times$ (tasks reachable by a fully charged UAV)
or
(UAV count) $\times$ (vehicle count) $\times$ (charging points reachable by a fully charged UAV),
which explains the approximate cubic behavior of graph construction, i.e., $O(N^3)$, where $N$ reflects these coverage-dependent combinations.Nevertheless, the worst-case $O(N^3)$ growth is unlikely to materialize in many practical, large-area disaster deployments: real scenarios typically exhibit relatively spatial density, substantially reducing the number of coverage relationships and thus keeping the actual graph size well below the worst-case bound. Moreover, queueing structures accelerate MWIS solving and further mitigate observed runtimes. Therefore, while we retain the theoretical caveat that initial graph construction is $O(N^3)$, both our experimental results and realistic deployment characteristics suggest that HoCs-MPQ’s runtime will not necessarily scale linearly or cubically with agent count in typical sparse, large-area scenarios.

\section{Conclusion and Future Work}
\label{section6}
\subsection{Conclusion}
\hspace{1em}Catastrophic natural disasters (such as earthquakes) are extremely destructive, and real-time, accurate sensing of the post-disaster environment situation is crucial for efficiently organizing post-disaster rescue operations. In this paper, we propose a Heterogeneous Multi-Agent Online Collaborative Scheduling Algorithm (HoCs-MPQ) for the collaborative scheduling problem of time-dependent UAVs, vehicles, and workers in post-disaster response crowd sensing.

Specifically, (1) we model the collaborative scheduling problem of UAVs, vehicles, and workers as a maximum weight independent set solution problem based on a weighted undirected graph. We capture the collaborative relationships between multiple elements through graph node design and embody the conflict relationships between multiple elements through edge connections between nodes. Then, by solving the maximum weight independent set of the weighted undirected graph, we maximize the collaboration between UAVs, workers, and vehicles to enhance sensing effects.(2) Since the weighted undirected graph constructed in this paper is a dense graph, solving its maximum weight independent set faces a huge computational burden. Under computational resource limitations and time sensitivity constraints, we significantly accelerate the solution process by utilizing the sequential indexing characteristics of multi-priority queues based on the graph structure characteristics.

Finally, we conducted detailed experiments based on rich real data and simulated data. Compared with baseline algorithms (HoCs-GREEDY, HoCs-KWTA, HoCs-MADL, HoCs-MARL), HoCs-MPQ shows significant improvement in task completion rate, with average increases of 54.13\%, 23.82\%, 14.12\%, and 12.89\%, respectively. Additionally, the computation time for a single online scheduling decision of HoCs-MPQ does not exceed 3 seconds, effectively meeting the low latency requirements of online scenario sensing decisions.

\subsection{Future Work}

\hspace{1em}However, there are still several aspects of this research worth further exploration:

(1) The weighted undirected graph constructed in this paper is a dense graph. Although we iteratively construct subgraphs through multi-priority queues to reduce its scale as much as possible, the time complexity of graph construction is still enormous. In the future, efficient weighted undirected graph construction algorithms will be the focus of our research.

(2) Multi-priority queues mainly sort nodes based on node weights, without fully utilizing the structural information of the graph and historical search experience. This may lead to repeated exploration of similar but suboptimal regions of the solution space, affecting the convergence efficiency of the algorithm. In the future, we should focus on exploring node evaluation methods based on self-learning, better guiding the selection of high-potential nodes by analyzing historical search trajectories and structural features of excellent solutions.

(3) To further enhance HoCs-MPQ's scalability for ultra-large-scale disaster response, future work will focus on optimizing candidate node management and search. We aim to apply Reinforcement Learning to optimize HoCs-ILS's internal search, learning to adaptively adjust parameters for faster convergence and reduced computation. Intelligent pre-filtering and pruning mechanisms will be developed to reduce the initial candidate pool for multi-priority queues, significantly lightening the computational burden. Furthermore, we will refine elastic expansion rules and queue prioritization through learned policies, enabling more adaptive and efficient exploration of the solution space.

\section{Acknowledgment}
This work was supported in part by the National Nat ural Science Foundation of China (No. 62502364, No. 62332014), the Postdoctoral Fellowship Program (Grade C) of China Postdoctoral Science Foundation (GZC20241315), the China Postdoctoral Science Foundation (2024M762554), the Key Research and Development Projects of Shaanxi Province (No. 2024GX-YBXM-107), the Natural Science Basic Research Program of Shaanxi (2025JC-YBQN-872) and the Fundamental Research Funds for the Central Universities (ZYTS25078). 

\bibliographystyle{ACM-Reference-Format}
\bibliography{sample}


\begin{thebibliography}{75}


\ifx \showCODEN    \undefined \def \showCODEN     #1{\unskip}     \fi
\ifx \showDOI      \undefined \def \showDOI       #1{#1}\fi
\ifx \showISBNx    \undefined \def \showISBNx     #1{\unskip}     \fi
\ifx \showISBNxiii \undefined \def \showISBNxiii  #1{\unskip}     \fi
\ifx \showISSN     \undefined \def \showISSN      #1{\unskip}     \fi
\ifx \showLCCN     \undefined \def \showLCCN      #1{\unskip}     \fi
\ifx \shownote     \undefined \def \shownote      #1{#1}          \fi
\ifx \showarticletitle \undefined \def \showarticletitle #1{#1}   \fi
\ifx \showURL      \undefined \def \showURL       {\relax}        \fi
\providecommand\bibfield[2]{#2}
\providecommand\bibinfo[2]{#2}
\providecommand\natexlab[1]{#1}
\providecommand\showeprint[2][]{arXiv:#2}

\bibitem[\protect\citeauthoryear{ADREM, SNSE, and et~al.}{ADREM
  et~al\mbox{.}}{2023}]%
        {ADREM2023}
\bibfield{author}{\bibinfo{person}{ADREM}, \bibinfo{person}{SNSE}, {and}
  \bibinfo{person}{NDRC et al.}} \bibinfo{year}{2023}\natexlab{}.
\newblock \bibinfo{title}{2023 Global Natural Disaster Assessment Report}.
\newblock
  \bibinfo{howpublished}{\url{https://www.gddat.cn/gw/micro-file-simple/api/file/showFile/920804e8-0192-89fdf283-0012-ff808081}}.
\newblock
\newblock
\shownote{Accessed: 2025-02-28, Published: 2024-10-13.}


\bibitem[\protect\citeauthoryear{Boltzmann}{Boltzmann}{1868}]%
        {boltzmann1868studien}
\bibfield{author}{\bibinfo{person}{Ludwig Boltzmann}.}
  \bibinfo{year}{1868}\natexlab{}.
\newblock \showarticletitle{Studien uber das gleichgewicht der lebenden kraft}.
\newblock \bibinfo{journal}{\emph{Wissenschafiliche Abhandlungen}}
  \bibinfo{volume}{1} (\bibinfo{year}{1868}), \bibinfo{pages}{49--96}.
\newblock


\bibitem[\protect\citeauthoryear{Bridle}{Bridle}{1990}]%
        {bridle1990probabilistic}
\bibfield{author}{\bibinfo{person}{John~S Bridle}.}
  \bibinfo{year}{1990}\natexlab{}.
\newblock \showarticletitle{Probabilistic interpretation of feedforward
  classification network outputs, with relationships to statistical pattern
  recognition}.
\newblock In \bibinfo{booktitle}{\emph{Neurocomputing: Algorithms,
  architectures and applications}}. \bibinfo{publisher}{Springer},
  \bibinfo{pages}{227--236}.
\newblock


\bibitem[\protect\citeauthoryear{Christie, Shoemaker, Kochersberger, Tokekar,
  McLean, and Leonessa}{Christie et~al\mbox{.}}{2016}]%
        {christie2016radiationsearchoperationsusing}
\bibfield{author}{\bibinfo{person}{Gordon Christie}, \bibinfo{person}{Adam
  Shoemaker}, \bibinfo{person}{Kevin Kochersberger}, \bibinfo{person}{Pratap
  Tokekar}, \bibinfo{person}{Lance McLean}, {and} \bibinfo{person}{Alexander
  Leonessa}.} \bibinfo{year}{2016}\natexlab{}.
\newblock \bibinfo{title}{Radiation Search Operations using Scene Understanding
  with Autonomous UAV and UGV}.
\newblock
\newblock
\showeprint[arxiv]{1609.00017}~[cs.RO]
\urldef\tempurl%
\url{https://arxiv.org/abs/1609.00017}
\showURL{%
\tempurl}


\bibitem[\protect\citeauthoryear{{CocaColaZero}}{{CocaColaZero}}{2025}]%
        {CocaColaZero}
\bibfield{author}{\bibinfo{person}{{CocaColaZero}}.}
  \bibinfo{year}{2025}\natexlab{}.
\newblock \bibinfo{title}{{Collaborative-Scheduling-of-Time-dependent-Agent}:
  GitHub Repository}.
\newblock
  \bibinfo{howpublished}{\url{https://github.com/CocaColaZero/Collaborative-Scheduling-of-Time-dependent-Agent}}.
\newblock
\newblock
\shownote{Accessed: 2025-04-26.}


\bibitem[\protect\citeauthoryear{DidiGAIA}{DidiGAIA}{2025}]%
        {gaia_didi}
DidiGAIA \bibinfo{year}{2025}\natexlab{}.
\newblock \bibinfo{title}{{GAIA: Didi’s Spatiotemporal Open Data Platform}}.
\newblock \bibinfo{howpublished}{\url{https://gaia.didichuxing.com}}.
\newblock
\newblock
\shownote{Accessed: 2025-04-25.}


\bibitem[\protect\citeauthoryear{Ding, Zhao, Cao, and Ma}{Ding
  et~al\mbox{.}}{2021}]%
        {ding2021crowdsourcing}
\bibfield{author}{\bibinfo{person}{Lige Ding}, \bibinfo{person}{Dong Zhao},
  \bibinfo{person}{Mingzhe Cao}, {and} \bibinfo{person}{Huadong Ma}.}
  \bibinfo{year}{2021}\natexlab{}.
\newblock \showarticletitle{When crowdsourcing meets unmanned vehicles: Toward
  cost-effective collaborative urban sensing via deep reinforcement learning}.
\newblock \bibinfo{journal}{\emph{IEEE Internet of Things Journal}}
  \bibinfo{volume}{8}, \bibinfo{number}{15} (\bibinfo{year}{2021}),
  \bibinfo{pages}{12150--12162}.
\newblock


\bibitem[\protect\citeauthoryear{{DJI Enterprise}}{{DJI Enterprise}}{2024}]%
        {dji_flighthub2}
\bibfield{author}{\bibinfo{person}{{DJI Enterprise}}.}
  \bibinfo{year}{2024}\natexlab{}.
\newblock \bibinfo{title}{{FlightHub 2 - Cloud-Based Drone Operations
  Management Platform}}.
\newblock
  \bibinfo{howpublished}{\url{https://enterprise.dji.com/cn/flighthub-2}}.
\newblock
\newblock
\shownote{Accessed: 2025-07-29.}


\bibitem[\protect\citeauthoryear{Dugas, Bengio, B{\'e}lisle, Nadeau, and
  Garcia}{Dugas et~al\mbox{.}}{2000}]%
        {dugas2000incorporating}
\bibfield{author}{\bibinfo{person}{Charles Dugas}, \bibinfo{person}{Yoshua
  Bengio}, \bibinfo{person}{Fran{\c{c}}ois B{\'e}lisle},
  \bibinfo{person}{Claude Nadeau}, {and} \bibinfo{person}{Ren{\'e} Garcia}.}
  \bibinfo{year}{2000}\natexlab{}.
\newblock \showarticletitle{Incorporating second-order functional knowledge for
  better option pricing}.
\newblock \bibinfo{journal}{\emph{Advances in neural information processing
  systems}}  \bibinfo{volume}{13} (\bibinfo{year}{2000}).
\newblock


\bibitem[\protect\citeauthoryear{Fu, Zhao, Min, Miao, Zhao, and Huang}{Fu
  et~al\mbox{.}}{2022}]%
        {fu2022energy}
\bibfield{author}{\bibinfo{person}{Luwei Fu}, \bibinfo{person}{Zhiwei Zhao},
  \bibinfo{person}{Geyong Min}, \bibinfo{person}{Wang Miao},
  \bibinfo{person}{Liang Zhao}, {and} \bibinfo{person}{Wenjie Huang}.}
  \bibinfo{year}{2022}\natexlab{}.
\newblock \showarticletitle{Energy-efficient 3-d data collection formulti-uav
  assisted mobile crowdsensing}.
\newblock \bibinfo{journal}{\emph{IEEE Trans. Comput.}} \bibinfo{volume}{72},
  \bibinfo{number}{7} (\bibinfo{year}{2022}), \bibinfo{pages}{2025--2038}.
\newblock


\bibitem[\protect\citeauthoryear{Ghassemi, DePauw, and Chowdhury}{Ghassemi
  et~al\mbox{.}}{2019}]%
        {ghassemi2019decentralized}
\bibfield{author}{\bibinfo{person}{Payam Ghassemi}, \bibinfo{person}{David
  DePauw}, {and} \bibinfo{person}{Souma Chowdhury}.}
  \bibinfo{year}{2019}\natexlab{}.
\newblock \showarticletitle{Decentralized dynamic task allocation in swarm
  robotic systems for disaster response}. In \bibinfo{booktitle}{\emph{2019
  international symposium on multi-robot and multi-agent systems (mrs)}}. IEEE,
  \bibinfo{pages}{83--85}.
\newblock


\bibitem[\protect\citeauthoryear{Gulati, {Kumar Boddu}, Kapila, Bangare,
  Chandnani, and Saravanan}{Gulati et~al\mbox{.}}{2022}]%
        {GULATI2022161}
\bibfield{author}{\bibinfo{person}{Kamal Gulati}, \bibinfo{person}{Raja~Sarath
  {Kumar Boddu}}, \bibinfo{person}{Dhiraj Kapila}, \bibinfo{person}{Sunil~L.
  Bangare}, \bibinfo{person}{Neeraj Chandnani}, {and} \bibinfo{person}{G.
  Saravanan}.} \bibinfo{year}{2022}\natexlab{}.
\newblock \showarticletitle{A review paper on wireless sensor network
  techniques in Internet of Things (IoT)}.
\newblock \bibinfo{journal}{\emph{Materials Today: Proceedings}}
  \bibinfo{volume}{51} (\bibinfo{year}{2022}), \bibinfo{pages}{161--165}.
\newblock
\showISSN{2214-7853}
\urldef\tempurl%
\url{https://doi.org/10.1016/j.matpr.2021.05.067}
\showDOI{\tempurl}
\newblock
\shownote{CMAE'21.}


\bibitem[\protect\citeauthoryear{Guo, Liu, and Yu}{Guo et~al\mbox{.}}{2022}]%
        {Guo2022HCPS}
\bibfield{author}{\bibinfo{person}{Bin Guo}, \bibinfo{person}{Sicong Liu},
  {and} \bibinfo{person}{Zhiwen Yu}.} \bibinfo{year}{2022}\natexlab{}.
\newblock \bibinfo{booktitle}{\emph{Cyber-Physical-Human Integrated Collective
  Intelligence Computing}}.
\newblock \bibinfo{publisher}{China Machine Press}.
\newblock
\newblock
\shownote{In Chinese.}


\bibitem[\protect\citeauthoryear{Han, Tu, Yu, Yu, Shan, Wang, and Guo}{Han
  et~al\mbox{.}}{2024}]%
        {10.1109/TNET.2024.3395493}
\bibfield{author}{\bibinfo{person}{Lei Han}, \bibinfo{person}{Chunyu Tu},
  \bibinfo{person}{Zhiwen Yu}, \bibinfo{person}{Zhiyong Yu},
  \bibinfo{person}{Weihua Shan}, \bibinfo{person}{Liang Wang}, {and}
  \bibinfo{person}{Bin Guo}.} \bibinfo{year}{2024}\natexlab{}.
\newblock \showarticletitle{Collaborative Route Planning of UAVs, Workers, and
  Cars for Crowdsensing in Disaster Response}.
\newblock \bibinfo{journal}{\emph{IEEE/ACM Trans. Netw.}} \bibinfo{volume}{32},
  \bibinfo{number}{4} (\bibinfo{date}{May} \bibinfo{year}{2024}),
  \bibinfo{pages}{3606–3621}.
\newblock
\showISSN{1063-6692}
\urldef\tempurl%
\url{https://doi.org/10.1109/TNET.2024.3395493}
\showDOI{\tempurl}


\bibitem[\protect\citeauthoryear{Herdel, Yamin, and Cauchard}{Herdel
  et~al\mbox{.}}{2022}]%
        {herdel2022above}
\bibfield{author}{\bibinfo{person}{Viviane Herdel}, \bibinfo{person}{Lee~J
  Yamin}, {and} \bibinfo{person}{Jessica~R Cauchard}.}
  \bibinfo{year}{2022}\natexlab{}.
\newblock \showarticletitle{Above and beyond: A scoping review of domains and
  applications for human-drone interaction}. In
  \bibinfo{booktitle}{\emph{Proceedings of the 2022 CHI Conference on Human
  Factors in Computing Systems}}. \bibinfo{pages}{1--22}.
\newblock


\bibitem[\protect\citeauthoryear{Hu, Lu, Xu, Xie, Chen, and Wang}{Hu
  et~al\mbox{.}}{2023}]%
        {hu2023collaboration}
\bibfield{author}{\bibinfo{person}{Yue Hu}, \bibinfo{person}{Yifan Lu},
  \bibinfo{person}{Runsheng Xu}, \bibinfo{person}{Weidi Xie},
  \bibinfo{person}{Siheng Chen}, {and} \bibinfo{person}{Yanfeng Wang}.}
  \bibinfo{year}{2023}\natexlab{}.
\newblock \showarticletitle{Collaboration helps camera overtake lidar in 3d
  detection}. In \bibinfo{booktitle}{\emph{Proceedings of the IEEE/CVF
  Conference on Computer Vision and Pattern Recognition}}.
  \bibinfo{pages}{9243--9252}.
\newblock


\bibitem[\protect\citeauthoryear{Iqbal, de~Witt, Peng, B{\"o}hmer, Whiteson,
  and Sha}{Iqbal et~al\mbox{.}}{2020}]%
        {iqbal2020ai}
\bibfield{author}{\bibinfo{person}{Shariq Iqbal}, \bibinfo{person}{Christian
  A~Schroeder de Witt}, \bibinfo{person}{Bei Peng}, \bibinfo{person}{Wendelin
  B{\"o}hmer}, \bibinfo{person}{Shimon Whiteson}, {and} \bibinfo{person}{Fei
  Sha}.} \bibinfo{year}{2020}\natexlab{}.
\newblock \showarticletitle{Ai-qmix: Attention and imagination for dynamic
  multi-agent reinforcement learning}.
\newblock \bibinfo{journal}{\emph{arXiv preprint arXiv:2006.04222}}
  (\bibinfo{year}{2020}).
\newblock


\bibitem[\protect\citeauthoryear{Kalaitzakis, Cain, Vitzilaios, Rekleitis, and
  Moulton}{Kalaitzakis et~al\mbox{.}}{2021}]%
        {kalaitzakis2021marsupial}
\bibfield{author}{\bibinfo{person}{Michail Kalaitzakis},
  \bibinfo{person}{Brennan Cain}, \bibinfo{person}{Nikolaos Vitzilaios},
  \bibinfo{person}{Ioannis Rekleitis}, {and} \bibinfo{person}{Jason Moulton}.}
  \bibinfo{year}{2021}\natexlab{}.
\newblock \showarticletitle{A marsupial robotic system for surveying and
  inspection of freshwater ecosystems}.
\newblock \bibinfo{journal}{\emph{Journal of Field Robotics}}
  \bibinfo{volume}{38}, \bibinfo{number}{1} (\bibinfo{year}{2021}),
  \bibinfo{pages}{121--138}.
\newblock


\bibitem[\protect\citeauthoryear{Kang, Song, Wang, Zhang, Dong, and Liu}{Kang
  et~al\mbox{.}}{2025}]%
        {KANG2025111189}
\bibfield{author}{\bibinfo{person}{Yunchuan Kang},
  \bibinfo{person}{Houbing~Herbert Song}, \bibinfo{person}{Tian Wang},
  \bibinfo{person}{Shaobo Zhang}, \bibinfo{person}{Mianxiong Dong}, {and}
  \bibinfo{person}{Anfeng Liu}.} \bibinfo{year}{2025}\natexlab{}.
\newblock \showarticletitle{A trust and bundling-based task allocation scheme
  to enhance completion rate and data quality for mobile crowdsensing}.
\newblock \bibinfo{journal}{\emph{Computer Networks}}  \bibinfo{volume}{262}
  (\bibinfo{year}{2025}), \bibinfo{pages}{111189}.
\newblock
\showISSN{1389-1286}
\urldef\tempurl%
\url{https://doi.org/10.1016/j.comnet.2025.111189}
\showDOI{\tempurl}


\bibitem[\protect\citeauthoryear{Krizmancic, Arbanas, Petrovic, Petric, and
  Bogdan}{Krizmancic et~al\mbox{.}}{2020}]%
        {krizmancic2020cooperative}
\bibfield{author}{\bibinfo{person}{Marko Krizmancic}, \bibinfo{person}{Barbara
  Arbanas}, \bibinfo{person}{Tamara Petrovic}, \bibinfo{person}{Frano Petric},
  {and} \bibinfo{person}{Stjepan Bogdan}.} \bibinfo{year}{2020}\natexlab{}.
\newblock \showarticletitle{Cooperative aerial-ground multi-robot system for
  automated construction tasks}.
\newblock \bibinfo{journal}{\emph{IEEE Robotics and Automation Letters}}
  \bibinfo{volume}{5}, \bibinfo{number}{2} (\bibinfo{year}{2020}),
  \bibinfo{pages}{798--805}.
\newblock


\bibitem[\protect\citeauthoryear{Kruijff, Kruijff-Korbayov{\'a}, Keshavdas,
  Larochelle, Jan{\'\i}{\v{c}}ek, Colas, Liu, Pomerleau, Siegwart, Neerincx,
  et~al\mbox{.}}{Kruijff et~al\mbox{.}}{2014}]%
        {kruijff2014designing}
\bibfield{author}{\bibinfo{person}{Geert-Jan~M Kruijff}, \bibinfo{person}{Ivana
  Kruijff-Korbayov{\'a}}, \bibinfo{person}{Shanker Keshavdas},
  \bibinfo{person}{Benoit Larochelle}, \bibinfo{person}{Miroslav
  Jan{\'\i}{\v{c}}ek}, \bibinfo{person}{Francis Colas}, \bibinfo{person}{Ming
  Liu}, \bibinfo{person}{Francois Pomerleau}, \bibinfo{person}{Roland
  Siegwart}, \bibinfo{person}{Mark~A Neerincx}, {et~al\mbox{.}}}
  \bibinfo{year}{2014}\natexlab{}.
\newblock \showarticletitle{Designing, developing, and deploying systems to
  support human--robot teams in disaster response}.
\newblock \bibinfo{journal}{\emph{Advanced Robotics}} \bibinfo{volume}{28},
  \bibinfo{number}{23} (\bibinfo{year}{2014}), \bibinfo{pages}{1547--1570}.
\newblock


\bibitem[\protect\citeauthoryear{Li, Cheng, Yuan, Wang, and Chen}{Li
  et~al\mbox{.}}{2019}]%
        {DBLP:conf/kdd/LiC0W019}
\bibfield{author}{\bibinfo{person}{Boyang Li}, \bibinfo{person}{Yurong Cheng},
  \bibinfo{person}{Ye Yuan}, \bibinfo{person}{Guoren Wang}, {and}
  \bibinfo{person}{Lei Chen}.} \bibinfo{year}{2019}\natexlab{}.
\newblock \showarticletitle{Three-Dimensional Stable Matching Problem for
  Spatial Crowdsourcing Platforms}. In \bibinfo{booktitle}{\emph{Proceedings of
  the 25th {ACM} {SIGKDD} International Conference on Knowledge Discovery {\&}
  Data Mining, {KDD} 2019, Anchorage, AK, USA, August 4-8, 2019}},
  \bibfield{editor}{\bibinfo{person}{Ankur Teredesai}, \bibinfo{person}{Vipin
  Kumar}, \bibinfo{person}{Ying Li}, \bibinfo{person}{R{\'{o}}mer Rosales},
  \bibinfo{person}{Evimaria Terzi}, {and} \bibinfo{person}{George Karypis}}
  (Eds.). \bibinfo{publisher}{{ACM}}, \bibinfo{pages}{1643--1653}.
\newblock
\urldef\tempurl%
\url{https://doi.org/10.1145/3292500.3330879}
\showDOI{\tempurl}


\bibitem[\protect\citeauthoryear{Li, Yuan, Cheng, Chai, and Lewis}{Li
  et~al\mbox{.}}{2024b}]%
        {li2024reinforcement}
\bibfield{author}{\bibinfo{person}{Jinna Li}, \bibinfo{person}{Lin Yuan},
  \bibinfo{person}{Weiran Cheng}, \bibinfo{person}{Tianyou Chai}, {and}
  \bibinfo{person}{Frank~L Lewis}.} \bibinfo{year}{2024}\natexlab{b}.
\newblock \showarticletitle{Reinforcement Learning for Synchronization of
  Heterogeneous Multiagent Systems by Improved $ Q $-Functions}.
\newblock \bibinfo{journal}{\emph{IEEE Transactions on Cybernetics}}
  (\bibinfo{year}{2024}).
\newblock


\bibitem[\protect\citeauthoryear{Li, Zheng, Zhang, and Chen}{Li
  et~al\mbox{.}}{2015}]%
        {li2015traffic}
\bibfield{author}{\bibinfo{person}{Yexin Li}, \bibinfo{person}{Yu Zheng},
  \bibinfo{person}{Huichu Zhang}, {and} \bibinfo{person}{Lei Chen}.}
  \bibinfo{year}{2015}\natexlab{}.
\newblock \showarticletitle{Traffic prediction in a bike-sharing system}. In
  \bibinfo{booktitle}{\emph{Proceedings of the 23rd SIGSPATIAL international
  conference on advances in geographic information systems}}.
  \bibinfo{pages}{1--10}.
\newblock


\bibitem[\protect\citeauthoryear{Li, Xia, Tang, Xu, Shi, Xia, Yin, and
  Huang}{Li et~al\mbox{.}}{2024a}]%
        {li2024urbangpt}
\bibfield{author}{\bibinfo{person}{Zhonghang Li}, \bibinfo{person}{Lianghao
  Xia}, \bibinfo{person}{Jiabin Tang}, \bibinfo{person}{Yong Xu},
  \bibinfo{person}{Lei Shi}, \bibinfo{person}{Long Xia}, \bibinfo{person}{Dawei
  Yin}, {and} \bibinfo{person}{Chao Huang}.} \bibinfo{year}{2024}\natexlab{a}.
\newblock \showarticletitle{Urbangpt: Spatio-temporal large language models}.
  In \bibinfo{booktitle}{\emph{Proceedings of the 30th ACM SIGKDD Conference on
  Knowledge Discovery and Data Mining}}. \bibinfo{pages}{5351--5362}.
\newblock


\bibitem[\protect\citeauthoryear{Liu, Liu, Stone, Garg, Zhu, and
  Anandkumar}{Liu et~al\mbox{.}}{2021}]%
        {liu2021coach}
\bibfield{author}{\bibinfo{person}{Bo Liu}, \bibinfo{person}{Qiang Liu},
  \bibinfo{person}{Peter Stone}, \bibinfo{person}{Animesh Garg},
  \bibinfo{person}{Yuke Zhu}, {and} \bibinfo{person}{Anima Anandkumar}.}
  \bibinfo{year}{2021}\natexlab{}.
\newblock \showarticletitle{Coach-player multi-agent reinforcement learning for
  dynamic team composition}. In \bibinfo{booktitle}{\emph{International
  Conference on Machine Learning}}. PMLR, \bibinfo{pages}{6860--6870}.
\newblock


\bibitem[\protect\citeauthoryear{Liu, Chen, and Zhan}{Liu
  et~al\mbox{.}}{2019}]%
        {liu2019energy}
\bibfield{author}{\bibinfo{person}{Chi~Harold Liu}, \bibinfo{person}{Zheyu
  Chen}, {and} \bibinfo{person}{Yufeng Zhan}.} \bibinfo{year}{2019}\natexlab{}.
\newblock \showarticletitle{Energy-efficient distributed mobile crowd sensing:
  A deep learning approach}.
\newblock \bibinfo{journal}{\emph{IEEE Journal on Selected Areas in
  Communications}} \bibinfo{volume}{37}, \bibinfo{number}{6}
  (\bibinfo{year}{2019}), \bibinfo{pages}{1262--1276}.
\newblock


\bibitem[\protect\citeauthoryear{Liu, Piao, and Tang}{Liu
  et~al\mbox{.}}{2020a}]%
        {liu2020energy}
\bibfield{author}{\bibinfo{person}{Chi~Harold Liu}, \bibinfo{person}{Chengzhe
  Piao}, {and} \bibinfo{person}{Jian Tang}.} \bibinfo{year}{2020}\natexlab{a}.
\newblock \showarticletitle{Energy-efficient UAV crowdsensing with multiple
  charging stations by deep learning}. In \bibinfo{booktitle}{\emph{IEEE
  INFOCOm 2020-IEEE conference on computer communications}}. IEEE,
  \bibinfo{pages}{199--208}.
\newblock


\bibitem[\protect\citeauthoryear{Liu, Zhao, Dai, Yuan, Wang, Wu, and Leung}{Liu
  et~al\mbox{.}}{2020b}]%
        {liu2020curiosity}
\bibfield{author}{\bibinfo{person}{Chi~Harold Liu}, \bibinfo{person}{Yinuo
  Zhao}, \bibinfo{person}{Zipeng Dai}, \bibinfo{person}{Ye Yuan},
  \bibinfo{person}{Guoren Wang}, \bibinfo{person}{Dapeng Wu}, {and}
  \bibinfo{person}{Kin~K Leung}.} \bibinfo{year}{2020}\natexlab{b}.
\newblock \showarticletitle{Curiosity-driven energy-efficient worker scheduling
  in vehicular crowdsourcing: A deep reinforcement learning approach}. In
  \bibinfo{booktitle}{\emph{2020 IEEE 36th International conference on data
  engineering (ICDE)}}. IEEE, \bibinfo{pages}{25--36}.
\newblock


\bibitem[\protect\citeauthoryear{Liu and Zhang}{Liu and Zhang}{2025}]%
        {DBLP:journals/tnn/LiuZ25}
\bibfield{author}{\bibinfo{person}{Kexin Liu} {and} \bibinfo{person}{Yinyan
  Zhang}.} \bibinfo{year}{2025}\natexlab{}.
\newblock \showarticletitle{Distributed Dynamic Task Allocation for Moving
  Target Tracking of Networked Mobile Robots Using k-WTA Network}.
\newblock \bibinfo{journal}{\emph{{IEEE} Trans. Neural Networks Learn. Syst.}}
  \bibinfo{volume}{36}, \bibinfo{number}{3} (\bibinfo{year}{2025}),
  \bibinfo{pages}{5795--5802}.
\newblock
\urldef\tempurl%
\url{https://doi.org/10.1109/TNNLS.2024.3377433}
\showDOI{\tempurl}


\bibitem[\protect\citeauthoryear{Liu, Li, Chen, Qi, and Jin}{Liu
  et~al\mbox{.}}{2024}]%
        {DBLP:journals/tmc/LiuLCQJ24}
\bibfield{author}{\bibinfo{person}{Mei Liu}, \bibinfo{person}{Yutong Li},
  \bibinfo{person}{Yingqi Chen}, \bibinfo{person}{Yimeng Qi}, {and}
  \bibinfo{person}{Long Jin}.} \bibinfo{year}{2024}\natexlab{}.
\newblock \showarticletitle{A Distributed Competitive and Collaborative
  Coordination for Multirobot Systems}.
\newblock \bibinfo{journal}{\emph{{IEEE} Trans. Mob. Comput.}}
  \bibinfo{volume}{23}, \bibinfo{number}{12} (\bibinfo{year}{2024}),
  \bibinfo{pages}{11436--11448}.
\newblock
\urldef\tempurl%
\url{https://doi.org/10.1109/TMC.2024.3397242}
\showDOI{\tempurl}


\bibitem[\protect\citeauthoryear{Louren{\c{c}}o, Martin, and
  St{\"u}tzle}{Louren{\c{c}}o et~al\mbox{.}}{2018}]%
        {lourencco2018iterated}
\bibfield{author}{\bibinfo{person}{Helena~Ramalhinho Louren{\c{c}}o},
  \bibinfo{person}{Olivier~C Martin}, {and} \bibinfo{person}{Thomas
  St{\"u}tzle}.} \bibinfo{year}{2018}\natexlab{}.
\newblock \showarticletitle{Iterated local search: Framework and applications}.
\newblock In \bibinfo{booktitle}{\emph{Handbook of metaheuristics}}.
  \bibinfo{publisher}{Springer}, \bibinfo{pages}{129--168}.
\newblock


\bibitem[\protect\citeauthoryear{Luo, Deng, Zhang, Ke, Wan, and Qian}{Luo
  et~al\mbox{.}}{2024}]%
        {luo2024collaborative}
\bibfield{author}{\bibinfo{person}{Yiwen Luo}, \bibinfo{person}{Xiaoheng Deng},
  \bibinfo{person}{Wendong Zhang}, \bibinfo{person}{Yan Ke},
  \bibinfo{person}{Shaohua Wan}, {and} \bibinfo{person}{Yurong Qian}.}
  \bibinfo{year}{2024}\natexlab{}.
\newblock \showarticletitle{Collaborative intelligent delivery with one truck
  and multiple heterogeneous drones in COVID-19 pandemic environment}.
\newblock \bibinfo{journal}{\emph{IEEE Transactions on Intelligent
  Transportation Systems}} \bibinfo{volume}{25}, \bibinfo{number}{7}
  (\bibinfo{year}{2024}), \bibinfo{pages}{7907--7920}.
\newblock


\bibitem[\protect\citeauthoryear{Manjunatha, Distefano, Jani, Ghassemi,
  Chowdhury, Dantu, Doermann, and Esfahani}{Manjunatha et~al\mbox{.}}{2020}]%
        {manjunatha2020using}
\bibfield{author}{\bibinfo{person}{Hemanth Manjunatha},
  \bibinfo{person}{Joseph~P Distefano}, \bibinfo{person}{Apurv Jani},
  \bibinfo{person}{Payam Ghassemi}, \bibinfo{person}{Souma Chowdhury},
  \bibinfo{person}{Karthik Dantu}, \bibinfo{person}{David Doermann}, {and}
  \bibinfo{person}{Ehsan~T Esfahani}.} \bibinfo{year}{2020}\natexlab{}.
\newblock \showarticletitle{Using physiological measurements to analyze the
  tactical decisions in human swarm teams}. In \bibinfo{booktitle}{\emph{2020
  IEEE International Conference on Systems, Man, and Cybernetics (SMC)}}. IEEE,
  \bibinfo{pages}{256--261}.
\newblock


\bibitem[\protect\citeauthoryear{McEntire}{McEntire}{2021}]%
        {McEntire2021}
\bibfield{author}{\bibinfo{person}{David~A. McEntire}.}
  \bibinfo{year}{2021}\natexlab{}.
\newblock \bibinfo{booktitle}{\emph{Disaster Response and Recovery: Strategies
  and Tactics for Resilience}}.
\newblock \bibinfo{publisher}{John Wiley \& Sons}.
\newblock


\bibitem[\protect\citeauthoryear{Michael, Shen, Mohta, Kumar, Nagatani, Okada,
  Kiribayashi, Otake, Yoshida, Ohno, et~al\mbox{.}}{Michael
  et~al\mbox{.}}{2014}]%
        {michael2014collaborative}
\bibfield{author}{\bibinfo{person}{Nathan Michael}, \bibinfo{person}{Shaojie
  Shen}, \bibinfo{person}{Kartik Mohta}, \bibinfo{person}{Vijay Kumar},
  \bibinfo{person}{Keiji Nagatani}, \bibinfo{person}{Yoshito Okada},
  \bibinfo{person}{Seiga Kiribayashi}, \bibinfo{person}{Kazuki Otake},
  \bibinfo{person}{Kazuya Yoshida}, \bibinfo{person}{Kazunori Ohno},
  {et~al\mbox{.}}} \bibinfo{year}{2014}\natexlab{}.
\newblock \showarticletitle{Collaborative mapping of an earthquake damaged
  building via ground and aerial robots}. In \bibinfo{booktitle}{\emph{Field
  and service robotics: results of the 8th international conference}}.
  Springer, \bibinfo{pages}{33--47}.
\newblock


\bibitem[\protect\citeauthoryear{Miller, Cladera, Smith, Taylor, and
  Kumar}{Miller et~al\mbox{.}}{2022}]%
        {miller2022stronger}
\bibfield{author}{\bibinfo{person}{Ian~D Miller}, \bibinfo{person}{Fernando
  Cladera}, \bibinfo{person}{Trey Smith}, \bibinfo{person}{Camillo~Jose
  Taylor}, {and} \bibinfo{person}{Vijay Kumar}.}
  \bibinfo{year}{2022}\natexlab{}.
\newblock \showarticletitle{Stronger together: Air-ground robotic collaboration
  using semantics}.
\newblock \bibinfo{journal}{\emph{IEEE Robotics and Automation Letters}}
  \bibinfo{volume}{7}, \bibinfo{number}{4} (\bibinfo{year}{2022}),
  \bibinfo{pages}{9643--9650}.
\newblock


\bibitem[\protect\citeauthoryear{Minaeian, Liu, and Son}{Minaeian
  et~al\mbox{.}}{2015}]%
        {minaeian2015vision}
\bibfield{author}{\bibinfo{person}{Sara Minaeian}, \bibinfo{person}{Jian Liu},
  {and} \bibinfo{person}{Young-Jun Son}.} \bibinfo{year}{2015}\natexlab{}.
\newblock \showarticletitle{Vision-based target detection and localization via
  a team of cooperative UAV and UGVs}.
\newblock \bibinfo{journal}{\emph{IEEE Transactions on systems, man, and
  cybernetics: systems}} \bibinfo{volume}{46}, \bibinfo{number}{7}
  (\bibinfo{year}{2015}), \bibinfo{pages}{1005--1016}.
\newblock


\bibitem[\protect\citeauthoryear{Mohsan, Khan, Noor, Ullah, and
  Alsharif}{Mohsan et~al\mbox{.}}{2022}]%
        {mohsan2022towards}
\bibfield{author}{\bibinfo{person}{Syed Agha~Hassnain Mohsan},
  \bibinfo{person}{Muhammad~Asghar Khan}, \bibinfo{person}{Fazal Noor},
  \bibinfo{person}{Insaf Ullah}, {and} \bibinfo{person}{Mohammed~H Alsharif}.}
  \bibinfo{year}{2022}\natexlab{}.
\newblock \showarticletitle{Towards the unmanned aerial vehicles (UAVs): A
  comprehensive review}.
\newblock \bibinfo{journal}{\emph{Drones}} \bibinfo{volume}{6},
  \bibinfo{number}{6} (\bibinfo{year}{2022}), \bibinfo{pages}{147}.
\newblock


\bibitem[\protect\citeauthoryear{Nemhauser, Wolsey, and Fisher}{Nemhauser
  et~al\mbox{.}}{1978}]%
        {nemhauser1978analysis}
\bibfield{author}{\bibinfo{person}{George~L Nemhauser},
  \bibinfo{person}{Laurence~A Wolsey}, {and} \bibinfo{person}{Marshall~L
  Fisher}.} \bibinfo{year}{1978}\natexlab{}.
\newblock \showarticletitle{An analysis of approximations for maximizing
  submodular set functions—I}.
\newblock \bibinfo{journal}{\emph{Mathematical programming}}
  \bibinfo{volume}{14} (\bibinfo{year}{1978}), \bibinfo{pages}{265--294}.
\newblock


\bibitem[\protect\citeauthoryear{Niroui, Zhang, Kashino, and Nejat}{Niroui
  et~al\mbox{.}}{2019}]%
        {niroui2019deep}
\bibfield{author}{\bibinfo{person}{Farzad Niroui}, \bibinfo{person}{Kaicheng
  Zhang}, \bibinfo{person}{Zendai Kashino}, {and} \bibinfo{person}{Goldie
  Nejat}.} \bibinfo{year}{2019}\natexlab{}.
\newblock \showarticletitle{Deep reinforcement learning robot for search and
  rescue applications: Exploration in unknown cluttered environments}.
\newblock \bibinfo{journal}{\emph{IEEE Robotics and Automation Letters}}
  \bibinfo{volume}{4}, \bibinfo{number}{2} (\bibinfo{year}{2019}),
  \bibinfo{pages}{610--617}.
\newblock


\bibitem[\protect\citeauthoryear{Plocher}{Plocher}{2017}]%
        {plocher2017german}
\bibfield{author}{\bibinfo{person}{Generalleutnant~Hermann Plocher}.}
  \bibinfo{year}{2017}\natexlab{}.
\newblock \bibinfo{booktitle}{\emph{The German Air Force Versus Russia, 1943}}.
\newblock \bibinfo{publisher}{Pickle Partners Publishing}.
\newblock


\bibitem[\protect\citeauthoryear{Qi, Jin, Luo, Shi, and Liu}{Qi
  et~al\mbox{.}}{2022}]%
        {DBLP:journals/tcyb/QiJLSL22}
\bibfield{author}{\bibinfo{person}{Yimeng Qi}, \bibinfo{person}{Long Jin},
  \bibinfo{person}{Xin Luo}, \bibinfo{person}{Yang Shi}, {and}
  \bibinfo{person}{Mei Liu}.} \bibinfo{year}{2022}\natexlab{}.
\newblock \showarticletitle{Robust k-WTA Network Generation, Analysis, and
  Applications to Multiagent Coordination}.
\newblock \bibinfo{journal}{\emph{{IEEE} Trans. Cybern.}} \bibinfo{volume}{52},
  \bibinfo{number}{8} (\bibinfo{year}{2022}), \bibinfo{pages}{8515--8527}.
\newblock
\urldef\tempurl%
\url{https://doi.org/10.1109/TCYB.2021.3079457}
\showDOI{\tempurl}


\bibitem[\protect\citeauthoryear{Rashid, Farquhar, Peng, and Whiteson}{Rashid
  et~al\mbox{.}}{2020}]%
        {NEURIPS2020_73a427ba}
\bibfield{author}{\bibinfo{person}{Tabish Rashid}, \bibinfo{person}{Gregory
  Farquhar}, \bibinfo{person}{Bei Peng}, {and} \bibinfo{person}{Shimon
  Whiteson}.} \bibinfo{year}{2020}\natexlab{}.
\newblock \showarticletitle{Weighted QMIX: Expanding Monotonic Value Function
  Factorisation for Deep Multi-Agent Reinforcement Learning}. In
  \bibinfo{booktitle}{\emph{Advances in Neural Information Processing
  Systems}}, \bibfield{editor}{\bibinfo{person}{H.~Larochelle},
  \bibinfo{person}{M.~Ranzato}, \bibinfo{person}{R.~Hadsell},
  \bibinfo{person}{M.F. Balcan}, {and} \bibinfo{person}{H.~Lin}} (Eds.),
  Vol.~\bibinfo{volume}{33}. \bibinfo{publisher}{Curran Associates, Inc.},
  \bibinfo{pages}{10199--10210}.
\newblock
\urldef\tempurl%
\url{https://proceedings.neurips.cc/paper_files/paper/2020/file/73a427badebe0e32caa2e1fc7530b7f3-Paper.pdf}
\showURL{%
\tempurl}


\bibitem[\protect\citeauthoryear{Russell and Norvig}{Russell and
  Norvig}{2016}]%
        {russell2016artificial}
\bibfield{author}{\bibinfo{person}{Stuart~J Russell} {and}
  \bibinfo{person}{Peter Norvig}.} \bibinfo{year}{2016}\natexlab{}.
\newblock \bibinfo{booktitle}{\emph{Artificial intelligence: a modern
  approach}}.
\newblock \bibinfo{publisher}{pearson}.
\newblock


\bibitem[\protect\citeauthoryear{Salimi, Azizi, and Dogani}{Salimi
  et~al\mbox{.}}{2025}]%
        {DBLP:journals/cn/SalimiAD25}
\bibfield{author}{\bibinfo{person}{Rezvan Salimi}, \bibinfo{person}{Sadoon
  Azizi}, {and} \bibinfo{person}{Javad Dogani}.}
  \bibinfo{year}{2025}\natexlab{}.
\newblock \showarticletitle{A hybrid priority-aware genetic algorithm and
  opposition-based learning for scheduling IoT tasks in green fog computing}.
\newblock \bibinfo{journal}{\emph{Comput. Networks}}  \bibinfo{volume}{267}
  (\bibinfo{year}{2025}), \bibinfo{pages}{111349}.
\newblock
\urldef\tempurl%
\url{https://doi.org/10.1016/J.COMNET.2025.111349}
\showDOI{\tempurl}


\bibitem[\protect\citeauthoryear{Son, Kim, Kang, Hostallero, and Yi}{Son
  et~al\mbox{.}}{2019}]%
        {son2019qtran}
\bibfield{author}{\bibinfo{person}{Kyunghwan Son}, \bibinfo{person}{Daewoo
  Kim}, \bibinfo{person}{Wan~Ju Kang}, \bibinfo{person}{David~Earl Hostallero},
  {and} \bibinfo{person}{Yung Yi}.} \bibinfo{year}{2019}\natexlab{}.
\newblock \showarticletitle{Qtran: Learning to factorize with transformation
  for cooperative multi-agent reinforcement learning}. In
  \bibinfo{booktitle}{\emph{International conference on machine learning}}.
  PMLR, \bibinfo{pages}{5887--5896}.
\newblock


\bibitem[\protect\citeauthoryear{Sun, Fei, Zhang, and Xie}{Sun
  et~al\mbox{.}}{2025}]%
        {SUN2025111081}
\bibfield{author}{\bibinfo{person}{Hao Sun}, \bibinfo{person}{Juntao Fei},
  \bibinfo{person}{Guoping Zhang}, {and} \bibinfo{person}{Mande Xie}.}
  \bibinfo{year}{2025}\natexlab{}.
\newblock \showarticletitle{A multi-agent-based dynamic charging strategy for
  UAV-assisted wireless rechargeable sensor networks}.
\newblock \bibinfo{journal}{\emph{Computer Networks}}  \bibinfo{volume}{259}
  (\bibinfo{year}{2025}), \bibinfo{pages}{111081}.
\newblock
\showISSN{1389-1286}
\urldef\tempurl%
\url{https://doi.org/10.1016/j.comnet.2025.111081}
\showDOI{\tempurl}


\bibitem[\protect\citeauthoryear{Tanha, Shirvani, and Rahmani}{Tanha
  et~al\mbox{.}}{2021}]%
        {DBLP:journals/nca/TanhaSR21}
\bibfield{author}{\bibinfo{person}{Mozhdeh Tanha},
  \bibinfo{person}{Mirsaeid~Hosseini Shirvani}, {and}
  \bibinfo{person}{Amir~Masoud Rahmani}.} \bibinfo{year}{2021}\natexlab{}.
\newblock \showarticletitle{A hybrid meta-heuristic task scheduling algorithm
  based on genetic and thermodynamic simulated annealing algorithms in cloud
  computing environments}.
\newblock \bibinfo{journal}{\emph{Neural Comput. Appl.}} \bibinfo{volume}{33},
  \bibinfo{number}{24} (\bibinfo{year}{2021}), \bibinfo{pages}{16951--16984}.
\newblock
\urldef\tempurl%
\url{https://doi.org/10.1007/S00521-021-06289-9}
\showDOI{\tempurl}


\bibitem[\protect\citeauthoryear{{Unity Technologies}}{{Unity
  Technologies}}{2023}]%
        {unity}
\bibfield{author}{\bibinfo{person}{{Unity Technologies}}.}
  \bibinfo{year}{2023}\natexlab{}.
\newblock \bibinfo{title}{{Unity} Game Engine}.
\newblock \bibinfo{howpublished}{\url{https://unity.com/}}.
\newblock


\bibitem[\protect\citeauthoryear{Wang, Chen, Cheng, Wu, Dang, and Chen}{Wang
  et~al\mbox{.}}{2022a}]%
        {wang2022h}
\bibfield{author}{\bibinfo{person}{Haoyang Wang}, \bibinfo{person}{Xuecheng
  Chen}, \bibinfo{person}{Yuhan Cheng}, \bibinfo{person}{Chenye Wu},
  \bibinfo{person}{Fan Dang}, {and} \bibinfo{person}{Xinlei Chen}.}
  \bibinfo{year}{2022}\natexlab{a}.
\newblock \showarticletitle{H-swarmloc: Efficient scheduling for localization
  of heterogeneous mav swarm with deep reinforcement learning}. In
  \bibinfo{booktitle}{\emph{Proceedings of the 20th ACM Conference on Embedded
  Networked Sensor Systems}}. \bibinfo{pages}{1148--1154}.
\newblock


\bibitem[\protect\citeauthoryear{Wang, Xu, Zhao, Lu, Cheng, Chen, Zhang, Liu,
  and Chen}{Wang et~al\mbox{.}}{2024b}]%
        {Wang_2024}
\bibfield{author}{\bibinfo{person}{Haoyang Wang}, \bibinfo{person}{Jingao Xu},
  \bibinfo{person}{Chenyu Zhao}, \bibinfo{person}{Zihong Lu},
  \bibinfo{person}{Yuhan Cheng}, \bibinfo{person}{Xuecheng Chen},
  \bibinfo{person}{Xiao-Ping Zhang}, \bibinfo{person}{Yunhao Liu}, {and}
  \bibinfo{person}{Xinlei Chen}.} \bibinfo{year}{2024}\natexlab{b}.
\newblock \showarticletitle{TransformLoc: Transforming MAVs into Mobile
  Localization Infrastructures in Heterogeneous Swarms}. In
  \bibinfo{booktitle}{\emph{IEEE INFOCOM 2024 - IEEE Conference on Computer
  Communications}}. \bibinfo{publisher}{IEEE}, \bibinfo{pages}{1101–1110}.
\newblock
\urldef\tempurl%
\url{https://doi.org/10.1109/infocom52122.2024.10621375}
\showDOI{\tempurl}


\bibitem[\protect\citeauthoryear{Wang, Ren, Liu, Yu, and Zhang}{Wang
  et~al\mbox{.}}{2020}]%
        {wang2020qplex}
\bibfield{author}{\bibinfo{person}{Jianhao Wang}, \bibinfo{person}{Zhizhou
  Ren}, \bibinfo{person}{Terry Liu}, \bibinfo{person}{Yang Yu}, {and}
  \bibinfo{person}{Chongjie Zhang}.} \bibinfo{year}{2020}\natexlab{}.
\newblock \showarticletitle{Qplex: Duplex dueling multi-agent q-learning}.
\newblock \bibinfo{journal}{\emph{arXiv preprint arXiv:2008.01062}}
  (\bibinfo{year}{2020}).
\newblock


\bibitem[\protect\citeauthoryear{Wang, Wu, Chen, and Ju}{Wang
  et~al\mbox{.}}{2022c}]%
        {wang2022multi}
\bibfield{author}{\bibinfo{person}{Jing Wang}, \bibinfo{person}{Yu~Xuan Wu},
  \bibinfo{person}{Yang-Quan Chen}, {and} \bibinfo{person}{Shuang Ju}.}
  \bibinfo{year}{2022}\natexlab{c}.
\newblock \showarticletitle{Multi-UAVs collaborative tracking of moving target
  with maximized visibility in urban environment}.
\newblock \bibinfo{journal}{\emph{Journal of the Franklin Institute}}
  \bibinfo{volume}{359}, \bibinfo{number}{11} (\bibinfo{year}{2022}),
  \bibinfo{pages}{5512--5532}.
\newblock


\bibitem[\protect\citeauthoryear{Wang, Yang, Yu, Xiong, Han, Pan, and Guo}{Wang
  et~al\mbox{.}}{2023b}]%
        {DBLP:journals/tmc/WangYYXHPG23}
\bibfield{author}{\bibinfo{person}{Liang Wang}, \bibinfo{person}{Dingqi Yang},
  \bibinfo{person}{Zhiwen Yu}, \bibinfo{person}{Fei Xiong},
  \bibinfo{person}{Lei Han}, \bibinfo{person}{Shirui Pan}, {and}
  \bibinfo{person}{Bin Guo}.} \bibinfo{year}{2023}\natexlab{b}.
\newblock \showarticletitle{Task Scheduling in Three-Dimensional Spatial
  Crowdsourcing: {A} Social Welfare Perspective}.
\newblock \bibinfo{journal}{\emph{{IEEE} Trans. Mob. Comput.}}
  \bibinfo{volume}{22}, \bibinfo{number}{9} (\bibinfo{year}{2023}),
  \bibinfo{pages}{5555--5567}.
\newblock
\urldef\tempurl%
\url{https://doi.org/10.1109/TMC.2022.3175305}
\showDOI{\tempurl}


\bibitem[\protect\citeauthoryear{Wang, Liu, Piao, Yuan, Han, Wang, and
  Tang}{Wang et~al\mbox{.}}{2022b}]%
        {wang2022human}
\bibfield{author}{\bibinfo{person}{Yu Wang}, \bibinfo{person}{Chi~Harold Liu},
  \bibinfo{person}{Chengzhe Piao}, \bibinfo{person}{Ye Yuan},
  \bibinfo{person}{Rui Han}, \bibinfo{person}{Guoren Wang}, {and}
  \bibinfo{person}{Jian Tang}.} \bibinfo{year}{2022}\natexlab{b}.
\newblock \showarticletitle{Human-drone collaborative spatial crowdsourcing by
  memory-augmented and distributed multi-agent deep reinforcement learning}. In
  \bibinfo{booktitle}{\emph{2022 IEEE 38th International Conference on Data
  Engineering (ICDE)}}. IEEE, \bibinfo{pages}{459--471}.
\newblock


\bibitem[\protect\citeauthoryear{Wang, Wu, Hua, Liu, Li, Zhao, Yuan, and
  Wang}{Wang et~al\mbox{.}}{2023a}]%
        {wang2023air}
\bibfield{author}{\bibinfo{person}{Yu Wang}, \bibinfo{person}{Jingfei Wu},
  \bibinfo{person}{Xingyuan Hua}, \bibinfo{person}{Chi~Harold Liu},
  \bibinfo{person}{Guozheng Li}, \bibinfo{person}{Jianxin Zhao},
  \bibinfo{person}{Ye Yuan}, {and} \bibinfo{person}{Guoren Wang}.}
  \bibinfo{year}{2023}\natexlab{a}.
\newblock \showarticletitle{Air-ground spatial crowdsourcing with UAV carriers
  by geometric graph convolutional multi-agent deep reinforcement learning}. In
  \bibinfo{booktitle}{\emph{2023 IEEE 39th International Conference on Data
  Engineering (ICDE)}}. IEEE, \bibinfo{pages}{1790--1802}.
\newblock


\bibitem[\protect\citeauthoryear{Wang, Cao, Jiang, Zhou, Kang, Zhuang, Tian,
  and Leung}{Wang et~al\mbox{.}}{2024a}]%
        {10529209}
\bibfield{author}{\bibinfo{person}{Zhenning Wang}, \bibinfo{person}{Yue Cao},
  \bibinfo{person}{Kai Jiang}, \bibinfo{person}{Huan Zhou},
  \bibinfo{person}{Jiawen Kang}, \bibinfo{person}{Yuan Zhuang},
  \bibinfo{person}{Daxin Tian}, {and} \bibinfo{person}{Victor C.~M. Leung}.}
  \bibinfo{year}{2024}\natexlab{a}.
\newblock \showarticletitle{When Crowdsensing Meets Smart Cities: A
  Comprehensive Survey and New Perspectives}.
\newblock \bibinfo{journal}{\emph{IEEE Communications Surveys \& Tutorials}}
  (\bibinfo{year}{2024}), \bibinfo{pages}{1--1}.
\newblock
\urldef\tempurl%
\url{https://doi.org/10.1109/COMST.2024.3400121}
\showDOI{\tempurl}


\bibitem[\protect\citeauthoryear{Wu}{Wu}{2022}]%
        {wu2022task}
\bibfield{author}{\bibinfo{person}{Yu Wu}.} \bibinfo{year}{2022}\natexlab{}.
\newblock \showarticletitle{Task scheduling of the collaborative aerial--ground
  system for the search and capture of multiple targets}.
\newblock \bibinfo{journal}{\emph{Knowledge-Based Systems}}
  \bibinfo{volume}{250} (\bibinfo{year}{2022}), \bibinfo{pages}{109031}.
\newblock


\bibitem[\protect\citeauthoryear{Wu, Wu, and Hu}{Wu et~al\mbox{.}}{2020}]%
        {wu2020cooperative}
\bibfield{author}{\bibinfo{person}{Yu Wu}, \bibinfo{person}{Shaobo Wu}, {and}
  \bibinfo{person}{Xinting Hu}.} \bibinfo{year}{2020}\natexlab{}.
\newblock \showarticletitle{Cooperative path planning of UAVs \& UGVs for a
  persistent surveillance task in urban environments}.
\newblock \bibinfo{journal}{\emph{IEEE Internet of Things Journal}}
  \bibinfo{volume}{8}, \bibinfo{number}{6} (\bibinfo{year}{2020}),
  \bibinfo{pages}{4906--4919}.
\newblock


\bibitem[\protect\citeauthoryear{Xu, Li, Meng, and Zhang}{Xu
  et~al\mbox{.}}{2024}]%
        {xu2024iterated}
\bibfield{author}{\bibinfo{person}{Ying Xu}, \bibinfo{person}{Xiaobo Li},
  \bibinfo{person}{Xiangpei Meng}, {and} \bibinfo{person}{Weipeng Zhang}.}
  \bibinfo{year}{2024}\natexlab{}.
\newblock \showarticletitle{An iterated greedy heuristic for collaborative
  human-uav search of missing tourists}.
\newblock \bibinfo{journal}{\emph{Knowledge-Based Systems}}
  \bibinfo{volume}{286} (\bibinfo{year}{2024}), \bibinfo{pages}{111409}.
\newblock


\bibitem[\protect\citeauthoryear{Yang, Yang, Wang, Liu, Xu, Yin, Zhai, and
  Zhang}{Yang et~al\mbox{.}}{2023}]%
        {yang2023how2comm}
\bibfield{author}{\bibinfo{person}{Dingkang Yang}, \bibinfo{person}{Kun Yang},
  \bibinfo{person}{Yuzheng Wang}, \bibinfo{person}{Jing Liu},
  \bibinfo{person}{Zhi Xu}, \bibinfo{person}{Rongbin Yin},
  \bibinfo{person}{Peng Zhai}, {and} \bibinfo{person}{Lihua Zhang}.}
  \bibinfo{year}{2023}\natexlab{}.
\newblock \showarticletitle{How2comm: Communication-efficient and
  collaboration-pragmatic multi-agent perception}.
\newblock \bibinfo{journal}{\emph{Advances in Neural Information Processing
  Systems}}  \bibinfo{volume}{36} (\bibinfo{year}{2023}),
  \bibinfo{pages}{25151--25164}.
\newblock


\bibitem[\protect\citeauthoryear{Ye, Liu, Dai, Zhao, Yuan, Wang, and Tang}{Ye
  et~al\mbox{.}}{2023}]%
        {ye2023exploring}
\bibfield{author}{\bibinfo{person}{Yuxiao Ye}, \bibinfo{person}{Chi~Harold
  Liu}, \bibinfo{person}{Zipeng Dai}, \bibinfo{person}{Jianxin Zhao},
  \bibinfo{person}{Ye Yuan}, \bibinfo{person}{Guoren Wang}, {and}
  \bibinfo{person}{Jian Tang}.} \bibinfo{year}{2023}\natexlab{}.
\newblock \showarticletitle{Exploring both individuality and cooperation for
  air-ground spatial crowdsourcing by multi-agent deep reinforcement learning}.
  In \bibinfo{booktitle}{\emph{2023 IEEE 39th International Conference on Data
  Engineering (ICDE)}}. IEEE, \bibinfo{pages}{205--217}.
\newblock


\bibitem[\protect\citeauthoryear{Yu, Han, Chen, Guo, and Yu}{Yu
  et~al\mbox{.}}{2021}]%
        {yu2021object}
\bibfield{author}{\bibinfo{person}{Zhiyong Yu}, \bibinfo{person}{Lei Han},
  \bibinfo{person}{Chao Chen}, \bibinfo{person}{Wenzhong Guo}, {and}
  \bibinfo{person}{Zhiwen Yu}.} \bibinfo{year}{2021}\natexlab{}.
\newblock \showarticletitle{Object tracking by the least spatiotemporal
  searches}.
\newblock \bibinfo{journal}{\emph{IEEE Internet of Things Journal}}
  \bibinfo{volume}{8}, \bibinfo{number}{16} (\bibinfo{year}{2021}),
  \bibinfo{pages}{12934--12946}.
\newblock


\bibitem[\protect\citeauthoryear{Yuan, Ding, Feng, Jin, and Li}{Yuan
  et~al\mbox{.}}{2024}]%
        {yuan2024unist}
\bibfield{author}{\bibinfo{person}{Yuan Yuan}, \bibinfo{person}{Jingtao Ding},
  \bibinfo{person}{Jie Feng}, \bibinfo{person}{Depeng Jin}, {and}
  \bibinfo{person}{Yong Li}.} \bibinfo{year}{2024}\natexlab{}.
\newblock \showarticletitle{Unist: A prompt-empowered universal model for urban
  spatio-temporal prediction}. In \bibinfo{booktitle}{\emph{Proceedings of the
  30th ACM SIGKDD Conference on Knowledge Discovery and Data Mining}}.
  \bibinfo{pages}{4095--4106}.
\newblock


\bibitem[\protect\citeauthoryear{Zang, Zhang, Song, Lu, Li, Dong, Li, Ju, and
  Gong}{Zang et~al\mbox{.}}{2024}]%
        {zang2024coordinated}
\bibfield{author}{\bibinfo{person}{Zheng Zang}, \bibinfo{person}{Xi Zhang},
  \bibinfo{person}{Jiarui Song}, \bibinfo{person}{Yaomin Lu},
  \bibinfo{person}{Zhiwei Li}, \bibinfo{person}{Haotian Dong},
  \bibinfo{person}{Yuanyuan Li}, \bibinfo{person}{Zhiyang Ju}, {and}
  \bibinfo{person}{Jianwei Gong}.} \bibinfo{year}{2024}\natexlab{}.
\newblock \showarticletitle{A coordinated behavior planning and trajectory
  planning framework for multi-ugvs in unstructured narrow interaction
  scenarios}.
\newblock \bibinfo{journal}{\emph{IEEE Transactions on Intelligent Vehicles}}
  (\bibinfo{year}{2024}).
\newblock


\bibitem[\protect\citeauthoryear{Zhang, Tang, Zhang, and Ding}{Zhang
  et~al\mbox{.}}{2025}]%
        {10887881}
\bibfield{author}{\bibinfo{person}{Enze Zhang}, \bibinfo{person}{Huaze Tang},
  \bibinfo{person}{Xiao-Ping Zhang}, {and} \bibinfo{person}{Wenbo Ding}.}
  \bibinfo{year}{2025}\natexlab{}.
\newblock \showarticletitle{Mean-Field Aided QMIX: A Scalable and Flexible
  Q-Learning Approach for Large-Scale Agent Groups}. In
  \bibinfo{booktitle}{\emph{ICASSP 2025 - 2025 IEEE International Conference on
  Acoustics, Speech and Signal Processing (ICASSP)}}. \bibinfo{pages}{1--5}.
\newblock
\urldef\tempurl%
\url{https://doi.org/10.1109/ICASSP49660.2025.10887881}
\showDOI{\tempurl}


\bibitem[\protect\citeauthoryear{Zhang, Tong, Zhu, Xu, and Wu}{Zhang
  et~al\mbox{.}}{2024}]%
        {zhang2024sqix}
\bibfield{author}{\bibinfo{person}{Miaomiao Zhang}, \bibinfo{person}{Wei Tong},
  \bibinfo{person}{Guangyu Zhu}, \bibinfo{person}{Xin Xu}, {and}
  \bibinfo{person}{Edmond~Q Wu}.} \bibinfo{year}{2024}\natexlab{}.
\newblock \showarticletitle{SQIX: QMIX algorithm activated by general softmax
  operator for cooperative multiagent reinforcement learning}.
\newblock \bibinfo{journal}{\emph{IEEE Transactions on Systems, Man, and
  Cybernetics: Systems}} \bibinfo{volume}{54}, \bibinfo{number}{11}
  (\bibinfo{year}{2024}), \bibinfo{pages}{6550--6560}.
\newblock


\bibitem[\protect\citeauthoryear{Zhao, Liu, Yi, Li, and Wu}{Zhao
  et~al\mbox{.}}{2024}]%
        {zhao2024energy}
\bibfield{author}{\bibinfo{person}{Yinuo Zhao}, \bibinfo{person}{Chi~Harold
  Liu}, \bibinfo{person}{Tianjiao Yi}, \bibinfo{person}{Guozheng Li}, {and}
  \bibinfo{person}{Dapeng Wu}.} \bibinfo{year}{2024}\natexlab{}.
\newblock \showarticletitle{Energy-Efficient Ground-Air-Space Vehicular
  Crowdsensing by Hierarchical Multi-Agent Deep Reinforcement Learning with
  Diffusion Models}.
\newblock \bibinfo{journal}{\emph{IEEE Journal on Selected Areas in
  Communications}} (\bibinfo{year}{2024}).
\newblock


\bibitem[\protect\citeauthoryear{Zhao, Wu, Xu, Che, Lu, Tang, and Liu}{Zhao
  et~al\mbox{.}}{2022}]%
        {zhao2022cadre}
\bibfield{author}{\bibinfo{person}{Yinuo Zhao}, \bibinfo{person}{Kun Wu},
  \bibinfo{person}{Zhiyuan Xu}, \bibinfo{person}{Zhengping Che},
  \bibinfo{person}{Qi Lu}, \bibinfo{person}{Jian Tang}, {and}
  \bibinfo{person}{Chi~Harold Liu}.} \bibinfo{year}{2022}\natexlab{}.
\newblock \showarticletitle{Cadre: A cascade deep reinforcement learning
  framework for vision-based autonomous urban driving}. In
  \bibinfo{booktitle}{\emph{Proceedings of the AAAI conference on artificial
  intelligence}}, Vol.~\bibinfo{volume}{36}. \bibinfo{pages}{3481--3489}.
\newblock


\bibitem[\protect\citeauthoryear{Zheng, Capra, Wolfson, and Yang}{Zheng
  et~al\mbox{.}}{2014}]%
        {zheng2014urban}
\bibfield{author}{\bibinfo{person}{Yu Zheng}, \bibinfo{person}{Licia Capra},
  \bibinfo{person}{Ouri Wolfson}, {and} \bibinfo{person}{Hai Yang}.}
  \bibinfo{year}{2014}\natexlab{}.
\newblock \showarticletitle{Urban computing: concepts, methodologies, and
  applications}.
\newblock \bibinfo{journal}{\emph{ACM Transactions on Intelligent Systems and
  Technology (TIST)}} \bibinfo{volume}{5}, \bibinfo{number}{3}
  (\bibinfo{year}{2014}), \bibinfo{pages}{1--55}.
\newblock


\bibitem[\protect\citeauthoryear{Zheng, Du, Ling, Sheng, and Chen}{Zheng
  et~al\mbox{.}}{2019}]%
        {zheng2019evolutionary}
\bibfield{author}{\bibinfo{person}{Yu-Jun Zheng}, \bibinfo{person}{Yi-Chen Du},
  \bibinfo{person}{Hai-Feng Ling}, \bibinfo{person}{Wei-Guo Sheng}, {and}
  \bibinfo{person}{Sheng-Yong Chen}.} \bibinfo{year}{2019}\natexlab{}.
\newblock \showarticletitle{Evolutionary collaborative human-UAV search for
  escaped criminals}.
\newblock \bibinfo{journal}{\emph{IEEE Transactions on Evolutionary
  Computation}} \bibinfo{volume}{24}, \bibinfo{number}{2}
  (\bibinfo{year}{2019}), \bibinfo{pages}{217--231}.
\newblock


\bibitem[\protect\citeauthoryear{Zheng, Du, Su, Ling, Zhang, and Chen}{Zheng
  et~al\mbox{.}}{2020}]%
        {zheng2020evolutionary}
\bibfield{author}{\bibinfo{person}{Yu-Jun Zheng}, \bibinfo{person}{Yi-Chen Du},
  \bibinfo{person}{Zheng-Lian Su}, \bibinfo{person}{Hai-Feng Ling},
  \bibinfo{person}{Min-Xia Zhang}, {and} \bibinfo{person}{Sheng-Yong Chen}.}
  \bibinfo{year}{2020}\natexlab{}.
\newblock \showarticletitle{Evolutionary human-UAV cooperation for transmission
  network restoration}.
\newblock \bibinfo{journal}{\emph{IEEE Transactions on Industrial Informatics}}
  \bibinfo{volume}{17}, \bibinfo{number}{3} (\bibinfo{year}{2020}),
  \bibinfo{pages}{1648--1657}.
\newblock


\bibitem[\protect\citeauthoryear{Zhou, Feng, Gu, Ai, Mumtaz, Rodriguez, and
  Guizani}{Zhou et~al\mbox{.}}{2018}]%
        {zhou2018mobile}
\bibfield{author}{\bibinfo{person}{Zhenyu Zhou}, \bibinfo{person}{Junhao Feng},
  \bibinfo{person}{Bo Gu}, \bibinfo{person}{Bo Ai}, \bibinfo{person}{Shahid
  Mumtaz}, \bibinfo{person}{Jonathan Rodriguez}, {and} \bibinfo{person}{Mohsen
  Guizani}.} \bibinfo{year}{2018}\natexlab{}.
\newblock \showarticletitle{When mobile crowd sensing meets UAV:
  Energy-efficient task assignment and route planning}.
\newblock \bibinfo{journal}{\emph{IEEE Transactions on Communications}}
  \bibinfo{volume}{66}, \bibinfo{number}{11} (\bibinfo{year}{2018}),
  \bibinfo{pages}{5526--5538}.
\newblock


\bibitem[\protect\citeauthoryear{Zuo, Zhou, Yang, Su, Zhu, and Li}{Zuo
  et~al\mbox{.}}{2023}]%
        {zuo2023real}
\bibfield{author}{\bibinfo{person}{Xinkai Zuo}, \bibinfo{person}{Jian Zhou},
  \bibinfo{person}{Fan Yang}, \bibinfo{person}{Fei Su},
  \bibinfo{person}{Haihong Zhu}, {and} \bibinfo{person}{Lin Li}.}
  \bibinfo{year}{2023}\natexlab{}.
\newblock \showarticletitle{Real-time global action planning for unmanned
  ground vehicle exploration in three-dimensional spaces}.
\newblock \bibinfo{journal}{\emph{Expert Systems with Applications}}
  \bibinfo{volume}{215} (\bibinfo{year}{2023}), \bibinfo{pages}{119264}.
\newblock


\end{thebibliography}

\appendix

\section{Np-hard Proof}
\textbf{Lemma 1:} \textbf{Problem 1 } is NP-Hard.

{\itshape Proof:}
To facilitate the NP-Hardness proof, we consider a restricted version of \textbf{Problem 1} by making the following assumptions :
UAVs have unlimited endurance and are not constrained by vehicles or charging points;UAVs can independently complete tasks without the need for workers;A task ${\rm task}_x$ is instantly completed when a UAV arrives at its location ${tLoc}_x$, with no additional cost:
\begin{equation*}
{\rm task}_x = \emptyset,  \text{if } {uLoc}_i = {tLoc}_x
\end{equation*}

Based on the above, \textbf{Problem 1 } simplifies to\textbf{Problem 2 } :
\begin{small}
\begin{equation*}
\begin{array}{l}
{\rm{\textbf{confirm}}}
\
\{ {A\_u}_i^{t \times Interval} \},
(t = 0, 1, 2, \ldots,\ t \times Interval \in [0, LimitTime]) \\
\qquad \qquad\qquad\qquad \qquad 
{\rm{\textbf{maximize}}}
\
Cplt\_Tasks \\
\qquad\qquad\qquad\qquad  \qquad 
s.t.\
\text{assumptions (3)}
\end{array}
\end{equation*}
\end{small}

Since \textbf{Problem 2} is a special case of \textbf{Problem 1 }with simpler constraints, \textbf{Problem 2} can be reduced to  \textbf{Problem 1}. \textbf{Problem 2} aims to determine the scheduling actions of multiple UAVs to maximize $Cplt\_Tasks$, which is a sequential optimal subset selection problem. Since the optimal subset selection problem is known to be NP-Hard~\cite{nemhauser1978analysis}, \textbf{Problem 1 } is also NP-Hard. 
\section{Detailed experimental setup}

\hspace{1em}We comprehensively evaluate the performance of the proposed algorithm based on real and simulated datasets. The detailed description of the datasets is as follows.

\subsection{Dataset Settings}

\subsubsection{Real Datasets}
Real data comes from actual scenarios, which can reflect task distribution, agent behavior patterns, and resource limitations in real environments. The real datasets include bicycle riding order data (BicycleOrder) \cite{li2015traffic, zheng2014urban},Didi order data (DidiOrders) \cite{gaia_didi}, and taxi trajectory data (TaxiTrajectories) \cite{yu2021object} . First, we discretize the coverage area of the real datasets. Then, we filter, classify, and annotate the data to obtain data that meets the experimental specifications. The specific processing procedures can be referred to in \cite{CocaColaZero}.

\subsubsection{Random Simulated Datasets}
Random simulated datasets are generated through parameterization, which can flexibly adjust area scale, task quantity, agent scale, etc., to verify the algorithm's performance in different scenarios. The introduction of random data compensates for the lack of diversity and controllability in real data. The detailed data generation process can be referred to in \cite{CocaColaZero}.

Specifically, the distribution of task points and charging points in the four types of datasets is shown in Fig. \ref{figure6}, with significant differences in their distribution patterns. In addition, the detailed dataset description is shown in Table \ref{table1}.
\begin{table}[htbp]
  \centering
  \setlength{\abovecaptionskip}{0.2cm}
  \setlength{\belowcaptionskip}{-0.25cm}
  \caption{Description of experimental datasets}
  \label{table1}
  \begin{tabular}{c|ccccccc}
    \hline
     Dataset 
      & \makecell[c]{Area\\scale} 
      & \makecell[c]{Tasks\\number} 
      & \makecell[c]{Charging point\\number} 
      & \makecell[c]{Agents online\\time} 
      & \makecell[c]{Agent\\number} 
      & \makecell[c]{Task cost\\power} 
      & \makecell[c]{Charging\\power} \\
    \hline
    BicycleOrder    & 25 $\times$ 16 & 99  & 30 & $>$70 min & (54,17,34) & 3 & 10 \\
    DidiOrders      & 10 $\times$ 9  & 34  & 10 & $>$50 min & (26,18,14) & 3 & 10 \\
    TaxiTrajectories& 30 $\times$ 30 & 102 & 60 & $>$90 min & (38,20,16) & 3 & 10 \\
    Random\_1       & 30 $\times$ 30 & 120 & 20 & 60 min   & (50,30,20) & 3 & 10 \\
    Random\_2       & 20 $\times$ 20 & 120 & 20 & 60 min   & (50,30,20) & 3 & 10 \\
    Random\_3       & 40 $\times$ 40 & 120 & 20 & 60 min   & (50,30,20) & 3 & 10 \\
    Random\_4       & 30 $\times$ 30 & 100 & 20 & 60 min   & (50,30,20) & 3 & 10 \\
    Random\_5       & 30 $\times$ 30 & 140 & 20 & 60 min   & (50,30,20) & 3 & 10 \\
    Random\_6       & 30 $\times$ 30 & 120 & 10 & 60 min   & (50,30,20) & 3 & 10 \\
    Random\_7       & 30 $\times$ 30 & 120 & 30 & 60 min   & (50,30,20) & 3 & 10 \\
    Random\_8       & 30 $\times$ 30 & 120 & 20 & 40 min   & (50,30,20) & 3 & 10 \\
    Random\_9       & 30 $\times$ 30 & 120 & 20 & 80 min   & (50,30,20) & 3 & 10 \\
    Random\_10      & 30 $\times$ 30 & 80  & 20 & 60 min   & (30,20,10) & 3 & 10 \\
    Random\_11      & 30 $\times$ 30 & 120 & 20 & 60 min   & (70,40,30) & 3 & 10 \\
    Random\_12      & 30 $\times$ 30 & 120 & 20 & 60 min   & (30,30,20) & 3 & 10 \\
    Random\_13      & 30 $\times$ 30 & 120 & 20 & 60 min   & (70,30,20) & 3 & 10 \\
    Random\_14      & 30 $\times$ 30 & 120 & 20 & 60 min   & (50,20,20) & 3 & 10 \\
    Random\_15      & 30 $\times$ 30 & 120 & 20 & 60 min   & (50,40,20) & 3 & 10 \\
    Random\_16      & 30 $\times$ 30 & 120 & 20 & 60 min   & (50,30,10) & 3 & 10 \\
    Random\_17      & 30 $\times$ 30 & 120 & 20 & 60 min   & (50,30,30) & 3 & 10 \\
    Random\_18      & 30 $\times$ 30 & 120 & 20 & 60 min   & (50,30,20) & 2 & 10 \\
    Random\_19      & 30 $\times$ 30 & 120 & 20 & 60 min   & (50,30,20) & 4 & 10 \\
    Random\_20      & 30 $\times$ 30 & 120 & 20 & 60 min   & (50,30,20) & [2,3] & 10 \\
    Random\_21      & 30 $\times$ 30 & 120 & 20 & 60 min   & (50,30,20) & [3,4] & 10 \\
    Random\_22      & 30 $\times$ 30 & 120 & 20 & 60 min   & (50,30,20) & [4,5] & 10 \\
    Random\_23      & 30 $\times$ 30 & 120 & 20 & 60 min   & (50,30,20) & 3 & 6  \\
    Random\_24      & 30 $\times$ 30 & 120 & 20 & 60 min   & (50,30,20) & 3 & 14 \\
    Random\_25      & 30 $\times$ 30 & 120 & 20 & 60 min   & (50,30,20) & 3 & [2,6]  \\
    Random\_26      & 30 $\times$ 30 & 120 & 20 & 60 min   & (50,30,20) & 3 & [6,10] \\
    Random\_27      & 30 $\times$ 30 & 120 & 20 & 60 min   & (50,30,20) & 3 & [10,14] \\
    Random\_28 & 40 $\times$ 40 & 240 & 40 & 60 min & (100,60,40) & 3 & 10 \\
    Random\_29 & 60 $\times$ 60 & 480 & 80 & 60 min & (200,120,80) & 3 & 10 \\
    \hline
  \end{tabular}
\end{table}
\begin{figure}[htbp]
  \centering
  \setlength{\abovecaptionskip}{0.2cm}
  \includegraphics[width=\linewidth]{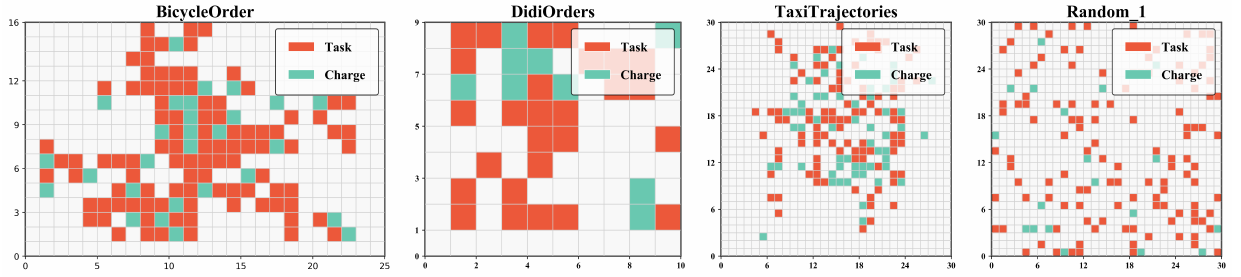}
  \caption{Distribution of task points and charging points in four types of datasets (the size of each grid is 1KM*1KM).}
  \Description{A figure showing the distribution of task points and charging points across four different datasets, with each grid representing a 1KM*1KM area.}
  \label{figure6}
\end{figure}

\subsection{Experiment Setup}

\hspace{1em}Based on the above datasets, we set up 5 groups of experiments to comprehensively evaluate the HoCs-MPQ algorithm., as shown in Table~\ref{table2}.

\begin{table}[htbp]
  \centering
  \setlength{\abovecaptionskip}{0.2cm}
  \setlength{\belowcaptionskip}{-0.25cm}
  \caption{Experimental groups and controlled variables}
  \label{table2}
  \begin{tabular}{c|cccc}
    \hline
    Group
      & Dataset
      & Interval
      & Limit time
      & Controlled variable \\
    \hline
    (1)
      & \makecell[c]{BicycleOrder /\\ DidiOrders /\\ TaxiTrajectories /\\ Random\_1}
      & \makecell[c]{5\,min/\\ 10\,min/\\ 15\,min}
      & 3\,h
      & Interval \\
      \hline
    (2)
      & \makecell[c]{BicycleOrder /\\ DidiOrders /\\ TaxiTrajectories /\\ Random\_1}
      & 5\,min
      & \makecell[c]{2\,h/\\ 3\,h/\\ 4\,h}
      & Limit time \\
      \hline
    (3)
      & \makecell[c]{Random\_1, \\ Random\_2--Random\_9}
      & 10\,min
      & 3\,h
      & \makecell[c]{Area scale\\ Tasks number\\ Charging point number\\ Agents online time} \\
      \hline
    (4)
      & \makecell[c]{Random\_1, \\ Random\_10--Random\_17}
      & 10\,min
      & 3\,h
      & \makecell[c]{Workers number\\ UAVs number\\ Vehicles number} \\
      \hline
    (5)
      & \makecell[c]{Random\_1, \\ Random\_18--Random\_27}
      & 10\,min
      & 3\,h
      & \makecell[c]{Task cost power\\ Charging power} \\
    \hline
  \end{tabular}
\end{table}

\subsection{Comparison Methods}

\subsubsection{HoCs-GREEDY (Heterogeneous Multi-Agent Online Collaborative Scheduling Algorithm Based on GREEDY)}

The greedy algorithm adopts a locally optimal strategy for real-time decision-making \cite{DBLP:conf/kdd/LiC0W019, DBLP:journals/tmc/WangYYXHPG23}. It optimizes the movement paths of UAVs, workers, and vehicles in stages. Its core process is divided into three steps: First, plan the path for UAVs, calculate the sum of Euclidean distances from all positions within the selectable moving range to various task points, and select the position with the minimum total distance as the target point for the next moment. Second, perform the same logic for workers, selecting the next position with the shortest total distance to task points. Finally, plan the path for vehicles, calculate the total distance from each position to all UAVs' next target points, and select the minimum value as the vehicle's moving target. All three types of agents complete the entire path planning through iterative calculation. This method achieves global path optimization through a cascade of locally optimal choices, suitable for collaborative scenarios of multiple agent types.

\subsubsection{HoCs-KWTA (Heterogeneous Multi-Agent Online Collaborative Scheduling Algorithm Based on K-Winners-Take-All)}

HoCs-KWTA extends from the selective activation mechanism of neural networks and can be used for multi-agent task matching \cite{DBLP:journals/tmc/LiuLCQJ24, DBLP:journals/tnn/LiuZ25, DBLP:journals/tcyb/QiJLSL22}. UAV matching is dimensionally divided into task UAVs and charging UAVs. Taking task matching as an example, task UAVs score  task points within their power range, and workers score all task points. The top K high-scoring tasks are retained, and the intersection is checked to determine the ternary matching for collaborative execution. If there is no intersection, suboptimal tasks are tried in descending order of scores. Charging point matching is similar.

\subsubsection{HoCs-MADL (Heterogeneous Multi-Agent Online Collaborative Scheduling Algorithm Based on Multi-Agent Deep Learning)}

Based on a multi-agent deep learning framework\cite{liu2019energy,liu2020energy}, HoCs-MADL models the state and action spaces of heterogeneous agents through shared neural networks. Global information (task point distribution, charging point distribution, etc.) is extracted by convolutional networks and then combined with local states (agent position, UAV power, etc.) as input to a deep network. This produces the immediate action value for each agent and uses end-to-end training to optimize the immediate task completion rate. Note that the detailed implementation processes of HoCs-MADL and HoCs-MARL heavily reference \cite{10.1109/TNET.2024.3395493}.

To facilitate this, we introduce the Heterogeneous Multi-Agent Network Framework (MANF), inspired by the factorization principles of the QMIX algorithm. MANF is specifically designed to handle diverse agent types—UAVs, workers, and cars—within a unified computational structure.

\begin{figure}[htbp]
  \centering
  \setlength{\abovecaptionskip}{0.2cm}
  \includegraphics[width=12cm]{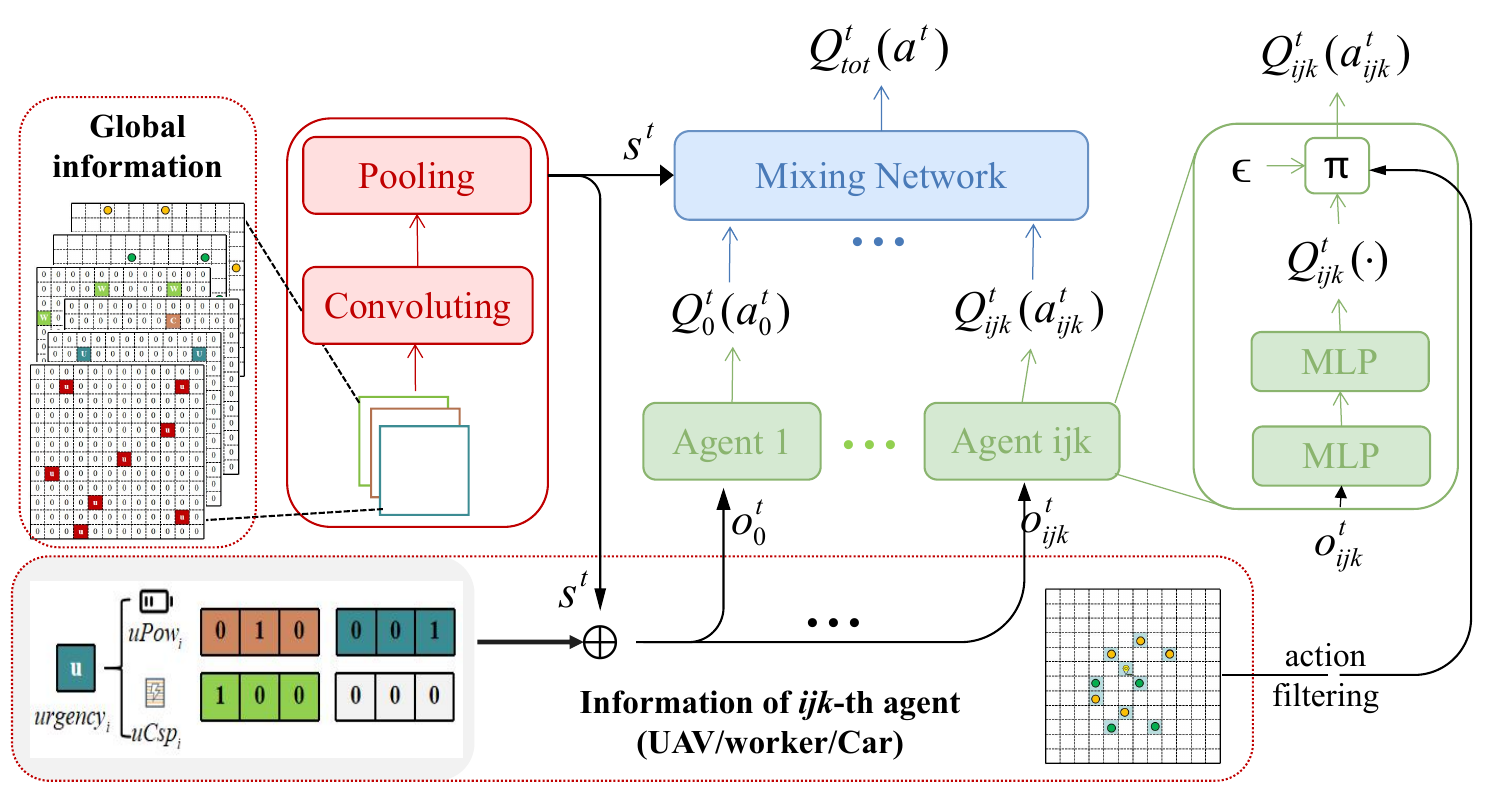}
  \caption{Heterogeneous multi-agent network framework (MANF).}
  \Description{A diagram illustrating the architecture of the Heterogeneous Multi-Agent Network Framework (MANF), showing the interaction between agent networks and a mixing network, and the flow of global and local information.}
  \label{MANF}
\end{figure}

As illustrated in Figure \ref{MANF}, MANF primarily consists of two interconnected neural networks: the agent network and the mixing network. The agent network processes observations from individual agents (UAVs, workers, or cars) and outputs their respective individual Q-values, denoted as $Q_{ijk}^t(a_{ijk}^t)$. The mixing network then takes these individual $Q_{ijk}^t(a_{ijk}^t)$ values as inputs and aggregates them to produce a joint Q-value, $Q_{tot}^t(\mathbf{a}^t)$, representing the value of the joint action for the entire multi-agent system. The parameters (weights and biases) of the mixing network are dynamically generated by a hypernetworks architecture. A crucial constraint for the mixing network's weights, derived from its design principles, is that they must always be greater than zero. Fundamentally, MANF processes information through a dual stream: global information, providing a broad environmental context, and local agent information, capturing specific details of each individual agent. This comprehensive information structure feeds into the subsequent route planning algorithms.

Delving deeper into this information structure, the state space within the MANF framework is meticulously defined by categorizing observed information into two distinct types: global environmental information and local agent-specific information.

\begin{figure}[htbp]
  \centering
  \setlength{\abovecaptionskip}{0.2cm}
  \includegraphics[width=10cm]{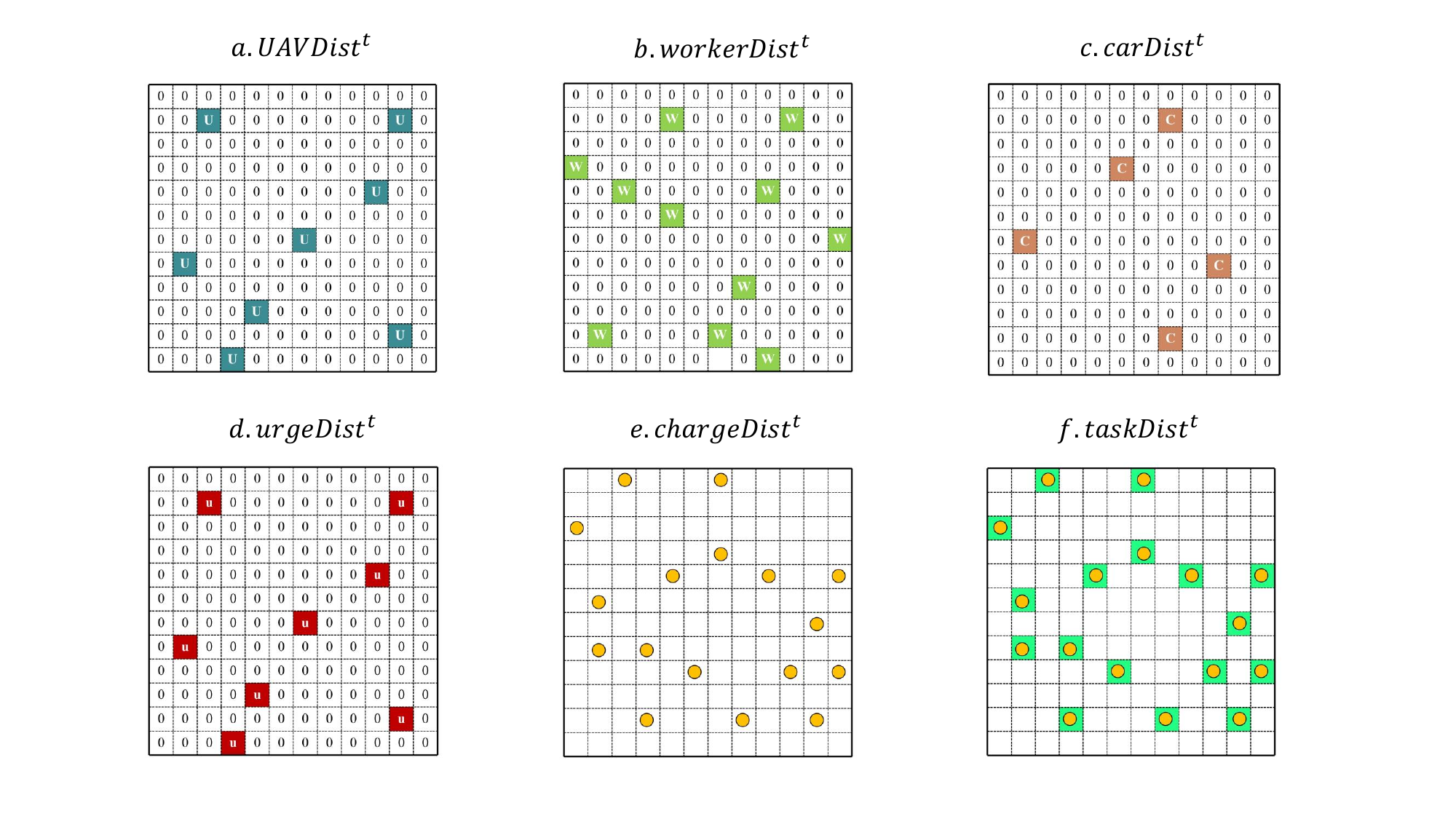}
  \caption{Global information at the moment t}
  \Description{Global information at the moment t}
  \label{Global information at the moment t}
\end{figure}
As conceptually depicted in Figure \ref{Global information at the moment t}, global information provides a system-wide view of the environment at time $t$. This includes:
\begin{itemize}
    \item $UAVDist^t$: Shows the distribution of UAVs at the moment t.
    \item $workDist^t$: Shows the distribution of workers at the moment t.
    \item $carDist^t$: Shows the distribution of cars at the moment t.
    \item $urgeDist^t$: Shows the urgency distribution to replace UAVs’ battery at the moment t, which measures the cumulative urgency of replacing batteries for all UAVs in different locations.
    \item $taskDist^t$: Shows the distribution of tasks at the moment t.
    \item $chargeDist^t$: Shows the distribution of chargepoints at the moment t.
\end{itemize}
These global features are processed by a spatial convolutional neural network (CNN), specifically `cnnSpace`, to extract meaningful spatial embeddings, which are then shared among all heterogeneous agents as part of the global state $s^t$.

\begin{figure}[htbp]
  \centering
  \setlength{\abovecaptionskip}{0.2cm}
  \includegraphics[width=10cm]{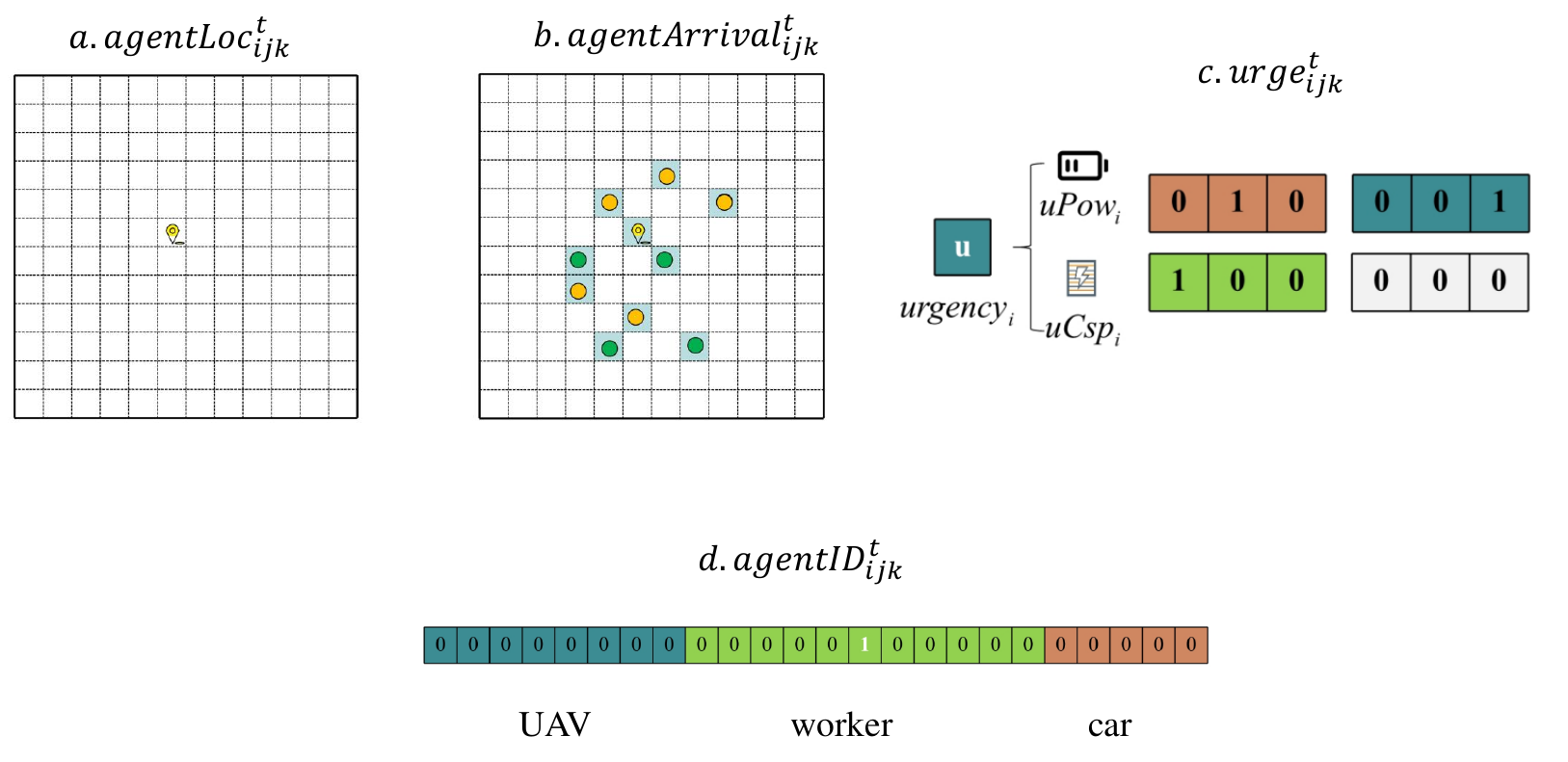}
  \caption{Local information of the ijk agent at the moment t}
  \Description{Local information of the ijk − th agent at the moment t}
  \label{Local information of the ijk − th agent at the moment t}
\end{figure}
As illustrated in Figure \ref{Local information of the ijk − th agent at the moment t}, local information provides specific details pertinent to individual agents (UAVs, workers, and cars) at time $t$. This comprises:
\begin{itemize}
    \item $agentLoc_{ijk}^t$: The precise geographical location of the $ijk$-th agent.
    \item $agentArrival_{ijk}^t$: Represents the set of reachable areas for the $ijk$-th agent within the forthcoming time step $[t, t+1)$. These movement ranges are predefined and already exclude physically unreachable locations.
    \item $urge_{ijk}^t$: The urgency of the $ijk$-th agent to replace its battery at time $t$. Workers and cars are assumed to have infinite power, so their urgency is zero. For UAVs, $urge_{ijk}^t$ is determined such that higher current power leads to lower urgency, and the rate of urgency decrease diminishes as power increases.
    \item $agentID_{ijk}^t$: The unique identification number of the $ijk$-th agent, typically implemented using one-hot encoding.
\end{itemize}
This local information, along with the global state $s^t$ derived from the CNN, is concatenated to form the comprehensive input $o_{ijk}^t$ for each agent's evaluation network (`evalAgent`).

With the state space clearly defined, the HoCs-MADL algorithm, effectively leveraging the MANF architecture, executes its core logic. This heterogeneous multi-agent route planning approach, built upon deep learning principles, is designed to identify the optimal action within a specific time step. The process begins by calculating the expected immediate reward for selecting a set of actions, such as $\{a_t^0, ..., a_{ijk}^t, ...\}$, given the current state, characterized by observations like $\{o_t^0, ..., o_{ijk}^t, ...\}$, all within the time interval $[t, t+1)$. This immediate reward is comprehensively composed of two essential elements: the successful completion of sensing tasks and the reduction in urgency. Following this calculation, the algorithm precisely establishes a ternary mapping relationship that effectively links the current state, the chosen actions, and their corresponding immediate reward. This established mapping then serves as the foundation for training the deep neural network. Through this training phase, the network learns to accurately predict the expected immediate reward when presented with a current state and potential actions. Finally, leveraging the capabilities of the trained model, the system can compare the anticipated immediate rewards for various actions under the existing state, ultimately selecting the action that yields the highest expected reward. This selection process is iterated until the designated target moment is achieved. It is important to note that the target values for rewards in this approach are directly derived from objective real-world environments, ensuring their stability and freedom from bias.

\subsubsection{HoCs-MARL (Heterogeneous Multi-Agent Online Collaborative Scheduling Algorithm Based on Multi-Agent Reinforcement Learning)}

Multi-agent reinforcement learning optimizes long-term cumulative rewards through collaborative strategies \cite{zhao2022cadre,wang2022human,ye2023exploring}. Referencing the QMIX algorithm, it designs a mixing network that fuses the local Q values of each agent, constraining weights to be non-negative to maintain monotonicity. Agents balance "exploration-exploitation" between immediate returns and future collaboration potential. The reward function focuses on the number of completed tasks, avoiding short-term interference from charging gains. Target networks are introduced to mitigate training instability issues, supporting complex spatiotemporal collaborative decision-making.

The HoCs-MARL algorithm extends the capabilities by incorporating the long-term implications of current observations on subsequent action choices. This algorithm estimates the total expected immediate reward over an extended time horizon. While the basic immediate reward can be derived, a critical adjustment is made to align it more closely with the overall optimization objective, which typically aims to maximize total task completion. This refinement is crucial because the impact of certain factors, such as car contributions to battery replacement, may have a delayed effect, influencing future task completion rates. To address potential training instability often seen with single neural networks, such as biased estimations and escalating discrepancies during updates, HoCs-MARL integrates a target network and an evaluation network. The evaluation network is continuously updated based on ongoing learning, while the target network, which provides more stable target values, is periodically synchronized with the evaluation network. This slower update mechanism for the target network helps to stabilize the learning process by ensuring that the target values change gradually, thereby facilitating more consistent and robust learning for the agents. Through these mechanisms, HoCs-MARL enables agents to make decisions that consider both immediate benefits and long-term strategic outcomes in complex routing scenarios.

Otherwise, HoCs MADL/MARL are implemented on the QMIX architecture and thus are only suitable for scenarios where the number and types of agents are fixed. This is because the structure and weights of QMIX’s mixing network directly depend on fixed parameters. When the number of agents changes (i.e., the coupling relationships among agents change), the mixing network cannot automatically adapt, causing the model to fail to varying degrees. Introducing attention mechanisms and hierarchical architectures can partially address the problem of dynamic agent composition. For example, Ai-qmix\cite{iqbal2020ai} employs a self-attention module to dynamically compute cooperation weights among agents, allowing the mixing network to adjust its focus on different agents’ Q-values according to the current scenario.The COPA\cite{liu2021coach} framework uses a hierarchical architecture—where the coach encodes the global state using global observations combined with multi-head attention and generates strategy vectors tailored for each player, and each player processes its own local observations with multi-head attention—to adapt to a variable number of agents, supporting coordination and zero-shot generalization when agents dynamically join or leave.

However, we did not further customize HoCs MADL/MARL based on related literature for two reasons. (1) Disasters are extreme and rare events that require immediate response, whereas reinforcement learning–based solutions require time to model the environment and train the model, making them unsuitable for disaster scenarios. (2) In practice, we do not yet know how to integrate attention mechanisms and hierarchical architectures into HoCs MADL/MARL in a way that fits our problem. Furthermore, developing a more deeply customized version of HoCs MADL/MARL for the specific problem addressed in this paper is beyond the scope of our research. We have instead identified and used the latest methods for comparison.
\subsubsection{LLM (Large Language Model)}

Given the excellent generalization properties and immediate reasoning capabilities demonstrated by large language models in handling combinatorial optimization problems, we exploratively designed an online scheduling mechanism based on LLMs \cite{li2024urbangpt, yuan2024unist}. The problem description is converted into natural language input, and the large model generates scheduling plans in real-time. For our problem, the detailed natural language prompts sent to the large model can be referred to in \cite{CocaColaZero}. 

However, after in-depth interaction with multiple mainstream large language models and repeated experiments, we observed obvious limitations in current models when handling multi-constraint scheduling tasks in complex post-disaster environments. Although the models perform well at the language understanding level, they still fall short in integrating task backgrounds, abstracting multi-source constraint conditions, and generating reasonable scheduling plans accordingly. The scheduling behaviors generated by the models often lack logical consistency, and deviations or even obvious errors easily occur in the reasoning chain, making it difficult to meet the actual requirements of such high-complexity tasks.

\begin{table}[htbp]
  \centering
  \setlength{\abovecaptionskip}{0.2cm}
  \setlength{\belowcaptionskip}{-0.25cm}
  \caption{Comparison and Characteristics of Common Combinatorial Optimization Algorithms}
  \label{table_comparison_algs_new}
  \begin{tabular}{
    >{\centering\arraybackslash}p{1.9cm}
    |>{\raggedright\arraybackslash}p{2.6cm}
    |>{\raggedright\arraybackslash}p{3.6cm}
    |>{\raggedright\arraybackslash}p{3.4cm}
    |>{\raggedright\arraybackslash}p{3.5cm}
  }
    \hline
    \textbf{Category} & \textbf{Algorithm} & \textbf{Brief Description} & \textbf{Advantages} & \textbf{Disadvantages} \\
    \hline
    Heuristic Methods & Greedy Algorithm (Greedy) & Selects the local optimum at each step to construct the overall solution. & Fast and simple to implement. & Prone to local optima; unstable solution quality. \\
    \cline{2-5}
    & k-Winner-Take-All (kWTA) & Selects the top k candidates based on response scores. & Efficient and easy to parallelize. & Static selection; does not consider global optimization. \\
    \cline{2-5}
    & Maximum Weighted Independent Set Heuristic Search (Ours) & Builds the optimal task subset by combining node weights and graph feasibility constraints. & Leverages graph structure; robust and suitable for complex constraints. & Depends on graph construction; limited scalability in large graphs. \\
    \hline
    Meta-Heuristic Methods & Genetic Algorithm (GA) & Simulates evolution through selection, crossover, and mutation in the solution space. & Escapes local optima; adaptable to complex spaces. & Slow convergence; many hyperparameters; computationally expensive. \\
    \cline{2-5}
    & Simulated Annealing (SA) & Simulates thermodynamic cooling process to escape local optima. & Strong global search capability; good theoretical foundation. & Slow convergence; sensitive to parameters like temperature schedule. \\
    \hline
    Learning-based Methods & Reinforcement Learning (RL) & Agents learn policies through interaction with the environment. & Suitable for dynamic, high-dimensional tasks; policies are continuously improvable. & Complex learning process; low sample efficiency; high data demand. \\
    \cline{2-5}
    & Deep Reinforcement Learning (Deep RL) & Uses neural networks to represent policies or value functions. & Can handle large state/action spaces; highly flexible. & Unstable convergence; high training cost; lacks interpretability. \\
    \cline{2-5}
    & Graph Neural Networks + RL / Search & Uses graph structure to model task dependencies and guide decision-making. & Captures inter-task relationships; suitable for graph-based combinatorial optimization. & Complex architecture; requires graph-based datasets and training. \\
    \hline
  \end{tabular}
\end{table}

\section{Comparison with Contemporary Deep and Reinforcement Learning Methods}

\subsection{Method Description}
\subsubsection{QMIX}
The QMIX algorithm  represents a significant advancement in multi-agent reinforcement learning (MARL), specifically designed to facilitate cooperation among multiple agents by factorizing the joint action-value function. A key characteristic of QMIX is its emphasis on monotonic factorization, which implies that the global Q-value monotonically increases with respect to each agent's individual Q-value. This property is crucial as it allows for decentralized policy execution while ensuring consistency with a centralized learner, thereby promoting efficient and stable learning in cooperative multi-agent environments.
In the comprehensive comparative experiments conducted against other QMIX-based algorithms, the HoCs-MARL algorithm is utilized for performance evaluation.

\subsubsection{QPLEX} This method augments QMIX's mixing network with a dual-stream architecture, comprising an advantage stream for individual advantage function computation and a state stream for global state encoding. These streams' outputs are connected via a learnable weight matrix.

QPLEX decomposes the joint Q-value into a state value and weighted advantage terms:
\begin{equation*}
Q_{tot}(s, \mathbf{a}) = V_{mix}(s) + \sum_{i=1}^{n} w_i(s, \mathbf{a}) \cdot A_i(o_i, a_i)
\end{equation*}
where:
\begin{itemize}
    \item $A_i(o_i, a_i) = Q_i(o_i, a_i) - V_i(o_i)$
    \item $w_i(s, \mathbf{a}) \ge 0$ represents the output of a weight network, dependent on state and action, satisfying normalization or non-negative constraints.
\end{itemize}

\subsubsection{QTRAN}This approach removes QMIX's monotonicity constraint and introduces a transformation network to map the joint Q-value into the individual Q-value space. Its implementation necessitates three distinct loss functions: 1) Mean Squared Error (MSE) loss between individual Q-values and the transformation network's output; 2) an optimal action selection constraint; and 3) a feasibility condition constraint.

QTRAN introduces an undecomposed joint Q-function $Q_{tot}^{\text{joint}}(s, \mathbf{a})$ and enforces consistency constraints to anchor the decomposed functions to it.

The training loss comprises three components:
\begin{equation*}
\mathcal{L}_{QTRAN} = \mathcal{L}_{TD} + \lambda_1 \mathcal{L}_{opt} + \lambda_2 \mathcal{L}_{nopt}
\end{equation*}
The hyperparameters are $\lambda_1 = 1$ and $\lambda_2 = 0.1$.

\begin{enumerate}
    \item TD Loss (consistent with QMIX) :
    \begin{equation*}
    \mathcal{L}_{TD} = (Q_{tot}(s, \mathbf{a}) - y)^2, \quad y = r + \gamma \cdot \max_{\mathbf{a}'} Q_{tot}^{\text{target}}(s', \mathbf{a}')
    \end{equation*}
    \item Optimality loss :
    \begin{equation*}
    \mathcal{L}_{opt} = \left\| Q_{tot}^{\text{joint}}(s, \mathbf{a}) - \sum_{i=1}^n Q_i(o_i, a_i) \right\|^2 \quad \text{for optimal } \mathbf{a}
    \end{equation*}
    \item Non-optimality loss :
    \begin{equation*}
    \mathcal{L}_{nopt} = \sum_{i} \left\| Q_{tot}^{\text{joint}}(s, (a^{-i},a_i^{\prime})) - \sum_{j \neq i} Q_j(o_j, a_j) - Q_i(o_i, a_i^{\prime}) \right\|^2
    \end{equation*}
\end{enumerate}
Intuitively, for optimal action combinations, local Q-values should approximate the globally optimal Q; conversely, for non-optimal combinations, they must not exceed the true joint Q.

\subsubsection{MF-QMIX} This method replaces QMIX's global state encoder with a mean-field approximation module, which utilizes the average values of neighboring agents as interactive input. Its implementation requires the design of a dynamic graph attention mechanism to accommodate varying numbers of agents and modifications to the experience replay buffer for local observation storage.

In QMIX, each agent's local Q-network primarily considers its own observation $o_i$ and action $a_i$. MF-QMIX, however, integrates a local mean-field mechanism, enabling each agent's Q-value to perceive the average policy behavior of its neighbors.

Consequently, each agent's Q-network in MF-QMIX is formulated as:
\begin{equation*}
Q_i(o_i, a_i, \mu_i)
\end{equation*}
where:
\begin{itemize}
    \item $o_i$: local observation of agent i
    \item $a_i$: action of agent i
    \item $\mu_i$: average policy behavior (mean field) of other agents within agent i's neighborhood
\end{itemize}
The specific form of $\mu_i$ is typically:
\begin{equation*}
\mu_i = \frac{1}{|N(i)|} \sum_{j \in N(i)} \pi_j(a_j | o_j)
\end{equation*}
For a fully connected structure, this simplifies to:
\begin{equation*}
\mu_i = \frac{1}{n-1} \sum_{j \ne i} \pi_j(a_j | o_j)
\end{equation*}
The original QMIX Mixing Network is defined as:
\begin{equation*}
Q_{tot} = f(Q_1, Q_2, ..., Q_n; s)
\end{equation*}
In MF-QMIX, the Mixing Network is modified to:
\begin{equation*}
Q_{tot} = f(Q_1(o_1, a_1, \mu_1), ..., Q_n(o_n, a_n, \mu_n); s)
\end{equation*}
Specifically, the input to each agent's Q-network is augmented with $\mu_i$, enabling the network to model inter-agent influences and thus improve the representation capability for cooperative group strategies.

\subsubsection{CW-QMIX}In its implementation, CW-QMIX preserves QMIX's original value function decomposition structure but incorporates a dynamic weight scheduling mechanism. This mechanism controls the learning weights of different subtasks or samples during training, facilitating curriculum learning. Specifically, when computing the TD loss, samples within the experience replay are assigned a weight $\omega_i$, which is automatically adjusted based on factors such as sample difficulty, training stage, or performance feedback. These dynamic weights can be allocated through a learnable weight network (e.g., a Multi-Layer Perceptron) or heuristic rules, with the weight factor integrated into the training objective:
\begin{equation*}
\mathcal{L} = \sum_{i} \omega_i \cdot (Q_{tot}(\mathbf{a}_i) - y_i)^2
\end{equation*}

\subsubsection{OW-QMIX} 
The central aspect of implementing OW-QMIX lies in adjusting QMIX's TD target term to enhance the efficient utilization of off-policy experiences. OW-QMIX continues to employ QMIX's value function mixing network, but it introduces a weight correction in the target value calculation. This correction accounts for the discrepancy between the behavior policy (behavior distribution) and the target policy, typically achieved using an importance sampling ratio:
\begin{equation*}
y_i = r + \gamma \cdot \rho_i \cdot Q_{tot}^{\text{target}}(s', \mathbf{a}') \quad \text{where } \rho_i = \frac{\pi(\mathbf{a}'|s')}{\mu(\mathbf{a}'|s')}
\end{equation*}
 
\subsubsection{SQIX}
Transitioning from QMIX to SQIX can be achieved through three core steps:
\begin{enumerate}
    \item Replacement of target action selection:
    QMIX's original target formulation:
    \begin{equation*}
    y = r + \gamma \cdot \max_{\mathbf{a}'} Q_{tot}^{\text{target}}(s', \mathbf{a}')
    \end{equation*}
    \item SQIX adopts a softmax-weighted form for the target:
    \begin{equation*}
    y = r + \gamma \cdot \sum_{\mathbf{a}' \in \mathcal{U}} P_{\beta}(\mathbf{a}'|s') Q_{tot}^{\text{target}}(s', \mathbf{a}'), \quad \text{where } P_{\beta}(\mathbf{a}'|s') \propto \exp(\beta Q_{tot}(s', \mathbf{a}')), \text{with parameter } \beta \text{controlling "softness".}
    \end{equation*}
    \item Definition of joint-subspace $\hat{U}^{**}$:
    As discussed in the original paper, this involves selecting only combinations proximate to the current $\text{argmax}^{**}$, such as top-k joint actions, to circumvent exhaustive traversal of the entire action space.
    \item Update of TD loss:
    Following the substitution of $y$, the TD loss continues to be computed as:
    \begin{equation*}
    \mathcal{L} = (Q_{tot} - y)^2
    \end{equation*}
    while the remaining architectural components are preserved.
\end{enumerate}

\subsection{Experiment resullt}

\begin{table}[htbp]
\centering
\setlength{\abovecaptionskip}{0.2cm}
\setlength{\belowcaptionskip}{-0.25cm}
\caption{Comparison of task completion rate across different algorithms.}
\label{tab:comparison_results}
\begin{tabular}{c|cccccccc}
\hline
Dataset & QMIX & CW-QMIX & OW-QMIX & QTRAN & QPLEX & MF-QMIX & SQIX & HoCs-MPQ \\
\hline
Random\_1 & 0.5917 & 0.5833 & 0.5917 & 0.6250 & 0.6333 & 0.5667 & 0.6083 & 0.7292 \\
BicycleOrder & 0.7071 & 0.7172 & 0.7172 & 0.6869 & 0.7475 & 0.7272 & 0.7374 & 0.8596 \\
TaxiTrajectories & 0.4412 & 0.4314 & 0.4510 & 0.4902 & 0.4706 & 0.4510 & 0.4412 & 0.6735 \\
DidiOrders & 0.5882 & 0.5588 & 0.6176 & 0.6765 & 0.6471 & 0.6176 & 0.6176 & 0.7941 \\
\hline
\end{tabular}
\end{table}

As evidenced by Table \ref{tab:comparison_results}, our HoCs-MPQ algorithm consistently achieves superior performance across all datasets, demonstrating a significant improvement in task completion rate compared to the benchmark QMIX-based algorithms. This marked advantage underscores the effectiveness of our approach in complex multi-agent scheduling environments.

Crucially, unlike the other evaluated algorithms which are rooted in deep learning and reinforcement learning paradigms, our method does not rely on extensive historical data for model training. Instead, our approach is fundamentally different: it meticulously models the collaborative and conflicting relationships between multiple heterogeneous agents by constructing a weighted undirected graph. Subsequently, it iteratively solves for the maximum weighted independent set within this graph, leveraging a multi-priority queue. This mechanism enables highly efficient collaborative scheduling. This distinct design provides significant advantages, particularly when addressing real-time problems in dynamic disaster scenarios. Learning-based algorithms typically necessitate large volumes of historical data for effective model training, a requirement that is inherently difficult to fulfill in the context of disaster events, which are by nature extreme and often unpredictable occurrences. Furthermore, deep learning and reinforcement learning methods exhibit a strong dependence on the intricate coupling relationships among multiple agents. Consequently, any changes in the number of agents or their online status can lead to a substantial degradation in the model's decision-making efficacy. Our graph-based approach, by contrast, is inherently more robust to such variations, allowing for better adaptability and real-time performance in unpredictable environments.

\section{Prompt template for interacting with LLM}

\subsection*{Prompt Template: Multi-Agent Scheduling for Post-Disaster Perception}

\subsubsection*{Problem Overview:}
A post-disaster environment requires coordination of three agent types ( UAVs, vehicles, and workers) to complete perception tasks at task points and being charged at charge points.

\subsubsection*{Agents \& Locations:}
\textbf{Agent Types}
\begin{itemize}
    \item UAVs: Primary task executors requiring periodic charging
    \item Workers: Task assistants (must collaborate with UAVs)
    \item Vehicles: Charge assistants (for UAVs)
\end{itemize}
\textbf{Location Types}
\begin{itemize}
    \item Task Points: Require UAV and Worker collaboration
    \item Charge Points: Provide UAV charging via Vehicles
\end{itemize}

\subsubsection*{Valid Combinations:}
\begin{itemize}
    \item UAV Stays Put 
    \item Worker Stays Put 
    \item Vehicle Stays Put 
    \item UAV Moves to Task Point 
    \item Worker Moves to Task Point 
    \item UAV Moves to Charge Point 
    \item Vehicle Moves to Charge Point 
    \item Task Execution: UAV + Worker move to Task Point 
    \item Charging Operation: UAV + Vehicle move to Charge Point 
\end{itemize}

\subsubsection*{Critical Constraints:}
\begin{itemize}
    \item Availability Window: Must operate within agent-specific [up, down] time ranges
    \item State Dependency: Only idle agents can be scheduled
\end{itemize}

\subsubsection*{State Transition Rules:}
\textbf{Task Completion:}
\begin{itemize}
    \item Position Update: UAV \& Worker move to task point
    \item Time Cost: max(movement time) + (T\_costPow/UAV\_speed)
    \item Power Update: UAV\_power -= (movement time + T\_costPow)
\end{itemize}
\textbf{Charging Completion:}
\begin{itemize}
    \item Position Update: UAV \& Vehicle move to charge point
    \item Time Cost: max(movement time) + (Fullpower - current\_power)/V\_chargePow
    \item Power Update: UAV\_power = Fullpower
\end{itemize}

\subsubsection*{Input Data Structure:}
\begin{itemize}
    \item UAV Data: id, x, y, uRge, Fullpower, uPow, U\_uptime, U\_downtime
    \item Worker Data: id, x, y, wRge, W\_uptime, W\_downtime
    \item Vehicle Data: id, x, y, vRge, V\_uptime, V\_downtime, V\_chargePow
    \item Task Data: id, x, y, T\_costPow
    \item Charging Data: id, x, y
\end{itemize}

\subsubsection*{Optimization Objective:}
\begin{itemize}
    \item Time Horizon: 0-180 units (36 decision points @ 5-unit intervals)
    \item Primary Goal: Maximize completed tasks while preventing UAV power depletion
    \item Required Output:
    \begin{itemize}
        \item a) Scheduling decisions at each interval
        \item b) Total completed tasks count
    \end{itemize}
\end{itemize}

\subsubsection*{Current System State:}
[Specific data tables would be inserted here]

\subsubsection*{Please answer:}
Which agent-target pairing combination should be prioritized at the current decision interval to maximize task completion while maintaining UAV survivability?

\subsubsection*{Requirements:}
\begin{itemize}
    \item Choose only ONE combination (Task or Charging)
    \item Follow analysis steps:
    \begin{itemize}
        \item Step 1: Determine whether the UAV is performing a task or going to charge.
        \item Step 2: Evaluate the assistant (worker, vehicle) situation of the UAVs.
        \item Step 3: Select optimal combination based on above priorities
    \end{itemize}
    \item Final answer must use: $<$combination$>$ $;$ $<$combination$>$ $;$ $\ldots$
\end{itemize}

\end{document}